\documentclass[prd,twocolumn,showpacs,floatfix,amsmath,nofootinbib,amssymb,floatfix]{revtex4}
\usepackage{graphicx,dcolumn,booktabs,bm}
\usepackage{longtable,lscape}
\usepackage{txfonts}
\usepackage{overpic}
\usepackage{multirow}
\usepackage{amssymb}
\usepackage{indentfirst}
\usepackage{feynmf}   %{feynmp}
\usepackage{slashed}  %for Feynman symbols
\usepackage{cases}
\usepackage{color,ulem}
\usepackage{graphicx}
\graphicspath{{Figures/}} % 设置图形文件的搜索路径

\begin{document}
%\begin{CJK}{GBK}{}

%%%%%%%%%%%%%%%%%%%%%%%%%%%%%%%%%%%%%%%%%%%%%%%%%%%%%%%%%%%%%%%%%%%%%%%%%%%%%%%%%%%%%%%%%%%%%%%%%%%%%%%%%
%                                    The document begins here                                           %
%%%%%%%%%%%%%%%%%%%%%%%%%%%%%%%%%%%%%%%%%%%%%%%%%%%%%%%%%%%%%%%%%%%%%%%%%%%%%%%%%%%%%%%%%%%%%%%%%%%%%%%%%

\title{Probing the $XYZ$ states through radiative decays}
\author{Li Ma$^{1}$}\email{lima@pku.edu.cn}
\author{Zhi-Feng Sun$^{2,3}$}\email{sunzhif09@lzu.edu.cn}
\author{Xiao-Hai Liu$^{1}$}\email{liuxiaohai@pku.edu.cn}
\author{Wei-Zhen Deng$^{1}$}\email{dwz@pku.edu.cn}
\author{Xiang Liu$^{2,3}$}\email{xiangliu@lzu.edu.cn}
\author{Shi-Lin Zhu$^{1,4}$}\email{zhusl@pku.edu.cn}

\affiliation{
$^1$Department of Physics and State Key Laboratory of Nuclear Physics and Technology and Center of High Energy Physics, Peking University, Beijing 100871, China\\
$^2$Research Center for Hadron and CSR Physics, Lanzhou University and Institute of Modern Physics of CAS, Lanzhou 730000, China\\
$^3$School of Physical Science and Technology, Lanzhou University, Lanzhou 730000, China\\
$^4$Collaborative Innovation Center of Quantum Matter, Beijing
100871, China}

\date{\today}% It is always \today, today,

\begin{abstract}

In this work, we have adopted the spin rearrangement scheme in the
heavy quark limit and extensively investigated three classes of the
radiative decays: $\mathfrak{M}\to (b\bar{b})+\gamma$,
$(b\bar{b})\to \mathfrak{M}+\gamma$, $ \mathfrak{M} \to
\mathfrak{M}^\prime+\gamma$, corresponding to the electromagnetic
transitions between one molecular(resonant) state and bottomonium,
one bottomonium and molecular(resonant) state, and two
molecular(resonant) states respectively. We also extend the same
formalism to study the radiative decays of the molecular(resonant)
states with hidden charm. We have derived some model independent
ratios when the initial or final states belong to the same spin
flavor multiplet. Future experimental measurement of these ratios
will test the molecular picture and explore the underlying
structures of the $XYZ$ states.

\end{abstract}

 \maketitle

%%%%%%%%%%%%%%%%%%%%%%%%%%%%%%%%%%%%%%%%
\section{Introduction}\label{sec1}
%%%%%%%%%%%%%%%%%%%%%%%%%%%%%%%%%%%%%%%%

In the past decade, many charmonium-like and bottomonium-like states
have been observed by the Belle, BaBar, CLEO-c, CDF, D0, CMS, LHCb
and BESIII collaborations \cite{Beringer:1900zz}. These states are
sometimes denoted as $XYZ$ states. Especially, some of them do not
fit into the conventional charmonium or bottomonium spectrum in the
quark model. Their underlying structure, production mechanism and
decay pattern are very intriguing. Up to now, there has been
extensive phenomenological study of these $XYZ$ states (for a recent
review see Refs. \cite{zhu-review,Liu:2013waa}).

In fact, the observation of the $XYZ$ states provides an ideal
platform to investigate the exotic states since those charged $XYZ$
states are clearly not the charmonium or bottomonium states. The
theoretical interpretations of the $XYZ$ states include the hybrid
charmonium, tetraquark states, molecular states, cusp effect, final
state interaction, interference effect or pure phase space effect
etc. Among all the above scenarios, the molecular scheme is
particularly interesting since some of these $XYZ$ states are very
close to the open-charm or open-bottom threshold or even charged.

Since Yukawa proposed the pion as the mediator of the nuclear force,
it's well-known that the deuteron is a very loosely bound state of
the proton and neutron. To a very good extent, the deuteron is a
hadronic molecular state. Therefore, it is very natural to look for
the other loosely bound systems composed of two hadrons. In the past
several decades, there have been lots of investigations of the
di-meson systems composed of two charmed/bottomed mesons. The
existence of the loosely bound hadronic molecular states depends on
the competition between the kinetic energy and potential in the
Hamiltonian of the system. The presence of the heavy quarks lowers
the kinetic energy while the chiral interaction between the two
light quarks could still provide strong enough attraction.

Voloshin and Okun studied the interaction between a pair of the
charmed mesons and the possible hadronic molecular states
\cite{Voloshin:1976ap}. Rujula, Geogi and Glashow proposed that
$\psi(4040)$ was a $D^*\bar{D}^*$ molecular state \cite{De
Rujula:1976qd}. T\"{o}rnqvist calculated the possible deuteron-like
two-meson bound states such as $D\bar{D}^*$ and $D^*\bar{D}^*$ using
the quark-pion interaction model
\cite{Tornqvist:1993vu,Tornqvist:1993ng}. Later, Dubynskiy and
Voloshin suggested that there might exist a possible new resonance
at the $D^*\bar{D}^*$ threshold
\cite{Voloshin:2006pz,Dubynskiy:2006sg}. In the light quark sector,
Weinstein and Isgur suggested the scalar resonances $f_0(980)$ and
$a_0(980)$ as the $K\bar K$ molecular states
\cite{Weinstein:1982gc,Weinstein:1983gd,Weinstein:1990gu}.

The $XYZ$ states stimulated new interest in the molecular states
composed of a pair of heavy mesons. The possibility of X(3872) as
the $D\bar{D}^\ast$ molecular state was discussed in Refs.
\cite{Close:2003sg,Voloshin:2003nt,Wong:2003xk,Swanson:2003tb,Tornqvist:2004qy,Suzuki:2005ha,Liu:2008fh,Thomas:2008ja,Lee:2009hy,Li:2012cs}.
Liu and Zhu proposed $Y(3940)$ and $Y(4140)$ as the $D^*\bar{D}^*$
and $D_s^*\bar{D}_s^*$ molecular candidates respectively
\cite{Liu:2009ei,Liu:2008tn}, which were further studied in Refs.
\cite{Mahajan:2009pj,Branz:2009yt,Albuquerque:2009ak,Ding:2009vd,Zhang:2009st,Liu:2009iw,Liu:2009pu}.
$Y(4274)$ as the S-wave $D_s\bar{D}_{s0}(2317)+h.c.$ molecular state
was proposed in Refs. \cite{Liu:2010hf,He:2011ed}. In Refs.
\cite{Meng:2007fu,Liu:2007bf,Liu:2008xz}, the authors discussed
whether $Z^+(4430)$ is a loosely bound S-wave state of
$D^*\bar{D}_1$ or $D^*\bar{D}_1^\prime$ with $J^P=0^-,1^-,2^-$.
Later, the observations of two charged charmonium-like states
$Z^+(4051)$ and $Z^+(4248)$ also inspired the discussion of whether
$Z^+(4051)$ and $Z^+(4248)$ can be the molecular states
\cite{Liu:2008mi,Ding:2008gr}. Before the experimental observation
of $Z_b(10610)$ and $Z_b(10650)$, it was pointed out that there
probably exist loosely bound S-wave $B\bar{B}^{*}$ or
$B^*\bar{B}^{*}$ molecular states \cite{Liu:2008fh,Liu:2008tn}.
Later, these two charged bottomonium-like states $Z_b(10610)$ and
$Z_b(10650)$ attracted lots of attention in the molecular framework
\cite{Bondar:2011ev,Zhang:2011jja,Sun:2011uh,Sun:2012zzd,Guo:2013sya}.
In Refs. \cite{Sun:2011uh,Sun:2012zzd}, the possible charged
charmonium-like molecular states were explored. Later, the
observation $Z_c(3900)$ \cite{Ablikim:2013mio} again inspired lots
of theoretical work
\cite{Wang:2013cya,Bondar:2013jxa,Guo:2013sya,Cui:2013yva,Zhang:2013aoa,Dias:2013xfa,Wang:2013vex,mali}.
In addition, $Y(4260)$ was proposed as a candidate of the
$D_1\bar{D}$ molecular state \cite{Ding:2008gr,Wang:2013kra} while
the newly observed $Z_c(4025)$ also as the hadronic molecule
candidate \cite{He:2013nwa}.

Generally speaking, the dynamical calculation using the
phenomenological models such as the one boson exchange model may
answer whether there exists the loosely bound state. On the other
hand, the production mechanism and decay pattern will also shed
light on the inner structure of the $XYZ$ states. There were some
discussions about the decay of $Z_c(3900)$ \cite{Braaten:2013boa},
and the decays and productions of $Z_b(10610)/Z_b(10650)$ using the
re-coupling formulae of angular momentum \cite{Ohkoda:2012rj}. Liu
studied the heavy quark spin selection rules in the meson-antimeson
states \cite{Liu:2013rxa}. In Ref. \cite{He:2013nwa}, the authors
discussed the decay behavior of $Z_c(4025)$ as the $D^*\bar{D}^*$
molecular state.

Besides dynamical calculation, spin rearrangement scheme based on
the heavy quark symmetry provides another approach to investigate
the decay and production behaviors of the XYZ states. For instance,
in Ref. \cite{Voloshin:2011qa}, the authors studied the relations
between the rates of the radiative transitions of $\Upsilon(5S)$ to
the hypothetical hidden-bottom isovector molecular resonances with
negative G parity through spin recoupling method in the heavy quark
limit. In this work, we adopt extend the spin rearrangement scheme
to discuss the systems containing a P-wave bottom meson.

We will perform a comprehensive study of the radiative decay pattern
of the heavy flavor molecular states or resonant states. In our
present work, we do not introduce any dynamical models. We will not
focus on whether there exist molecular states composed of
heavy-light mesons and how two components form the molecular states.
This issue can only be solved by the concrete dynamical calculation.
For example, one can extract the effective potential and solve the
Schrodinger equation to see whether a loosely bound molecular state
exists or not. In fact, the main task of this work is to study the
decay or production behavior of the systems composed of one P-wave
bottom meson and one S-wave bottom meson or two S-wave bottom
mesons. The symmetry analysis presented in this manuscript is based
the heavy quark spin-flavor symmetry only. Therefore, such an
analysis does not depend on the details whether this system is a
bound molecular state or mixture of the molecule and two meson
states.

We choose the heavy flavor molecular states or resonant states
composed of a pair of S-wave and P-wave bottom mesons to illustrate
the formalism. We use the notations $\mathfrak{M}^{(\prime)}$ and
$(b\bar{b})$ to denote molecular(resonant) states and bottomonium,
respectively. We will consider the following three classes of
radiative decays
\begin{eqnarray*}
\mathfrak{M}&\to &(b\bar{b})+\gamma,\\
(b\bar{b})&\to& \mathfrak{M}+\gamma,\\
\mathfrak{M} &\to& \mathfrak{M}^\prime+\gamma
\end{eqnarray*}
corresponding to the electromagnetic transitions between one
molecular(resonant)state and bottomonium, one bottomonium and
molecular(resonant) state, and two molecular(resonant) states
respectively.

We also extend the same formalism to study the radiative decays of
the molecular(resonant) states with hidden charm. We will discuss
the radiative decay patterns of the hidden-charm molecular states.
These results are helpful to further test the molecular assignment
of some charmonium-like states such as $X(3872)$, $Y(4260)$ and
$Y(4360)$.

Radiative decays can be used to probe the inner structure of $XYZ$
states and identify their quantum numbers. For instance, the BaBar
Collaboration observed the decay mode $X(3872)\rightarrow
J/\psi\gamma$ and determined the $C$ parity of $X(3872)$
\cite{Aubert:2006aj}. 
Radiative decays can be used to probe the inner structure of $XYZ$
states and identify their quantum numbers. For instance, BABAR
Collaboration observed the decay mode $X(3872)\rightarrow
J/\psi\gamma$ and determined the $C$ parity of $X(3872)$
\cite{Aubert:2006aj}. The radiative decay behaviors of the $XYZ$
states can provide useful information to distinguish different
theoretical explanations for the $XYZ$ states. For example, $Y(4260)$
was suggested to be a $\frac{1}{\sqrt{2}}(D_1\bar{D}-D\bar{D}_1)$
molecular state \cite{Ding:2008gr} or a conventional $c\bar{c}$
state $\psi(4S)$ \cite{Li:2009zu}. Its decay modes
$\eta_c\gamma(M1)$ and $\eta_{c2}(1^1D_2)\gamma(M1)$ can help us to
judge whether the molecule explanation of $Y(4260)$ is favored or
not.

This paper is organized as follows. After introduction, we give the
calculation details of the radiative decays of the heavy molecular
states in Sec. \ref{sec2}. In Sec. \ref{sec3}, we present the
numerical results of the above three classes of radiative
transitions associated with the hidden-bottom molecular(resonant)
states. We collect the numerical results of the radiative decays of
the possible hidden-charm molecules(resonances) in Sec. \ref{sec4}.
The last section is the summary.

%%%%%%%%%%%%%%%%%%%%%%%%%%%%%%%%%%%%%%%%%%%%%%%%
\section{The formalism of the radiative decays in the heavy quark limit}
\label{sec2}
%%%%%%%%%%%%%%%%%%%%%%%%%%%%%%%%%%%%%%%

Heavy quark symmetry plays an important role in studying the
properties of hadrons containing heavy quarks. In the heavy quark
limit $m_Q\to \infty$, heavy quarks can be seen as a static color
source and only interact with gluons via its chromoelectric charge
in Quantum Chromodynamics (QCD). This kind of interaction is
invariant under the spin transformation of heavy quarks. The
spin-dependent interaction is proportional to the chromomagnetic
moment of heavy quarks. However, this potentially spin-flipping
chromomagnetic interaction is suppressed by $1/m_{Q_i}$. When
$m_Q\to \infty$, the spin of heavy quarks $S_H$ is conserved.
Similar conclusions hold for the electromagnetic interaction of the
heavy quarks in quantum electrodynamics (QED). The heavy quark spin
symmetry is very useful in the study of heavy hadron decays.

For the heavy hadron, its total angular momentum $J$ can be
decomposed into the sum of the heavy quark spin $S_H$ and the spin
of the light degrees of freedom, which is defined as
$\vec{S}_l\equiv \vec{J}-\vec{S}_H$. $\vec{S}_l$ is also a conserved
operator and includes the spin of light quark and all orbital
angular momenta within a hadron. In the following we simply denote
this operator as the "light spin" since it includes all degrees of
freedom except the spin of heavy quarks, which is named as "heavy
spin".

When considering the heavy flavor molecular(resonant) state system,
its light degrees of freedom are quite complicated. For instance, an
S-wave molecular(resonant) state $B_1\bar{B}$ is composed of a
P-wave $\bar{b}q$ meson and an S-wave $b\bar{q}$ antimeson. Here,
the spin of light quarks $q$ and $\bar{q}$, and the orbital angular
momentum between the heavy quark $\bar{b}$ and light quark $q$ are
not conserved separately. However, their sum is the conserved light
spin $\vec{S}_l$.

This work covers two kinds of hidden beauty systems. The first one
is composed of a P-wave bottom meson $B_0$, $B_1'$, $B_1$, $B_2$ and
a S-wave bottom meson $\bar{B}$, $\bar{B}^\ast$. The second one is
composed of two S-wave bottom mesons. $B$ and $B^\ast$ belong to the
$(0^-,1^-)$ doublet and $B_0$ and $B_1'$ belong to the $(0^+,1^+)$
doublet while $B_1$ and $B_2$ belong to the $(1^+,2^+)$ doublet in
the heavy quark limit. Following the definition of the $C$-parity
eigenstate of the molecular(resonant) states in Ref.
\cite{Liu:2013rxa}, we list all the hidden beauty
molecular(resonant) states covered in this work in Table
\ref{state}. In the following we will apply the heavy quark symmetry
to discuss the radiative decays of the $XYZ$ states.

\renewcommand{\arraystretch}{1.5}
\begin{table}[htbp]
  \caption{The hidden beauty molecular(resonant) states and their $J^{PC}$ quantum numbers.\label{state}}
\begin{center}
   \begin{tabular}{ c c c c } \toprule[1pt]
   $J^{PC}$&\multicolumn{3}{c}{States}\\
   \midrule[1pt]

     \multirow{2}{*}{$1^{--}$} & $\frac{1}{\sqrt{2}}(B_0\bar{B}^\ast-B^\ast\bar{B}_0)$ &$\frac{1}{\sqrt{2}}(B_1'\bar{B}-B\bar{B}_1')$ & $\frac{1}{\sqrt{2}}(B_1\bar{B}-B\bar{B}_1)$ \\
        &$\frac{1}{\sqrt{2}}(B_1'\bar{B}^\ast+B^\ast\bar{B}_1')$ & $\frac{1}{\sqrt{2}}(B_1\bar{B}^\ast+B^\ast\bar{B}_1)$  & $\frac{1}{\sqrt{2}}(B_2\bar{B}^\ast-B^\ast\bar{B}_2)$ \\
      \multirow{2}{*}{$1^{-+}$} &  $\frac{1}{\sqrt{2}}(B_0\bar{B}^\ast+B^\ast\bar{B}_0)$ & $\frac{1}{\sqrt{2}}(B_1'\bar{B}+B\bar{B}_1')$ & $\frac{1}{\sqrt{2}}(B_1\bar{B}+B\bar{B}_1)$ \\
      & $\frac{1}{\sqrt{2}}(B_1'\bar{B}^\ast-B^\ast\bar{B}_1')$
      & $\frac{1}{\sqrt{2}}(B_1\bar{B}^\ast-B^\ast\bar{B}_1)$ & $\frac{1}{\sqrt{2}}(B_2\bar{B}^\ast+B^\ast\bar{B}_2)$ \\
      $1^{++}$ &   $\frac{1}{\sqrt{2}}(B\bar{B}^\ast+B^\ast\bar{B})$ &  &  \\
      $1^{+-}$ &  $\frac{1}{\sqrt{2}}(B\bar{B}^\ast-B^\ast\bar{B})$ & $B^\ast\bar{B}^\ast$ &   \\
      {$0^{--}$} & $\frac{1}{\sqrt{2}}(B_0\bar{B}-B\bar{B}_0)$ &$\frac{1}{\sqrt{2}}(B_1'\bar{B}^\ast-B^\ast\bar{B}_1')$ & $\frac{1}{\sqrt{2}}(B_1\bar{B}^\ast-B^\ast\bar{B}_1)$ \\
      $0^{-+}$ &  $\frac{1}{\sqrt{2}}(B_0\bar{B}+B\bar{B}_0)$ &$\frac{1}{\sqrt{2}}(B_1'\bar{B}^\ast+B^\ast\bar{B}_1')$ & $\frac{1}{\sqrt{2}}(B_1\bar{B}^\ast+B^\ast\bar{B}_1)$ \\
      $0^{++}$ &  $B\bar{B}$ & $B^\ast\bar{B}^\ast$ &   \\
      \multirow{2}{*}{$2^{--}$} & $\frac{1}{\sqrt{2}}(B_1'\bar{B}^\ast-B^\ast\bar{B}_1')$ &$\frac{1}{\sqrt{2}}(B_1\bar{B}^\ast-B^\ast\bar{B}_1)$ & $\frac{1}{\sqrt{2}}(B_2\bar{B}-B\bar{B}_2)$ \\
        &$\frac{1}{\sqrt{2}}(B_2\bar{B}^\ast+B^\ast\bar{B}_2)$ &  &  \\
      \multirow{2}{*}{$2^{-+}$} &  $\frac{1}{\sqrt{2}}(B_1'\bar{B}^\ast+B^\ast\bar{B}_1')$ &$\frac{1}{\sqrt{2}}(B_1\bar{B}^\ast+B^\ast\bar{B}_1)$ & $\frac{1}{\sqrt{2}}(B_2\bar{B}+B\bar{B}_2)$ \\
        &$\frac{1}{\sqrt{2}}(B_2\bar{B}^\ast-B^\ast\bar{B}_2)$ &  &  \\
      $2^{++}$ &   $B^\ast\bar{B}^\ast$ &  &  \\\bottomrule[1pt]
       \end{tabular}
 \end{center}
\end{table}

%%%%%%%%%%%%%%%%%%%%%%%%%%%%%%%%%
\subsection{The spin structures of the hidden beauty molecular(resonant) states and bottomonia}\label{sec21}
%%%%%%%%%%%%%%%%%%%%%%%%%%%%%%%%%

In the radiative decays of the hidden beauty molecules, the heavy
spin, light spin and total angular momentum are good quantum numbers
in the heavy quark limit, which are separately conserved. Therefore
we can decompose the total angular momentum of the initial and final
states according to their heavy spin and light spin.

With the heavy quark spin symmetry, the heavy and light spins of
molecular(resonant) state can be re-coupled separately. We adopt the
spin re-coupling formula with 6-$j$ or 9-$j$ symbols in analyzing
the general spin structure. Let's take the molecular(resonant)
states composed of an S-wave bottom antimeson and a P-wave bottom
meson as an example. We explicitly write out its spin structure
\begin{eqnarray}
 && |B_{1(2)}\bar{B}^{(\ast)}\rangle\nonumber\\& &= \bigg[[ \bar{b}\otimes(q\otimes 1)_s]_K\otimes[b\otimes \bar{q}]_L\bigg]_J |(\bar{b}q)(b\bar{q})\rangle  \nonumber\\
  &&= \sum_{g=0}^1 \sum_{m=0}^1 \sum_{h=|s-\frac{1}{2}|}^{s+\frac{1}{2}} (-1)^{s+\frac{3}{2}+m}\Big[(2s+1)(2m+1)\nonumber\\
  &&\quad\times (2K+1)(2L+1)(2g+1)(2h+1)\Big]^{1/2}\nonumber\\
  &&\quad\times \left\{
             \begin{array}{ccc}
               \frac{1}{2} & s & K \\
               \frac{1}{2} & \frac{1}{2} & L \\
               g & h & J \\
             \end{array}
           \right\}
  \left\{
    \begin{array}{ccc}
      \frac{1}{2} & 1 & s \\
      h & \frac{1}{2} & m \\
    \end{array}
  \right\} \nonumber\\
  &&\quad\times\Bigg|\left[\left[\bar{b}b\right]_g\otimes\left(\left[q\bar{q}\right]_m\otimes1\right)_h\right]_J \Bigg\rangle |(\bar{b}q)(b\bar{q})\rangle\nonumber\\
  &&=\sum_{g=0}^1 \sum_{m=0}^1 \sum_{h=|s-\frac{1}{2}|}^{s+\frac{1}{2}} \mathcal{B}^{s,L,K,J}_{g,m,h}\Bigg|\left[\left[\bar{b}b\right]_g\otimes\left(\left[q\bar{q}\right]_m\otimes1\right)_h\right]_J \Bigg\rangle |(\bar{b}q)(b\bar{q})\rangle.\nonumber\\\label{h1}
 \end{eqnarray}
In the above equation, $s$ and $K$ denote the light spin and total
angular momentum of the P-wave bottom meson respectively. $L$ is the
total angular momentum of the S-wave bottom meson. $J$ is the total
angular momentum of the molecular(resonant) state. We are only
interested in the S-wave molecular(resonant) systems in this work.
The orbital angular momentum between the two bottom mesons is zero.

In addition, the indices $b$, $\bar{b}$, $q$ and $\bar{q}$ in the
square brackets in Eq. (\ref{h1}) represent the corresponding quark
spin wave functions. The notation
$[[\bar{b}b]_g\otimes([q\bar{q}]_m\otimes1)_h]_J$ means that the
spins of the $\bar{b}$ and $b$ quarks are coupled into the heavy
quark spin $g$ and the spins of the $q$ and $\bar{q}$ quarks are
coupled into the light quark spin $m$. And then $m$ couples with the
orbital angular momentum to form the light spin $h$. We need to
emphasize that we explicitly include the flavor wave function
$|(\bar{b}q)(b\bar{q})\rangle$ in Eq. (\ref{h1}).

The spin structure of the $|B^{(\ast)}\bar{B}_{1(2)}\rangle$ can be
obtained by applying the $C$-parity transformation to
$|B_{1(2)}\bar{B}^{(\ast)}\rangle$, i.e.,
\begin{eqnarray}
 && |B^{(\ast)}\bar{B}_{1,2}\rangle\nonumber\\ &&= \bigg[[\bar{b}\otimes q]_L\otimes[b\otimes(\bar{q}\otimes 1)_s]_K\bigg]_J |(b\bar{q})(\bar{b}q)\rangle\nonumber\\
  &&= (-1)^{L+K-J} \sum_{g=0}^1 \sum_{m=0}^1 \sum_{h=|s-\frac{1}{2}|}^{s+\frac{1}{2}} (-1)^{s+\frac{3}{2}+g}[(2s+1)(2m+1)\nonumber\\
  &&\quad\times(2K+1)(2L+1)(2g+1)(2h+1)]^{1/2}\nonumber\\
  &&\quad\times \left\{
             \begin{array}{ccc}
               \frac{1}{2} & s & K \\
               \frac{1}{2} & \frac{1}{2} & L \\
               g & h & J \\
             \end{array}
           \right\}
  \left\{
    \begin{array}{ccc}
      \frac{1}{2} & 1 & s \\
      h & \frac{1}{2} & m \\
    \end{array}
  \right\} \nonumber\\
  &&\quad\times\Bigg|\left[\left[\bar{b}b\right]_g\otimes\left(\left[q\bar{q}\right]_m\otimes1\right)_h\right]_J\Bigg\rangle  |(b\bar{q})(\bar{b}q)\rangle\nonumber\\
  &&=\sum_{g=0}^1 \sum_{m=0}^1 \sum_{h=|s-\frac{1}{2}|}^{s+\frac{1}{2}} \mathcal{C}^{s,L,K,J}_{g,m,h}\Bigg|\left[\left[\bar{b}b\right]_g\otimes\left(\left[q\bar{q}\right]_m\otimes1\right)_h\right]_J \Bigg\rangle |({b}\bar{q})(\bar{b}{q})\rangle,\nonumber\\\label{h2}
 \end{eqnarray}
where the factor $(-1)^{L+K-J}$ arises from the exchange of the two
bottoms in the molecular(resonant) state.

With the above approach, we obtain the re-coupled spin structures of
these hidden beauty molecular(resonant) states, where the
coefficients $\mathcal{B}^{s,L,K,J}_{g,m,h}$ and
$\mathcal{C}^{s,L,K,J}_{g,m,h}$ in Eqs. (\ref{h1})-(\ref{h2}) are
collected in Table \ref{state} (see Table \ref{recouple} for more
details).

\renewcommand{\arraystretch}{1.4}
\begin{table*}[htbp]
\caption{The coefficients $\mathcal{B}^{s,L,K,J}_{g,m,h}$ and
$\mathcal{C}^{s,L,K,J}_{g,m,h}$ in Eqs. (\ref{h1})-(\ref{h2}), which
depend on the values $[g,m,h]$ and the corresponding
molecular(resonant) states. The combination $[g,m,h]$ corresponds to
the subscripts in
$[\bar{b}b]_g\otimes([q\bar{q}]_m\otimes1)_h$.\label{recouple}}
\begin{center}
   \begin{tabular}{cccc|ccccccc|cccccccccccccccc} \toprule[1pt]
      \multicolumn{4}{c|}{$J^P=0^-$}&\multicolumn{7}{c|}{$J^P=1^-$}&\multicolumn{5}{c}{$J^P=2^-$}&\\ \midrule[1pt]
                 &  $[0,1,0]$ &  $[1,0,1]$  & $[1,1,1]$&& $[0,0,1]$  & $[0,1,1]$ & $[1,1,0]$ & $[1,0,1]$  & $[1,1,1]$ & $[1,1,2]$&& $[0,1,2]$  & $[1,0,1]$  & $[1,1,1]$ & $[1,1,2]$ \\
      \midrule[1pt]
       $B_1'\bar{B}^\ast$      & $-\frac{\sqrt{3}}{2}$    & $-\frac{\sqrt{3}}{6}$ & $-\frac{\sqrt{6}}{6}$&$B_1^\prime\bar{B}$  &   $-\frac{\sqrt{3}}{6}$    & $-\frac{\sqrt{6}}{6}$  & $-\frac{1}{2}$  & $\frac{\sqrt{6}}{6}$&  $\frac{\sqrt{3}}{3}$ & 0 &$B_1'\bar{B}^\ast$     & $0$    & $\frac{\sqrt{3}}{3}$  & $\frac{\sqrt{6}}{3}$  & $0$\\
       $B^\ast\bar{B}_1'$     & $\frac{\sqrt{3}}{2}$    & $\frac{\sqrt{3}}{6}$ & $-\frac{\sqrt{6}}{6}$&$B\bar{B}_1^\prime$  &   $-\frac{\sqrt{3}}{6}$    & $\frac{\sqrt{6}}{6}$  & $-\frac{1}{2}$  & $-\frac{\sqrt{6}}{6}$ & $\frac{\sqrt{3}}{3}$ & 0 &$B^\ast\bar{B}_1'$     & $0$    & $-\frac{\sqrt{3}}{3}$  & $\frac{\sqrt{6}}{3}$  & $0$ \\
       $B_1\bar{B}^\ast$     & $0$    & $-\frac{\sqrt{6}}{3}$ & $\frac{\sqrt{3}}{3}$&$B_1\bar{B}$   &   $\frac{\sqrt{6}}{6}$    & $-\frac{\sqrt{3}}{6}$  & 0 & $\frac{\sqrt{3}}{6}$ & $-\frac{\sqrt{6}}{12}$ & $\frac{\sqrt{10}}{4}$&$B_1\bar{B}^\ast$     & $\frac{\sqrt{6}}{4}$    & $-\frac{\sqrt{6}}{12}$  & $\frac{\sqrt{3}}{12}$  & $\frac{3}{4}$  \\
       $B^\ast\bar{B}_1$     & $0$    & $\frac{\sqrt{6}}{3}$ & $\frac{\sqrt{3}}{3}$&$B\bar{B}_1$  &   $\frac{\sqrt{6}}{6}$    & $\frac{\sqrt{3}}{6}$  & 0 & $-\frac{\sqrt{3}}{6}$ & $-\frac{\sqrt{6}}{12}$ & $\frac{\sqrt{10}}{4}$&$B^\ast\bar{B}_1$     & $-\frac{\sqrt{6}}{4}$    & $\frac{\sqrt{6}}{12}$  & $\frac{\sqrt{3}}{12}$  & $\frac{3}{4}$  \\
       $B_0\bar{B}$     & $-\frac{1}{2}$    & $\frac{1}{2}$ & $\frac{\sqrt{2}}{2}$&$B_0\bar{B}^\ast$   &   $\frac{\sqrt{3}}{6}$    & $\frac{\sqrt{6}}{6}$  & $\frac{1}{2}$  & $\frac{\sqrt{6}}{6}$ & $\frac{\sqrt{3}}{3}$ & 0 &$B_2\bar{B}$     & $-\frac{1}{2}$    & $\frac{1}{2}$  & $-\frac{\sqrt{2}}{4}$  & $\frac{\sqrt{6}}{4}$  \\
       $B\bar{B}_0$     & $\frac{1}{2}$    & $-\frac{1}{2}$ & $\frac{\sqrt{2}}{2}$&$B^\ast\bar{B}_0$  &   $\frac{\sqrt{3}}{6}$    & $-\frac{\sqrt{6}}{6}$  & $\frac{1}{2}$  & $-\frac{\sqrt{6}}{6}$ & $\frac{\sqrt{3}}{3}$ & 0 &$B\bar{B}_2$    & $\frac{1}{2}$    & $-\frac{1}{2}$  & $-\frac{\sqrt{2}}{4}$  & $\frac{\sqrt{6}}{4}$ \\
       &&&&$B_1^\prime\bar{B}^\ast$ &   $\frac{\sqrt{6}}{6}$    & $\frac{\sqrt{3}}{3}$  & $-\frac{\sqrt{2}}{2}$  & 0 & 0 & 0  &$B_2\bar{B}^\ast$     & $\frac{\sqrt{6}}{4}$    & $\frac{\sqrt{6}}{4}$  & $-\frac{\sqrt{3}}{4}$  & $-\frac{1}{4}$\\
       &&&&$B^\ast\bar{B}_1^\prime$  &   $-\frac{\sqrt{6}}{6}$    & $\frac{\sqrt{3}}{3}$  & $\frac{\sqrt{2}}{2}$  & 0 & 0 & 0  &$B^\ast\bar{B}_2$     & $\frac{\sqrt{6}}{4}$    & $\frac{\sqrt{6}}{4}$  & $\frac{\sqrt{3}}{4}$  & $\frac{1}{4}$\\
       &&&&$B_1\bar{B}^\ast$  &   $\frac{\sqrt{3}}{6}$    & $-\frac{\sqrt{6}}{12}$  & 0  & $\frac{\sqrt{6}}{4}$ & $-\frac{\sqrt{3}}{4}$ & $-\frac{\sqrt{5}}{4}$&&&&& \\
       &&&&$B^\ast\bar{B}_1$  &   $-\frac{\sqrt{3}}{6}$    & $-\frac{\sqrt{6}}{12}$  & 0  & $\frac{\sqrt{6}}{4}$ & $\frac{\sqrt{3}}{4}$ & $\frac{\sqrt{5}}{4}$&&&&& \\
       &&&&$B_2\bar{B}^\ast$  &   $\frac{\sqrt{15}}{6}$    & $-\frac{\sqrt{30}}{12}$  & 0  & $-\frac{\sqrt{30}}{12}$ & $\frac{\sqrt{15}}{12}$ & $-\frac{1}{4}$&&&&& \\
       &&&&$B^\ast\bar{B}_2$  &   $\frac{\sqrt{15}}{6}$    & $\frac{\sqrt{30}}{12}$  & 0  & $\frac{\sqrt{30}}{12}$ & $\frac{\sqrt{15}}{12}$ & $-\frac{1}{4}$&&&&& \\\bottomrule[1pt]
       \end{tabular}
       \begin{tabular}{ccccc|ccccc|cccccccccccccccc} \toprule[1pt]
       \multicolumn{5}{c|}{$J^P=0^+$}&\multicolumn{5}{c|}{$J^P=1^+$}&\multicolumn{5}{c}{$J^P=2^+$}&\\
        \midrule[1pt]
         & $[0,0,0]$ & $[0,1,1]$  &  $[1,0,0]$ &  $[1,1,1]$ && $[0,0,0]$ & $[0,1,1]$  &  $[1,0,0]$ &  $[1,1,1]$& & $[0,0,0]$ & $[0,1,1]$  &  $[1,0,0]$ &  $[1,1,1]$\\
        \midrule[1pt]
        $B\bar{B}$     & $\frac{1}{2}$   & 0 & 0   & $\frac{\sqrt{3}}{2}$&$B\bar{B}^\ast$   &0 & $\frac{1}{2}$    & $-\frac{1}{2}$  & $\frac{\sqrt{2}}{2}$&$B^\ast\bar{B}^\ast$     & $0$    & $0$  & $0$ & $1$ \\
      $B^\ast\bar{B}^\ast$     & $\frac{\sqrt{3}}{2}$  & 0 & 0  & $-\frac{1}{2}$&$B^\ast\bar{B}$   &0  & $-\frac{1}{2}$    & $\frac{1}{2}$  & $\frac{\sqrt{2}}{2}$   \\
      &&&&&$B^\ast\bar{B}^\ast$  &0  & $\frac{\sqrt{2}}{2}$    & $\frac{\sqrt{2}}{2}$  & 0&&&&& \\
       \bottomrule[1pt]
   \end{tabular}
   \end{center}
\end{table*}

Besides the spin structures of these hidden beauty
molecular(resonant) states, we also need the spin structures of the
bottomonia relevant to the radiative decays, i.e.,
 \begin{eqnarray}
   |\eta_b(1^1S_0) \rangle &=& |(0_H^-\otimes0_l^+)_0^{-+}\rangle|(b\bar{b})\rangle,\label{k1}\\
  | \Upsilon(1^3S_1)\rangle &=& |(1_H^-\otimes0_l^+)_1^{--}\rangle| |(b\bar{b})\rangle,\\
  | h_b(1^1P_1)\rangle &=&| (0_H^-\otimes1_l^-)_1^{+-}\rangle|(b\bar{b})\rangle,\\
  | \chi_{b0}(1^3P_0) \rangle&=&| (1_H^-\otimes1_l^-)_0^{++}\rangle|(b\bar{b})\rangle,\\
   |\chi_{b1}(1^3P_1) \rangle&=&| (1_H^-\otimes1_l^-)_1^{++}\rangle|(b\bar{b})\rangle,\\
  | \chi_{b2}(1^3P_2)\rangle &=&| (1_H^-\otimes1_l^-)_2^{++}\rangle|(b\bar{b})\rangle,\\
   |\eta_{b2}(1^1D_2)\rangle &=&| (0_H^-\otimes2_l^+)_2^{-+}\rangle|(b\bar{b})\rangle,\\
   |\Upsilon(1^3D_1)\rangle &=&| (1_H^-\otimes2_l^+)_1^{--}\rangle|(b\bar{b})\rangle,\\
  | \Upsilon(1^3D_2)\rangle &=&| (1_H^-\otimes2_l^+)_2^{--}\rangle|(b\bar{b})\rangle,\\
  | \Upsilon(1^3D_3)\rangle &=&| (1_H^-\otimes2_l^+)_3^{--}\rangle|(b\bar{b})\rangle,\label{k2}
  \end{eqnarray}
where we use the subscripts $H$ and $l$ to distinguish the heavy and
light spins of bottomonia. The superscripts $+$ and $-$ inside the
parentheses denote the positive and negative parity of the
corresponding parts, respectively,  while the superscripts $-+$ and
subscripts $0,1,2,3$ correspond to the quantum numbers $PC$ and $J$
of $J^{PC}$ of the bottomonium. When calculating the radiative
decay, we also need the bottomonium flavor wave function
$|(b\bar{b})\rangle\equiv\frac{1}{\sqrt{2}}(|\bar{b}b\rangle+|b\bar{b}\rangle)$,
which is invariant under the $C$ transformation. The $C$ parity of
the bottomonium is reflected through its spin and orbital wave
function, i.e., $C=(-1)^{S_H+S_l}$.

For the spin wave function, the orthogonalization requires
  \begin{eqnarray}
   \langle(a_H\otimes b_L)_J^{pc}|(c_H\otimes d_L)_{J'}^{p'c'}\rangle =\delta_{ac}\delta_{bd}\delta_{JJ'}\delta_{pp'}\delta_{cc'},
  \end{eqnarray}
where the superscripts $p^{(\prime)}$ and $c^{(\prime)}$ denotes the
parity and $C$ parity, respectively. The requirement of the
orthogonalization of the spin wave function results in the
conservation of the parity, $C$ parity, the total angular momentum,
heavy spins, and light spins respectively.

In addition, the orthogonalization of the flavor wave function
requires
\begin{eqnarray*}
    \langle (\bar{b}q)(b\bar{q})|(\bar{b}q)(b\bar{q})\rangle &=& 1, \quad
    \langle (b\bar{q})(\bar{b}q)|(\bar{b}q)(b\bar{q})\rangle = 0, \\
    \langle (b\bar{q})(\bar{b}q)|(b\bar{q})(\bar{b}q)\rangle &=& 1, \quad
    \langle( \bar{b}q)(b\bar{q})|(b\bar{q})(\bar{b}q)\rangle = 0,
\end{eqnarray*}
which are adopted in the calculation of the electromagnetic
transitions between two hidden beauty molecular(resonant) states.

In the following sections, we will employ the above spin structures
and flavor wave functions of the hidden beauty molecular(resonant)
states and bottomonia to calculate the radiative decays. The spin
structure of the photon will be given later.

%%%%%%%%%%%%%%%%%%%%%%%%%%%%%%%%%%%%%%%%%%%%%%%%%%%%%%%%%%%
\subsection{Reduced matrix elements of the radiative decays}
%%%%%%%%%%%%%%%%%%%%%%%%%%%%%%%%%%%%%%%%%%%%%%%%%%%%%%%%%%

In Sec. \ref{sec21}, we present the definitions of the spin
structures of the initial and final hadrons in the radiative decays.
In the following, we give the general expression of the reduced
radiative decay matrix element. In the heavy quark limit, both the
heavy and light spins are conserved. Thus, it is convenient to
calculate the corresponding process with the uncoupled
representation. Denote the total angular momentum of the initial
state hadron as $j$ and the third component of $j$ as $j_z$. We have
\begin{eqnarray*}
 |j,j_z\rangle = \sum_{S_{H_z},S_{l_z}} \langle S_H, S_{H_z}; S_l, S_{l_z}| j,j_z\rangle |S_H, S_{H_z}\rangle|S_l, S_{l_z}\rangle,
\end{eqnarray*}
which is a unitary transformation between the coupled and uncoupled
representations. For the final state hadron, we denote the total
angular momentum of the final state hadron as $j'$ and its third
component as $j_z'$. In addition, $Q$ is the light spin of the
photon and $Q_z$ is its third component. Thus, the matrix element
for the radiative decay $A\to B+\gamma$ is written as
\begin{eqnarray}
   &&\mathcal{M}[A(j,j_z)\rightarrow B(j',j'_z)+\gamma(Q,Q_z)] \nonumber\\
   &&= \langle\gamma(Q,Q_z);j',j'_z|H_{eff}|j,j_z\rangle  \nonumber\\
  &&= \sum_{S_{H_z},S_{l_z};S'_{H_z},S'_{l_z}} \langle \gamma(Q,Q_z); S'_H, S'_{H_z}; S'_l, S'_{l_z}|H_{eff}|S_H, S_{H_z}; S_l, S_{l_z}\rangle \nonumber\\
  &&\quad\times \langle S'_H, S'_{H_z}; S'_l, S'_{l_z}| j',j'_z\rangle \langle S_H, S_{H_z}; S_l, S_{l_z}| j,j_z\rangle \nonumber\\
  &&= \sum_{S_{H_z},S_{l_z}}\langle S_H, S_{H_z}; S_l, S_{l_z}|j,j_z\rangle \langle Q,S'_l\| H_{eff}\|S_l\rangle \nonumber\\
  &&\quad\times \langle S_H, S_{H_z}; S'_l, S'_{l_z}| j',j'_z\rangle \langle Q,Q_z;S'_l, S'_{l_z}|S_l, S_{l_z}\rangle \nonumber\\
  &&= (-1)^{Q+S_H+S'_l+j}\sqrt{(2S_l+1)(2j'+1)}
  \left\{
    \begin{array}{ccc}
      Q & S_l' & S_l \\
      S_H & j & j' \\
    \end{array}
  \right\}\nonumber\\
  &&\quad\times
\langle Q,(j_z-j_z');j',j_z'|j,j_z\rangle \langle Q,S_l'\|
H_{eff}\|S_l\rangle,\nonumber\\\label{14}
\end{eqnarray}
where $S_H$ and $S_H^\prime$ stand for the heavy spin of initial and
final hadrons, respectively. The third components of $S_H$ and
$S_{H}^\prime$ are $S_{H_z}$ and $S_{H_z}'$, respectively. $S_l$ and
$S_l^\prime$ denote the light spins of the initial and final
hadrons, respectively. The third components of $S_l$ and
$S_l^\prime$ are $S_{l_z}$ and $S_{l_z}'$, respectively.

The effective Hamiltonian $H_{eff}$ conserves both the heavy and
light spins and can be decomposed into
\begin{equation}
  H_{eff}=H_{eff}^{spatial}\otimes H_{eff}^{flavor},
\end{equation}
where $H_{eff}^{spatial}$ and $H_{eff}^{flavor}$ denote the spatial
and flavor parts, respectively. With the help of the Wigner-Eckart
theorem, we get
\begin{eqnarray*}
  &&\langle\gamma(Q,Q_z); S'_H, S'_{H_z}; S'_l, S'_{l_z}|H_{eff}|S_H, S_{H_z}; S_l, S_{l_z}\rangle  \\
  &&= \langle Q,Q_z;S'_l, S'_{l_z}|S_l, S_{l_z}\rangle \langle Q,S'_l\| H_{eff}\|S_l\rangle \delta_{S_{H_{z}},S_{H_{z}}'},
\end{eqnarray*}
where $\langle Q,S'_l\| H_{eff}\|S_l\rangle$ is the reduced matrix
element. The 6-$j$ symbol and the factors in the front of 6-$j$
symbol in Eq. (\ref{14}) come from the spin rearrangement of the
initial and final states.

Let's consider two different radiative decay channels with the same
final state. If the initial state hadrons happen to have the same
radial and orbital quantum numbers, the ratio of the decay widths of
these two channels can be expressed as
\begin{eqnarray}
&&  \frac{\Gamma(H_1\rightarrow H_2+\gamma)}{\Gamma(H_3\rightarrow
H_2+\gamma)}\nonumber\\&& \sim \frac{|c_1\langle Q,S'_{l_1}\|
H_{eff}\|S_{l_1}\rangle+c_2\langle Q,S'_{l_2}\|
H_{eff}\|S_{l_2}\rangle+\cdots|^2}{|d_1\langle Q,S'_{l_1}\|
H_{eff}\|S_{l_1}\rangle+d_2\langle Q,S'_{l_2}\|
H_{eff}\|S_{l_2}\rangle+\cdots|^2}, \nonumber\\\label{15}
\end{eqnarray}
where $c_1$, $c_2$, $d_1$ and $d_2$ represent the corresponding
expansion coefficients of the spin configuration. We also consider
two different radiative decay channels with the same initial state.
If the two final state hadrons have the same radial and orbital
quantum numbers, we can obtain similar expressions as Eq.
(\ref{15}). We need to specify that the ratio listed in Eq.
(\ref{15}) does not include the phase space contribution.

%%%%%%%%%%%%%%%%%%%%%%%%%%%%%%%%%%%%%%%%%%%%%%%%%%%%%%%%%%
\subsection{$\mathfrak{M}\to (b\bar{b})+\gamma$}
%%%%%%%%%%%%%%%%%%%%%%%%%%%%%%%%%%%%%%%%%%%%%%%%%%%%%%%%%%

For the radiative decay $\mathfrak{M}\to (b\bar{b})+\gamma$, where
the initial state is a hadronic molecule and final state is a
bottomonium, the photon is from the $q\bar{q}$ annihilation. Now we
need to introduce the spin structure of the involved photon. We
assume that the photon has a so-called "flavor" wave function
equivalently, which is denoted by $|\gamma\rangle$ in the following.
Its spin structure is defined as
 \begin{eqnarray}
   |\gamma(E1) \rangle&=&| (0_H^+\otimes 1_l^-)_1^{--}\rangle|\gamma  \rangle, \label{m1}\\
   |\gamma(M1)\rangle &=&| (0_H^+\otimes 1_l^+)_1^{+-}\rangle|\gamma  \rangle, \\
  | \gamma(E2)\rangle &=& |(0_H^+\otimes 2_l^+)_2^{+-}\rangle|\gamma  \rangle,\label{m2}
 \end{eqnarray}
which correspond to the $E1$, $M1$ and $E2$ transitions,
respectively. Here, the notation of subscripts and subscripts are
similar to those in Eqs. (\ref{k1})-(\ref{k2}). The transition
matrix elements related to the flavor wave functions are
\begin{eqnarray*}
    \langle\gamma;\bar{b}b|H_{eff}^{flavor}|(\bar{b}q)(b\bar{q})\rangle &=& 1, \quad
    \langle\gamma;b\bar{b}|H_{eff}^{flavor}|(\bar{b}q)(b\bar{q})\rangle = 0,\\
    \langle\gamma; b\bar{b}|H_{eff}^{flavor}|(b\bar{q})(\bar{b}q)\rangle &=& 1, \quad
    \langle\gamma; \bar{b}b|H_{eff}^{flavor}|(b\bar{q})(\bar{b}q)\rangle = 0.
\end{eqnarray*}

With the above preparation, we can calculate the rearranged spin
structures of the final states in the $\mathfrak{M}\to
(b\bar{b})+\gamma$ decays. The general expression is
\begin{eqnarray}
 && |Bottomionia\rangle\otimes|\gamma\rangle\nonumber\\
 & &= \left[[(\bar{b}b)_g\otimes L ]_K\otimes Q\right]_J |(\bar{b}b)\rangle|\gamma\rangle  \nonumber\nonumber\\
  &&= \sum_{h=|L-Q|}^{L+Q} (-1)^{g+L+Q+J}\Big[(2K+1)(2h+1)\Big]^{1/2}\nonumber\\
  &&\quad\times
  \left\{
    \begin{array}{ccc}
      L & g & K \nonumber\\
      J & Q & h \nonumber\\
    \end{array}
  \right\} \Bigg|\left[\left(\bar{b}b\right)_g\otimes\left[L\otimes Q\right]_h\right]_J \Bigg\rangle |(\bar{b}b)\rangle|\gamma\rangle\nonumber\\
  &&=\sum_{h=|L-Q|}^{L+Q}  \mathcal{D}^{g,L,K,J}_{g,h}\Bigg|\left[\left(\bar{b}b\right)_g\otimes\left[L\otimes Q\right]_h\right]_J \Bigg\rangle |(\bar{b}b)\rangle|\gamma\rangle,\label{11}%\nonumber\\
%  &&=\sum_{h=|L-Q|}^{L+Q}  \mathcal{D}^{g,L,K,J}_{g,h}(g_H\otimes h_l)_J  (\bar{b}b)\rangle|\gamma\rangle,
   \end{eqnarray}
where the $g$ and $L$ denote the heavy and light spins of the
bottomionium, respectively. $Q$ stands for the light spin of the
photon. The indices $b$, $\bar{b}$ and $\gamma$ in the square
brackets represent the corresponding spin wave functions. We collect
the coefficients $\mathcal{D}^{L,K,J}_{g,h}$ in Table \ref{final
state}.

\renewcommand{\arraystretch}{1.4}
\begin{table*}[htbp]
 \caption{The coefficient $\mathcal{D}^{L,K,J}_{g,h}$ in Eq. (\ref{11}) corresponding to different combinations of $[g,h]$.}\label{final state}
\begin{center}
   \begin{tabular}{c|cc|c c c c|cccc} \toprule[1pt]
      %\multicolumn{2}{|c|}
     &\multicolumn{2}{c|}{$J=0$}  &&\multicolumn{2}{c}{$J=1$}&  &&\multicolumn{2}{c}{$J=2$}&\\\midrule[1pt]

    & $[0,0]$   &  $[1,1]$
      & $[0,1]$  &  $[1,0]$ &  $[1,1]$ &  $[1,2]$ & $[0,2]$   &  $[1,1]$ &  $[1,2]$  &  $[1,3]$\\\midrule[1pt]

      $|\eta_b(1^1S_0)\gamma(E1 / M1)\rangle$ &--&--
      & 1 & 0 & 0 & 0 &--&--&--&--\\

      $|\Upsilon(1^3S_1)\gamma(E1 / M1)\rangle$& 0 & 1
       & 0 & 0 & 1 & 0
       & 0 & 1 & 0 & 0\\

      $|h_b(1^1P_1)\gamma(E1 / M1)\rangle$& 1 & 0
       & 1 & 0 & 0 & 0
       & 1 & 0 & 0 & 0\\
      $|\chi_{b0}(1^3P_0)\gamma(E1 / M1)\rangle$&--&--
       & 0 & $\frac{1}{3}$ & $-\frac{\sqrt{3}}{3}$ & $\frac{\sqrt{5}}{3}$&--&--&--&-- \\

      $|\chi_{b1}(1^3P_1)\gamma(E1 / M1)\rangle$& 0 & 1
       & 0 & $-\frac{\sqrt{3}}{3}$ & $\frac{1}{2}$ & $\frac{\sqrt{15}}{6}$
       & 0 & $-\frac{1}{2}$ & $\frac{\sqrt{3}}{2}$ & 0\\

      $|\chi_{b2}(1^3P_2)\gamma(E1 / M1)\rangle$&--&--
       & 0 & $\frac{\sqrt{5}}{3}$ & $\frac{\sqrt{15}}{6}$ & $\frac{1}{6}$
       & 0 & $\frac{\sqrt{3}}{2}$ & $\frac{1}{2}$ & 0\\

      $|\eta_{b2}(1^1D_2)\gamma(E1 / M1)\rangle$&--&--
       & 1 & 0 & 0 & 0
       & 1 & 0 & 0 & 0\\

      $|\Upsilon(1^3D_1)\gamma(E1 / M1)\rangle$& 0 &1
       & 0 & 0 & $-\frac{1}{2}$ & $\frac{\sqrt{3}}{2}$
       & 0 & $\frac{1}{10}$ & $-\frac{\sqrt{15}}{10}$ & $\frac{\sqrt{21}}{5}$\\

      $|\Upsilon(1^3D_2)\gamma(E1 / M1)\rangle$&--&--
       & 0 & 0 & $\frac{\sqrt{3}}{2}$ & $\frac{1}{2}$
       & 0 & $-\frac{\sqrt{15}}{10}$ & $\frac{5}{6}$ & $\frac{\sqrt{35}}{15}$\\

      $|\Upsilon(1^3S_1)\gamma(E2)\rangle$&--&--
       & 0 & 0 & 0 & 1
       & 0 & 0 & 1 & 0\\

      $|h_b(1^1P_1)\gamma(E2)\rangle$&--&--
       & 1 & 0 & 0 & 0
       & 1 & 0 & 0 & 0\\

      $|\chi_{b1}(1^3P_0)\gamma(E2)\rangle$&--&--
       & 0 & 0 & $-\frac{1}{2}$ & $\frac{\sqrt{3}}{2}$&--&--&--&-- \\

      $|\chi_{b2}(1^3P_0)\gamma(E2)\rangle$&--&--
       & 0 & 0 & $\frac{\sqrt{3}}{2}$ & $\frac{1}{2}$ &--&--&--&--\\
      $|\Upsilon(1^3D_1)\gamma(E2)\rangle$&--&--
       & 0 & $\frac{\sqrt{5}}{5}$ & $-\frac{3\sqrt{5}}{10}$ & $\frac{\sqrt{35}}{10}$
       & 0 & $\frac{3}{10}$ & $-\frac{\sqrt{35}}{10}$ & $\frac{\sqrt{14}}{5}$\\

      $|\Upsilon(1^3D_2)\gamma(E2)\rangle$& 0 & 1
       & 0 & $-\frac{\sqrt{3}}{3}$ & $\frac{\sqrt{3}}{6}$ & $\frac{\sqrt{21}}{6}$
       & 0 & $-\frac{\sqrt{35}}{10}$ & $\frac{1}{2}$ & $\frac{\sqrt{10}}{5}$\\

      $|\Upsilon(1^3D_3)\gamma(E2)\rangle$&--&--
       & 0 & $\frac{\sqrt{105}}{15}$ & $\frac{\sqrt{105}}{15}$ & $\frac{\sqrt{15}}{15}$
       & 0 & $\frac{\sqrt{14}}{5}$ & $\frac{\sqrt{10}}{5}$ & $\frac{1}{5}$\\

      $|\eta_{b2}(1^1D_2)\gamma(E2)\rangle$ & 1 &0
       &--&--&--&--
       & 1 & 0 & 0 & 0\\

      $|\Upsilon(1^3D_3)\gamma(E1 / M1)\rangle$&--&--
      &--&--&--&--
      & 0 & $\frac{\sqrt{21}}{5}$ & $\frac{\sqrt{35}}{15}$ & $\frac{1}{15}$ \\

      $|\eta_b(1^3S_1)\gamma(E2)\rangle$&--&--
      &--&--&--&--
       & 1 & 0 & 0 & 0 \\\bottomrule[1pt]
      \end{tabular}
  \end{center}
\end{table*}

In this work, we mainly focus on the ratios of the decay widths in
the heavy quark limit, where the spatial matrix elements of some
decays are the same. For example, the molecular(resonant) state
$\frac{1}{\sqrt{2}}(B_0\bar{B}^\ast+B^\ast\bar{B}_0)$ can decay into
$\Upsilon(1^3D_1)+\gamma$ and $\Upsilon(1^3D_2)\gamma$, which are
typical $M1$ transitions and have the same spatial matrix elements.
In this case, there exist some model independent predictions.

%%%%%%%%%%%%%%%%%%%%%%%%%%%%%%%%%%%%%%%%%%%%%%%%%%%%%%%%%
\subsection{$(b\bar{b})\to \mathfrak{M}+\gamma$}
%%%%%%%%%%%%%%%%%%%%%%%%%%%%%%%%%%%%%%%%%%%%%%%%%%%%%%%%%

In the radiative decay $(b\bar{b})\to \mathfrak{M}+\gamma$,  a
bottomonium state emits a photon and decays into a hidden beauty
molecule state, where a $q\bar{q}$ pair is created from the vacuum.
Thus, this process can be expressed as
\begin{eqnarray*}
   &&|Heavy \ \ Quarkonium\rangle\otimes|VAC\rangle \\
   &&\rightarrow\frac{1}{\sqrt{2}}(|B_{1,2}\bar{B}^{(\ast)}\rangle\pm |B^{(\ast)}\bar{B_{1,2}}\rangle)\otimes(0_H^+\otimes 1_l^\pm)|\gamma  \rangle,
  \end{eqnarray*}
where $|VAC\rangle$ is the hadronic vacuum, which has the spin
structure
  \begin{eqnarray*}
    |VAC\rangle=|(0_H^+\otimes 0_l^+)\rangle |(q\bar{q})\rangle,
  \end{eqnarray*}
where $|(q\bar{q})\rangle=
\frac{1}{\sqrt{2}}(|q\bar{q}\rangle+|\bar{q}q\rangle)$ is the
so-called "flavor" wave function.

The flavor part of the final states in $(b\bar{b})\to
\mathfrak{M}+\gamma$ can be written as
  \begin{eqnarray*}
    \langle\gamma|\langle (\bar{b}q)(b\bar{q})| &=& \langle (\bar{b}(q\gamma))(b\bar{q})|+\langle (\bar{b}q)(b(\bar{q}\gamma))|\\
    &&+\langle ((\bar{b}\gamma)q)(b\bar{q})|+\langle (\bar{b}q)((b\gamma)\bar{q})|,
  \end{eqnarray*}
   \begin{eqnarray*}
    \langle\gamma|\langle (b\bar{q})(\bar{b}q)| &=& \langle (b(\bar{q}\gamma))(\bar{b}q)|+\langle (b\bar{q})(\bar{b}(q\gamma))|\\
    &&+\langle ((b\gamma)\bar{q})(\bar{b}q)|+\langle (b\bar{q})((\bar{b}\gamma)q)|.
  \end{eqnarray*}
Thus, the nonzero transition matrix elements of $(b\bar{b})\to
\mathfrak{M}+\gamma$ from the flavor wave functions only are
  \begin{eqnarray*}
    \langle (b(\bar{q}\gamma))(\bar{b}q)|H_{eff}^{flavor}|b\bar{b}\rangle|\bar{q}q\rangle &=& 1, \\
    \langle (b\bar{q})(\bar{b}(q\gamma))|H_{eff}^{flavor}|b\bar{b}\rangle|\bar{q}q\rangle &=& 1,\\
    \langle (\bar{b}(q\gamma))(b\bar{q})|H_{eff}^{flavor}|\bar{b}b\rangle|q\bar{q}\rangle &=& 1, \\
    \langle (\bar{b}q)(b(\bar{q}\gamma))|H_{eff}^{flavor}|\bar{b}b\rangle|q\bar{q}\rangle &=& 1, \\
    \langle ((b\gamma)\bar{q})(\bar{b}q)|H_{eff}^{flavor}|b\bar{b}\rangle|\bar{q}q\rangle &=& 1, \\
    \langle (b\bar{q})((\bar{b}\gamma)q)|H_{eff}^{flavor}|b\bar{b}\rangle|\bar{q}q\rangle &=& 1, \\
    \langle ((\bar{b}\gamma)q)(b\bar{q})|H_{eff}^{flavor}|\bar{b}b\rangle|q\bar{q}\rangle &=& 1, \\
    \langle (\bar{b}q)((b\gamma)\bar{q})|H_{eff}^{flavor}|\bar{b}b\rangle|q\bar{q}\rangle &=& 1.
  \end{eqnarray*}
Similarly, we list the spin structure of the final states in
$(b\bar{b})\to \mathfrak{M}+\gamma$:
\begin{eqnarray*}
 && |B_{1,2}\bar{B}^{(\ast)}\rangle\otimes|\gamma\rangle\nonumber\\ &&= [[\bar{b}\otimes(q\otimes 1)_s]_K\otimes[b\otimes \bar{q}]_L]_J\otimes(0_H^+\otimes 1_L^\pm)\\
  &&= \sum_{g=0}^1 \sum_{m=0}^1 \sum_{h=|s-\frac{1}{2}|}^{s+\frac{1}{2}} \sum_{h_0=|h-1|}^{h+1} (-1)^{s+\frac{3}{2}+m}\\
  &&\quad\times\sqrt{(2s+1)(2m+1)(2K+1)(2L+1)(2h+1)}\\
  &&\quad\times(2g+1)\sqrt{3(2J+1)(2h_0+1)}\\
  &&\quad\times \left\{
             \begin{array}{ccc}
               g & h & J \\
               0 & 1 & 1 \\
               g & h_0 & J_0 \\
             \end{array}
           \right\}
  \left\{
             \begin{array}{ccc}
               \frac{1}{2} & s & K \\
               \frac{1}{2} & \frac{1}{2} & L \\
               g & h & J \\
             \end{array}
           \right\}
  \left\{
    \begin{array}{ccc}
      \frac{1}{2} & 1 & s \\
      h & \frac{1}{2} & m \\
    \end{array}
  \right\} \\
  &&\quad\times\{[\bar{b}b]_g\otimes[([q\bar{q}]_m\otimes1)_h\otimes 1]_{h_0}\}_{J_0},
\end{eqnarray*}
and
\begin{eqnarray*}
 && |B^{(\ast)}\bar{B_{1,2}}\rangle\otimes|\gamma\rangle\nonumber\\ &&= [[b\otimes \bar{q}]_L\otimes[\bar{b}\otimes(q\otimes 1)_s]_K]_J\otimes(0_H^+\otimes 1_L^\pm)\\
  &&= (-1)^{L+K-J}\sum_{g=0}^1 \sum_{m=0}^1 \sum_{h=|s-\frac{1}{2}|}^{s+\frac{1}{2}} \sum_{h_0=|h-1|}^{h+1} (-1)^{s+\frac{3}{2}+g}\\
  &&\quad\times\sqrt{(2s+1)(2m+1)(2K+1)(2L+1)(2h+1)}\\
  &&\quad\times(2g+1)\sqrt{3(2J+1)(2h_0+1)}\\
  &&\quad\times \left\{
             \begin{array}{ccc}
               g & h & J \\
               0 & 1 & 1 \\
               g & h_0 & J_0 \\
             \end{array}
           \right\}
  \left\{
             \begin{array}{ccc}
               \frac{1}{2} & s & K \\
               \frac{1}{2} & \frac{1}{2} & L \\
               g & h & J \\
             \end{array}
           \right\}
  \left\{
    \begin{array}{ccc}
      \frac{1}{2} & 1 & s \\
      h & \frac{1}{2} & m \\
    \end{array}
  \right\} \\
  &&\quad\times\{[\bar{b}b]_g\otimes[([q\bar{q}]_m\otimes1)_h\otimes 1]_{h_0}\}_{J_0},
\end{eqnarray*}
which will be applied in the calculation.

%%%%%%%%%%%%%%%%%%%%%%%%%%%%%%%%%%%%%%%%%%%%%%%%%%%%%%%%%%%%%%%%%%
\subsection{$\mathfrak{M} \to \mathfrak{M}^\prime+\gamma$}
%%%%%%%%%%%%%%%%%%%%%%%%%%%%%%%%%%%%%%%%%%%%%%%%%%%%%%%%%%%%%%%%%%

For the radiative transition between two hidden beauty
molecular(resonant) states, its dynamics is different from that of
$(b\bar{b})\to \mathfrak{M}+\gamma$. The photon is emitted from
either $q/\bar q$ quark or $b/\bar b$ quark. The spin structure of
the photon is similar to that given in Eqs. (\ref{m1})-(\ref{m2}).

The nonzero transition matrix elements of $\mathfrak{M} \to
\mathfrak{M}^\prime+\gamma$ from the flavor wave functions are
 \begin{eqnarray*}
    \langle (b(\bar{q}\gamma))(\bar{b}q)|H_{eff}^{flavor}|(b\bar{q})(\bar{b}q)\rangle &=& 1, \\
    \langle (b\bar{q})(\bar{b}(q\gamma))|H_{eff}^{flavor}|(b\bar{q})(\bar{b}q)\rangle &=& 1, \\
    \langle (\bar{b}(q\gamma))(b\bar{q})|H_{eff}^{flavor}|(\bar{b}q)(b\bar{q})\rangle &=& 1, \\
    \langle (\bar{b}q)(b(\bar{q}\gamma))|H_{eff}^{flavor}|(\bar{b}q)(b\bar{q})\rangle &=& 1, \\
    \langle ((b\gamma)\bar{q})(\bar{b}q)|H_{eff}^{flavor}|(b\bar{q})(\bar{b}q)\rangle &=& 1, \\
    \langle (b\bar{q})((\bar{b}\gamma)q)|H_{eff}^{flavor}|(b\bar{q})(\bar{b}q)\rangle &=& 1, \\
    \langle ((\bar{b}\gamma)q)(b\bar{q})|H_{eff}^{flavor}|(\bar{b}q)(b\bar{q})\rangle &=& 1, \\
    \langle (\bar{b}q)((b\gamma)\bar{q})|H_{eff}^{flavor}|(\bar{b}q)(b\bar{q})\rangle &=&
    1.
\end{eqnarray*}

%%%%%%%%%%%%%%%%%%%%%%%%%%%%%%%%%%%%%%%%%%%
\section{Numerical results}\label{sec3}
%%%%%%%%%%%%%%%%%%%%%%%%%%%%%%%%%%%%%%%%%%%

With the above preparation, we are ready to calculate the typical
ratios of the radiative decays relevant to the hidden beauty
molecular(resonant) states. In the following, we present the results
of $\mathfrak{M}\to (b\bar{b})+\gamma$, $(b\bar{b})\to
\mathfrak{M}+\gamma$ and $\mathfrak{M}\to
\mathfrak{M}^\prime+\gamma$.

\subsection{$\mathfrak{M}\to (b\bar{b})+\gamma$}

Using the rearranged spin structures and the orthogonalization of
the spin and flavor wave functions, we first obtain some typical
ratios of the $\mathfrak{M}\to (b\bar{b})+\gamma$ decays. We only
consider the final states containing the S-wave, P-wave and D-wave
bottomonia.

If the bottomonia belong to the same spin multiplet, the spatial
matrix elements of these radiative decays are the same, which leads
to quite simple ratios between their decay widths. We collect the
typical ratios in Tables \ref{tab:2} and \ref{tab:3}, where the
values in the brackets are the values with the phase space
correction. Since the $B_0$ meson and D-wave bottomonia are still
absent experimentally, we do not consider the contribution of the
phase space factors when calculating the corresponding ratios.

\begin{table*}[htbp]
\begin{center}
\caption{\label{tab:2} The typical ratios of the $\mathfrak{M}\to
(b\bar{b})+\gamma$ decay widths. The parameters $x, x', \alpha$ are
defined as $x=H_{22}(M1)/H_{21}(M1)$, $x'=H_{22}(E2)/H_{21}(E2)$,
and $\alpha=H_{21}(E2)/H_{20}(E2)$, respectively.}
   \begin{tabular}{c|ccccccccccc} \toprule[1pt]
      %\multicolumn{2}{|c|}
      &$J^{PC}$ &  & Final state\\\cline{2-4}
   \multirow{46}{*}{\rotatebox{90}{Initial \,\, state}}&&& $\Gamma(\chi_{b0}\gamma(E1)):\Gamma(\chi_{b1}\gamma(E1)):\Gamma(\chi_{b2}\gamma(E1))$  \\
     % \midrule[1pt]
      &\multirow{6}{*}{$1^{--}$} & $\frac{1}{\sqrt{2}}(B_0\bar{B}^\ast-B^\ast\bar{B}_0)$     & $4:3:5$   \\%\hline
      &&$\frac{1}{\sqrt{2}}(B_1'\bar{B}-B\bar{B}_1')$     & $4:3:5$ $(1.5:1:1.6)$   \\%\hline
      &&$\frac{1}{\sqrt{2}}(B_1\bar{B}-B\bar{B}_1)$     & $4:3:5$ $(1.4:1:1.6)$   \\%\hline
      &&$\frac{1}{\sqrt{2}}(B_1'\bar{B}^\ast+B^\ast\bar{B}_1')$     & $0:0:0$   \\%\hline
      &&$\frac{1}{\sqrt{2}}(B_1\bar{B}^\ast+B^\ast\bar{B}_1)$     & $4:3:5$ $(1.4:1:1.6)$    \\%\hline
      &&$\frac{1}{\sqrt{2}}(B_2\bar{B}^\ast-B^\ast\bar{B}_2)$     & $4:3:5$ $(1.4:1:1.6)$  \\\cline{2-4}
     &&& $\Gamma(\chi_{b0}\gamma(M1)):\Gamma(\chi_{b1}\gamma(M1)):\Gamma(\chi_{b2}\gamma(M1))$  \\
     % \midrule[1pt]
     &\multirow{2}{*}{$1^{+-}$} & $\frac{1}{\sqrt{2}}(B\bar{B}^\ast-B^\ast\bar{B})$     & $1:3:5$ $(1:2.6:4.1)$  \\%\hline
      &&$B^\ast\bar{B}^\ast$     & $1:3:5$ $(1:2.7:4.1)$  \\\cline{2-4}
     &&& $\Gamma(\Upsilon(1^3D_1)\gamma(M1)):\Gamma(\Upsilon(1^3D_2)\gamma(M1))$  \\
      %\midrule[1pt]
&\multirow{6}{*}{$1^{-+}$}     &  $\frac{1}{\sqrt{2}}(B_0\bar{B}^\ast+B^\ast\bar{B}_0)$     & $1:3$   \\%\hline
     && $\frac{1}{\sqrt{2}}(B_1'\bar{B}+B\bar{B}_1')$     & $1:3$   \\%\hline
      &&$\frac{1}{\sqrt{2}}(B_1\bar{B}+B\bar{B}_1)$     & $\frac{(1+3\sqrt{5}x)^2}{3(1-\sqrt{5}x)^2}$   \\%\hline
      &&$\frac{1}{\sqrt{2}}(B_1'\bar{B}^\ast-B^\ast\bar{B}_1')$     & $0:0$   \\%\hline
      &&$\frac{1}{\sqrt{2}}(B_1\bar{B}^\ast-B^\ast\bar{B}_1)$     & $\frac{3(1-\sqrt{5}x)^2}{(3+\sqrt{5}x)^2}$   \\%\hline
      &&$\frac{1}{\sqrt{2}}(B_2\bar{B}^\ast+B^\ast\bar{B}_2)$     & $\frac{3(\sqrt{5}+3x)^2}{(\sqrt{15}-3x)^2}$   \\\cline{2-4}
    &&& $\Gamma(\Upsilon(1^3D_1)\gamma(E1)):\Gamma(\Upsilon(1^3D_2)\gamma(E1))$  \\
      %\midrule[1pt]
     &\multirow{1}{*}{$1^{++}$} & $\frac{1}{\sqrt{2}}(B\bar{B}^\ast+B^\ast\bar{B})$     & $1:3$   \\\cline{2-4}
      && & $\Gamma(\chi_{b1}\gamma(E2)):\Gamma(\chi_{b2}\gamma(E2))$  \\
     % \midrule[1pt]
    &\multirow{2}{*}{$1^{+-}$} &  $\frac{1}{\sqrt{2}}(B\bar{B}^\ast-B^\ast\bar{B})$     & $0:0$   \\%\hline
     && $B^\ast\bar{B}^\ast$     & $0:0$   \\\cline{2-4}
    &&& $\Gamma(\Upsilon(1^3D_1)\gamma(E2)):\Gamma(\Upsilon(1^3D_2)\gamma(E2)):\Gamma(\Upsilon(1^3D_3)\gamma(E2))$  \\
     % \midrule[1pt]
      &\multirow{6}{*}{$1^{-+}$} &$\frac{1}{\sqrt{2}}(B_0\bar{B}^\ast+B^\ast\bar{B}_0)$     & $9(1-\sqrt{3}\alpha)^2:5(\sqrt{3}-\alpha)^2:7(\sqrt{3}+2\alpha)^2$   \\ %\hline
     && $\frac{1}{\sqrt{2}}(B_1'\bar{B}+B\bar{B}_1')$     & $9(1+\sqrt{3}\alpha)^2:5(\sqrt{3}+\alpha)^2:7(\sqrt{3}-2\alpha)^2$   \\%\hline
      &&$\frac{1}{\sqrt{2}}(B_1\bar{B}+B\bar{B}_1)$     & $9(\sqrt{3}+\sqrt{35}x')^2:5(1+\sqrt{105}x')^2:4(\sqrt{7}-\sqrt{15}x')^2$   \\%\hline
    && $\frac{1}{\sqrt{2}}(B_1'\bar{B}^\ast-B^\ast\bar{B}_1')$     & $3:5:7$   \\%\hline
    &&  $\frac{1}{\sqrt{2}}(B_1\bar{B}^\ast-B^\ast\bar{B}_1)$     & $9(3\sqrt{3}-\sqrt{35}x')^2:15(\sqrt{3}+\sqrt{35}x')^2:4(3\sqrt{7}+\sqrt{15}x')^2$  \\%\hline
     && $\frac{1}{\sqrt{2}}(B_2\bar{B}^\ast+B^\ast\bar{B}_2)$     & $9(5\sqrt{3}+\sqrt{35}x')^2:25(\sqrt{5}-\sqrt{21}x)^2:4(5\sqrt{7}-\sqrt{15}x)^2$\\\cline{2-4}
      &&& $\Gamma(\chi_{b1}(1^3P_1)\gamma(E1)):\Gamma(\chi_{b2}(1^3P_2)\gamma(E1))$   \\
      %\midrule[1pt]
    &\multirow{4}{*}{$2^{--}$} & $\frac{1}{\sqrt{2}}(B_1\bar{B}^\ast-B^\ast\bar{B}_1)$     & $1:3$ $(1:2.9)$   \\%\hline
      &&$\frac{1}{\sqrt{2}}(B_1'\bar{B}^\ast-B^\ast\bar{B}_1')$     & $1:3$ $(1:2.8)$   \\%\hline
      &&$\frac{1}{\sqrt{2}}(B_2\bar{B}-B\bar{B}_2)$     & $1:3$ $(1:2.9)$  \\%\hline
      &&$\frac{1}{\sqrt{2}}(B_2\bar{B}^\ast+B^\ast\bar{B}_2)$     & $1:3$ $(1:2.9)$ \\\cline{2-4}
      && & $\Upsilon(1^3D_1)\gamma(E1):\Upsilon(1^3D_2)\gamma(E1):\Upsilon(1^3D_3)\gamma(E1)$   \\
      %\midrule[1pt]
    &$2^{++}$&  $B^\ast\bar{B}^\ast$     & $1:15:84$    \\\cline{2-4}
      &&& $\Gamma(\Upsilon(1^3D_1)\gamma(M1)):\Gamma(\Upsilon(1^3D_2)\gamma(M1)):\Gamma(\Upsilon(1^3D_3)\gamma(M1))$  \\
      %\midrule[1pt]
     &\multirow{9}{*}{$2^{-+}$}&\multirow{1}{*}{$\frac{1}{\sqrt{2}}(B_1'\bar{B}^\ast+B^\ast\bar{B}_1')$ }    & $1:15:84$   \\%\midrule[1pt]
     && $\frac{1}{\sqrt{2}}(B_1\bar{B}^\ast+B^\ast\bar{B}_1)$ & $(1-9\sqrt{5}x)^2:3(\sqrt{5}-25x)^2:84(1+\sqrt{5}x)^2$\\
     && $\frac{1}{\sqrt{2}}(B_2\bar{B}+B\bar{B}_2)$ & $3(1+\sqrt{45}x)^2:(3\sqrt{5}-25x)^2:28(3-\sqrt{5}x)^2$\\
     && $\frac{1}{\sqrt{2}}(B_2\bar{B}^\ast-B^\ast\bar{B}_2)$ & $27(1-\sqrt{5}x)^2:(9\sqrt{5}-25x)^2:28(9+\sqrt{5}x)^2$\\
      &&& $\Gamma(\Upsilon(1^3D_1)\gamma(E2)):\Gamma(\Upsilon(1^3D_2)\gamma(E2)):\Gamma(\Upsilon(1^3D_3)\gamma(E2))$  \\%\midrule[1pt]
     &&$\frac{1}{\sqrt{2}}(B_1'\bar{B}^\ast+B^\ast\bar{B}_1')$        & $9:35:56$  \\
     && $\frac{1}{\sqrt{2}}(B_1\bar{B}^\ast+B^\ast\bar{B}_1)$ & $9(\sqrt{3}-3\sqrt{35}x')^2:(\sqrt{105}-45x')^2:8(\sqrt{21}+9\sqrt{5}x')^2$\\
     && $\frac{1}{\sqrt{2}}(B_2\bar{B}+B\bar{B}_2)$ & $(3+\sqrt{105}x')^2:(\sqrt{35}+5\sqrt{3}x')^2:8(\sqrt{7}-\sqrt{15}x')^2$\\
     && $\frac{1}{\sqrt{2}}(B_2\bar{B}^\ast-B^\ast\bar{B}_2)$ & $(3\sqrt{3}-\sqrt{35}x')^2:(\sqrt{105}-5x')^2:8(\sqrt{21}+\sqrt{5}x')^2$\\\bottomrule[1pt]
       \end{tabular}
\end{center}
\end{table*}

\begin{table*}[htbp]
\scriptsize
\begin{center}
   \caption{\label{tab:3}
The typical ratios $\frac{\Gamma(\mathfrak{M} \to
(b\bar{b})+\gamma)}{ \Gamma(\mathfrak{M} ^\prime\to
(b\bar{b})+\gamma)}$, where the initial molecular(resonant) states
are different while the final states are same. The parameters are
defined as $x=H_{22}(M1)/H_{21}(M1)$, $x'=H_{22}(E2)/H_{21}(E2)$,
and $\alpha=H_{21}(E2)/H_{20}(E2)$.}
   \begin{tabular}{c|c|cccccccccc} \toprule[1pt]
     \multirow{46}{*}{\rotatebox{90}{Initial \,\, state}}&  & \multicolumn{7}{c}{Final state} \\\cline{2-12}
      && & $\eta_b\gamma(M1)$ & $\chi_{b0}\gamma(E1)$ & $\chi_{b1}\gamma(E1)$  & $\chi_{b2}\gamma(E1)$ & $\eta_{b2}(1^1D_2)\gamma(M1)$ & $\eta_{b}\gamma(E2)$ & $\eta_{b2}(1^1D_2)\gamma(E2)$ \\
      %\midrule[1pt]
      &\multirow{10}{*}{$1^{--}$} & $\frac{\frac{1}{\sqrt{2}}(B_0\bar{B}^\ast-B^\ast\bar{B}_0)}{\frac{1}{\sqrt{2}}(B_1'\bar{B}^\ast+B^\ast\bar{B}_1')}$     &$1:2$    &$4:0$   & $4:0$   & $4:0$ &$1:2$  & --& --  \\
      &&$\frac{\frac{1}{\sqrt{2}}(B_1\bar{B}^\ast+B^\ast\bar{B}_1)}{\frac{1}{\sqrt{2}}(B_2\bar{B}^\ast-B^\ast\bar{B}_2)}$ &$1:5$ $(1:5.2)$    &$9:5$ $(1.7:1)$  & $9:5$ $(1.7:1)$  & $9:5$ $(1.7:1)$ &$1:5$ & --& -- \\
      &&$\frac{\frac{1}{\sqrt{2}}(B_1'\bar{B}-B\bar{B}_1')}{\frac{1}{\sqrt{2}}(B_1'\bar{B}^\ast+B^\ast\bar{B}_1')}$      &$1:2$ $(1:2.1)$   &$4:0$   & $4:0$  &  $4:0$ &$1:2$ & --& --  \\
      &&$\frac{\frac{1}{\sqrt{2}}(B_1\bar{B}-B\bar{B}_1)}{\frac{1}{\sqrt{2}}(B_1\bar{B}^\ast+B^\ast\bar{B}_1)}$           &$2:1$ $(1.9:1)$    &$2:9$ $(1:5.0)$  & $2:9$ $(1:5.0)$ & $2:9$ $(1:5.0)$  &$2:1$  & --& -- \\
      &&$\frac{\frac{1}{\sqrt{2}}(B_0\bar{B}^\ast-B^\ast\bar{B}_0)}{\frac{1}{\sqrt{2}}(B_1'\bar{B}-B\bar{B}_1')}$ &$1:1$ &$1:1$ &$1:1$ &$1:1$ &$1:1$ & --& -- \\
      &&$\frac{\frac{1}{\sqrt{2}}(B_1\bar{B}-B\bar{B}_1)}{\frac{1}{\sqrt{2}}(B_2\bar{B}^\ast-B^\ast\bar{B}_2)}$ & $2:5$ $(1:2.8)$ &$2:5$ $(1:2.9)$ &$2:5$ $(1:2.9)$ &$2:5$ $(1:3.0)$ &$2:5$ & --& -- \\
                  %{Initial state$2^{--}$}
      &--&--&--&--& -- & --& --  & --& -- \\
      % \midrule[1pt]
      &\multirow{5}{*}{$2^{--}$} &$\frac{\frac{1}{\sqrt{2}}(B_1\bar{B}^\ast-B^\ast\bar{B}_1)}{\frac{1}{\sqrt{2}}(B_2\bar{B}^\ast+B^\ast\bar{B}_2)}$    &--&-- & $1:9$ $(1:9.5)$  & $1:9$ $(1:9.6)$ & $1:1$ & $1:1$ $(1:1.1)$ & $1:1$ \\
      &&$\frac{\frac{1}{\sqrt{2}}(B_2\bar{B}-B\bar{B}_2)}{\frac{1}{\sqrt{2}}(B_2\bar{B}^\ast+B^\ast\bar{B}_2)}$  &--&--   & $2:3$ $(1:1.7)$  & $2:3$ $(1:1.7)$ & $2:3$ & $2:3$ $(1:1.7)$ & $2:3$ \\
      &&$\frac{\frac{1}{\sqrt{2}}(B_1\bar{B}^\ast-B^\ast\bar{B}_1)}{\frac{1}{\sqrt{2}}(B_2\bar{B}-B\bar{B}_2)}$    &--&-- & $1:6$ $(1:5.7)$  & $1:6$ $(1:5.7)$ & $3:2$ & $3:2$ $(1.6:1)$ & $3:2$ \\\cline{2-12}
      && & $\eta_b\gamma(E1)$ & $\chi_{b0}\gamma(M1)$ & $\chi_{b1}\gamma(M1)$  & $\chi_{b2}\gamma(M1)$ & $\eta_{b2}(1^1D_2)\gamma(E1)$ \\
      &$1^{+-}$ & $\frac{\frac{1}{\sqrt{2}}(B\bar{B}^\ast-B^\ast\bar{B})}{B^\ast\bar{B}^\ast}$     &$1:1$ $(1:1.1)$   &$1:1$ $(1:1.2)$  & $1:1$ $(1:1.2)$  & $1:1$ $(1:1.2)$ &$1:1$    \\\cline{2-12}

      %{Initial state $1^{-+}$}
      && & $\Upsilon\gamma(M1)$ & $h_b\gamma(E1)$  &$\Upsilon\gamma(E2)$ & $\Upsilon(1^3D_1)\gamma(M1)$ & $\Upsilon(1^3D_2)\gamma(M1)$ & $\Upsilon(1^3D_3)\gamma(M1)$ & $\Upsilon(1^3D_1)\gamma(E2)$ & $\Upsilon(1^3D_2)\gamma(E2)$ & $\Upsilon(1^3D_3)\gamma(E2)$ \\
      %\midrule[1pt]
      &\multirow{9}{*}{$1^{-+}$} & $\frac{\frac{1}{\sqrt{2}}(B_0\bar{B}^\ast+B^\ast\bar{B}_0)}{\frac{1}{\sqrt{2}}(B_1'\bar{B}^\ast-B^\ast\bar{B}_1')}$     &$16:0$    &$1:2$   &$0:0$ & $16:0$ & $16:0$ &--& $1:2$ & $1:2$ & $1:2$ \\
      &&$\frac{\frac{1}{\sqrt{2}}(B_1\bar{B}^\ast-B^\ast\bar{B}_1)}{\frac{1}{\sqrt{2}}(B_2\bar{B}^\ast+B^\ast\bar{B}_2)}$ &$9:5$ $(1.7:1)$    &$1:5$ $(1:5.3)$ &$5:1$ $(4.7:1)$ & $\frac{9(1-\sqrt{5}x)^2}{(\sqrt{5}+3x)^2}$ & $\frac{(3+\sqrt{5}x)^2}{(\sqrt{5}-x)^2}$ &--& $\frac{(3\sqrt{3}-\sqrt{35}x')^2}{(\sqrt{15}+\sqrt{7}x')^2}$ & $\frac{(3+\sqrt{105}x')^2}{(\sqrt{5}-\sqrt{21}x')^2}$ & $\frac{3(\sqrt{21}+\sqrt{5}x')^2}{(\sqrt{35}-\sqrt{3}x')^2}$ \\
      &&$\frac{\frac{1}{\sqrt{2}}(B_1'\bar{B}+B\bar{B}_1')}{\frac{1}{\sqrt{2}}(B_1'\bar{B}^\ast-B^\ast\bar{B}_1')}$      &$16:0$    &$1:2$ $(1:2.2)$ &$0:0$ & $16:0$ & $16:0$ &--& $1:2$ & $1:2$ & $1:2$  \\%\hline
      &&$\frac{\frac{1}{\sqrt{2}}(B_1\bar{B}+B\bar{B}_1)}{\frac{1}{\sqrt{2}}(B_1\bar{B}^\ast-B^\ast\bar{B}_1)}$           &$2:9$ $(1:4.8)$   &$2:1$ $(1.8:1)$ &$2:1$ $(1.8:1)$ & $\frac{2(1+3\sqrt{5}x)^2}{9(1-\sqrt{5}x)^2}$ & $\frac{2(1-\sqrt{5}x)^2}{(3+\sqrt{5}x)^2}$ &--& $\frac{2(\sqrt{3}+\sqrt{35}x')^2}{(3\sqrt{3}-\sqrt{35}x')^2}$ & $\frac{2(\sqrt{3}-3\sqrt{35}x')^2}{9(\sqrt{3}+\sqrt{35}x')^2}$ & $\frac{2(\sqrt{7}-\sqrt{15}x')^2}{3(\sqrt{21}+\sqrt{5}x')^2}$ \\%\hline
      &&$\frac{\frac{1}{\sqrt{2}}(B_0\bar{B}^\ast+B^\ast\bar{B}_0)}{\frac{1}{\sqrt{2}}(B_1'\bar{B}+B\bar{B}_1')}$ &$1:1$ &$1:1$ &$0:0$ & $1:1$ & $1:1$ &--& $\frac{(1-\sqrt{3}\alpha)^2}{(1+\sqrt{3}\alpha)^2}$ & $\frac{(\sqrt{3}-\alpha)^2}{(\sqrt{3}+\alpha)^2}$ & $\frac{(\sqrt{3}+2\alpha)^2}{(\sqrt{3}-2\alpha)^2}$\\%\hline
      &&$\frac{\frac{1}{\sqrt{2}}(B_1\bar{B}+B\bar{B}_1)}{\frac{1}{\sqrt{2}}(B_2\bar{B}^\ast+B^\ast\bar{B}_2)}$ &$2:5$ $(1:2.8)$ &$2:5$ $(1:2.9)$ &$10:1$ (8.2:1) & $\frac{2(1+3\sqrt{5}x)^2}{(\sqrt{5}+3x)^2}$ & $\frac{2(1-\sqrt{5}x)^2}{(\sqrt{5}-x)^2}$ &--& $\frac{2(\sqrt{3}+\sqrt{35}x')^2}{(\sqrt{15}+\sqrt{7}x')^2}$ & $\frac{2(1-\sqrt{105}x')^2}{(\sqrt{5}-\sqrt{21}x')^2}$ & $\frac{2(\sqrt{21}-3\sqrt{5}x')^2}{3(\sqrt{35}-\sqrt{3}x')^2}$\\%\hline
      % {Initial state$2^{-+}$} & $\Upsilon(1^3S_1)\gamma(M1)$ & $h_b(1^1P_1)\gamma(E1)$ & $\Upsilon(1^3S_1)\gamma(E2)$   \\
      %\hline
      &--&--&--&--& -- &--&--&--&--&--&--  \\
      &\multirow{5}{*}{$2^{-+}$}& $\frac{\frac{1}{\sqrt{2}}(B_1\bar{B}^\ast+B^\ast\bar{B}_1)}{\frac{1}{\sqrt{2}}(B_2\bar{B}^\ast-B^\ast\bar{B}_2)}$     & $1:9$ $(1:9.4)$ & $0:0$ & $9:1$ $(8.4:1)$ & $\frac{(1-9\sqrt{5}x)^2}{9(1-\sqrt{5}x)^2}$ & $\frac{9(\sqrt{5}-25x)^2}{(9\sqrt{5}-25x)^2}$ & $\frac{9(1+\sqrt{5}x)^2}{(9+\sqrt{5}x)^2}$ & $\frac{(\sqrt{3}-3\sqrt{35}x')^2}{(3\sqrt{3}-\sqrt{35}x')^2}$ & $\frac{(\sqrt{105}-45x')^2}{9(\sqrt{105}-5x')^2}$ & $\frac{(\sqrt{21}+9\sqrt{5}x')^2}{9(\sqrt{21}+\sqrt{5}x')^2}$ \\%\hline
      &&$\frac{\frac{1}{\sqrt{2}}(B_2\bar{B}+B\bar{B}_2)}{\frac{1}{\sqrt{2}}(B_2\bar{B}^\ast-B^\ast\bar{B}_2)}$     & $2:3$ $(1:1.6)$  & $0:0$ & $6:1$ $(5.3:1)$ & $\frac{2(1+\sqrt{45}x)^2}{3(1-\sqrt{5}x)^2}$ & $\frac{6(3\sqrt{5}-25x)^2}{(9\sqrt{5}-25x)^2}$ & $\frac{6(3-\sqrt{5}x)^2}{(9+\sqrt{5}x)^2}$ & $\frac{2(3+\sqrt{105}x')^2}{(3\sqrt{3}-\sqrt{35}x')^2}$ & $\frac{2(\sqrt{35}+5\sqrt{3}x')^2}{(\sqrt{105}-5x')^2}$ & $\frac{2(\sqrt{7}-\sqrt{15}x')^2}{(\sqrt{21}+\sqrt{5}x')^2}$ \\%\hline
      &&$\frac{\frac{1}{\sqrt{2}}(B_1\bar{B}^\ast+B^\ast\bar{B}_1)}{\frac{1}{\sqrt{2}}(B_2\bar{B}+B\bar{B}_2)}$     & $1:6$ $(1:5.8)$ & $0:0$ & $3:2$ $(1.6:1)$ & $\frac{(1-9\sqrt{5}x)^2}{6(1+\sqrt{45}x)^2}$ & $\frac{3(\sqrt{5}-25x)^2}{2(3\sqrt{5}-25x)^2}$ & $\frac{3(\sqrt{5}-25x)^2}{2(3\sqrt{5}-25x)^2}$ & $\frac{(\sqrt{3}-3\sqrt{35}x')^2}{2(3+\sqrt{105}x')^2}$ & $\frac{(\sqrt{105}-45x')^2}{18(\sqrt{35}+5\sqrt{3}x')^2}$ & $\frac{(\sqrt{21}+9\sqrt{5}x')^2}{18(\sqrt{7}-\sqrt{15}x')^2}$\\
      &--&--&--&--& -- &--& --&--& --&--&-- \\
      &$0^{-+}$ & $\frac{\frac{1}{\sqrt{2}}(B_0\bar{B}+B\bar{B}_0)}{\frac{1}{\sqrt{2}}(B_1'\bar{B}^\ast+B^\ast\bar{B}_1')}$  & $3:1$  & $0:0$ &--& $3:1$  &--&--&--& $3:1$ &-- \\\cline{2-12}
      && &  $\Upsilon\gamma(E1)$ & $h_b\gamma(M1)$ & $\Upsilon(1^3D_1)\gamma(E1)$ \\
      &$0^{++}$ & $\frac{B\bar{B}}{B^\ast\bar{B}^\ast}$  & $3:1$ $(2.4:1)$  & $1:3$ $(1:4.3)$ & $3:1$ \\\cline{2-12}
      && &  $\chi_{b1}\gamma(E1)$ & $\eta_{b2}(1^1D_2)\gamma(E2)$ \\
      &$0^{--}$ & $\frac{\frac{1}{\sqrt{2}}(B_0\bar{B}-B\bar{B}_0)}{\frac{1}{\sqrt{2}}(B_1'\bar{B}^\ast-B^\ast\bar{B}_1')}$     & $3:1$  & $0:0$ \\\bottomrule[1pt]

      \end{tabular}
\end{center}
\end{table*}

There are six hidden beauty molecular(resonant) states with
$J^{PC}=1^{--}$, which can decay into $\chi_{bJ}(1^3P_J)\gamma$
($J=0,1,2$). Except
$\frac{1}{\sqrt{2}}(B_1'\bar{B}^\ast+B^\ast\bar{B}_1')$, the
remaining five molecular(resonant) states have the same ratio
$\Gamma(\chi_{b0}\gamma(E1)):\Gamma(\chi_{b1}\gamma(E1)):\Gamma(\chi_{b2}\gamma(E1))=4:3:5$
without considering the contribution of the phase space factor. This
phenomena can be understood well since these decay widths are
related to the spin configuration $(1_H^-\otimes
1_l^+)|_{J=1}^{--}$. For
$\frac{1}{\sqrt{2}}(B_1'\bar{B}^\ast+B^\ast\bar{B}_1')$, its decays
into $\chi_{bJ}\gamma(E1)$ are strongly suppressed due to heavy
quark symmetry.

Both $\frac{1}{\sqrt{2}}(B\bar{B}^\ast-B^\ast\bar{B})$ and
$B^\ast\bar{B}^\ast$ decay into $\chi_{bJ}(1^3P_J)\gamma$ with the
typical ratios are
$\Gamma(\chi_{b0}\gamma(M1)):\Gamma(\chi_{b1}\gamma(M1)):\Gamma(\chi_{b2}\gamma(M1))=1:3:5$
as listed in Table \ref{tab:2} if the phase space difference is
neglected. These decays are only governed by the spin configuration
$(1_H^-\otimes 0_l^-)|_{J=1}^{+-}$. We notice that the ratio $1:3:5$
is consistent with that given in Refs.
\cite{Ohkoda:2012rj,He:2013nwa}.

$\Upsilon(1^3D_1)\gamma$ and $\Upsilon(1^3D_2)\gamma$ are two
typical $M1$ radiative decays of
$\frac{1}{\sqrt{2}}(B_0\bar{B}^\ast+B^\ast\bar{B}_0)$ and
$\frac{1}{\sqrt{2}}(B_1'\bar{B}+B\bar{B}_1')$, where the spin
configuration $(1_H^-\otimes 1_l^+)_{J=1}^{-+}$ is dominant. Our
result indicates that
$\Gamma(\Upsilon(1^3D_1)\gamma(M1)):\Gamma(\Upsilon(1^3D_2)\gamma(M1))=1:3$
for both $\frac{1}{\sqrt{2}}(B_0\bar{B}^\ast+B^\ast\bar{B}_0)$ and
$\frac{1}{\sqrt{2}}(B_1'\bar{B}+B\bar{B}_1')$.

For $\frac{1}{\sqrt{2}}(B\bar{B}^\ast+B^\ast\bar{B})$ with
$J^{PC}=1^{++}$, it decays into $\Upsilon(1^3D_1)\gamma$ and
$\Upsilon(1^3D_2)\gamma$  via the $E1$ transition, where the spin
configuration is $(1_H^-\otimes 1_l^-)_{J=1}^{++}$. Its ratio
$\Gamma(\Upsilon(1^3D_1)\gamma(E1)):\Gamma(\Upsilon(1^3D_2)\gamma(E1))$
is also $1:3$.

When the initial state is
$\frac{1}{\sqrt{2}}(B_1\bar{B}+B\bar{B}_1)$,
$\frac{1}{\sqrt{2}}(B_1\bar{B}^\ast-B^\ast\bar{B}_1)$ and
$\frac{1}{\sqrt{2}}(B_2\bar{B}^\ast+B^\ast\bar{B}_2)$ with
$J^{PC}=1^{-+}$, we find that their decays into $\Upsilon(1^3D_J)$
$(J=1,2,3)$ depend on the two spin configurations $(1_H^-\otimes
1_l^+)_{J=1}^{-+}$ and $(1_H^-\otimes 2_l^+)_{J=1}^{-+}$. Thus, we
define
$$
 x=\frac{H_{22}(M1)}{H_{21}(M1)},\,\,x'=\frac{H_{22}(E2)}{H_{21}(E2)},\,\,\alpha=\frac{H_{21}(E2)}{H_{20}(E2)},
$$
where $H_{21}(M1)=\langle 1,2\|H_{eff}(M1)\|1\rangle$ and
$H_{22}(M1)=\langle 1,2\|H_{eff}(M1)\|2\rangle$ are the reduced
matrix elements for the $M1$ transition. $H_{20}(E2)=\langle
2,2\|H_{eff}(E2)\|0\rangle$, $H_{21}(E2)=\langle
2,2\|H_{eff}(E2)\|1\rangle$ and $H_{22}(E2)=\langle
2,2\|H_{eff}(E2)\|2\rangle$ are the reduced matrix elements for the
$E2$ transition.

The $E2$ decay modes $\chi_{bJ}(1^3P_J)\gamma$ $(J=1,2)$ of the two
states $\frac{1}{\sqrt{2}}(B\bar{B}^\ast-B^\ast\bar{B})$ and
$B^\ast\bar{B}^\ast$ with $J^{PC}=1^{+-}$ depend on the spin
configurations $(0_H^-\otimes 1_l^-)_{J=1}^{+-}$ and $(1_H^-\otimes
0_l^-)_{J=1}^{+-}$. From Table \ref{final state}, we conclude that
these $\chi_{b1}\gamma(E2)$ and $\chi_{b2}\gamma(E2)$ modes are
suppressed due to heavy quark symmetry.

We can see from Table \ref{recouple}, the $B_1'\bar{B}^\ast$ and
$B^\ast\bar{B}_1'$ molecular(resonant) states with $J=1$ contain the
spin configurations $(0_H^-\otimes 1_l^+)_{J=1}^{-+}$,
$(0_H^-\otimes 1_l^+)_{J=1}^{--}$ and $(1_H^-\otimes
0_l^+)_{J=1}^{-+}$. On the other hand, we know from Table \ref{final
state} that the recoupled final state $\Upsilon(1^3S_1)\gamma(M1)$
with $J=1$ only contains the component of $(1_H^-\otimes
1_l^+)_{J=1}^{-+}$. So both $B_1'\bar{B}^\ast$ and
$B^\ast\bar{B}_1'$ components with $J=1$ can independently decay
into $\Upsilon(1^3S_1)\gamma(M1)$. Unfortunately, these two parts
have the opposite relative phase for the component of $(1_H^-\otimes
1_l^+)_{J=1}^{-+}$. When they constitute the C-parity eigenstate
$\frac{1}{\sqrt{2}}(B_1'\bar{B}^\ast-B^\ast\bar{B}_1')$, the
radiative decay into $\Upsilon(1^3S_1)$ $\gamma(M1)$ is suppressed
in the heavy quark limit, as listed in Table.\ref{tab:3}.

Since the $0^{--}$ states
$\frac{1}{\sqrt{2}}(B_0\bar{B}-B\bar{B}_0)$ and
$\frac{1}{\sqrt{2}}(B_1'\bar{B}^\ast-B^\ast\bar{B}_1')$ do not
contain the spin configuration $(0_H^-\otimes 2_l^+)_{J=1}^{--}$,
their decays into $\eta_{b2}(1^1D_2)\gamma(E2)$ are suppressed.

Similarly, the $0^{-+}$ states
$\frac{1}{\sqrt{2}}(B_0\bar{B}+B\bar{B}_0)$ and
$\frac{1}{\sqrt{2}}(B_1'\bar{B}^\ast+B^\ast\bar{B}_1')$ do not
contain the spin configuration $(0_H^-\otimes 0_l^+)_{J=1}^{-+}$.
Thus, their decays into $\eta_{b2}(1^1D_2)\gamma(E2)$ are also
suppressed.

From Table \ref{tab:2} we see that the ratio
$\Gamma(\chi_{b1}(1^3P_1)\gamma(E1)):\Gamma(\chi_{b2}(1^3P_2)\gamma(E1))=1:3$
is the same for all the $2^{--}$ molecular(resonant) states. These
decay widths are due to the contribution of the spin configuration
$(1_H^-\otimes 1_l^+)|_{J=2}^{--}$.

Since $\frac{1}{\sqrt{2}}(B_1\bar{B}^\ast+B^\ast\bar{B}_1)$,
$\frac{1}{\sqrt{2}}(B_2\bar{B}^\ast-B^\ast\bar{B}_2)$ and
$\frac{1}{\sqrt{2}}(B_2\bar{B}+B\bar{B}_2)$ do not contain the spin
configuration $(0_H^-\otimes 2_l^+)_{J=2}^{-+}$, their decays into
$h_b(1^1P_1)\gamma(E1)$ are suppressed in the heavy quark limit.

We need to specify that the ratios shown in Tables \ref{tab:2} and
\ref{tab:3} are also suitable for the radiative decays involving the
higher radially excited bottomonium.

%%%%%%%%%%%%%%%%%%%%%%%%%%%%%%%%%%%%%%%%%%%%%%%%%%%%%%
\subsection{$(b\bar{b})\to \mathfrak{M}+\gamma$}
%%%%%%%%%%%%%%%%%%%%%%%%%%%%%%%%%%%%%%%%%%%%%%%%%%%%%%

Adopting the same formalism, we obtain the typical ratios of the
$(b\bar{b})\to \mathfrak{M}+\gamma$ decay widths, which are listed
in Tables \ref{tab:4} and \ref{tab:5}.

Among all the possible hidden beauty molecular(resonant) states, the
mass of the lowest state
$\frac{1}{\sqrt{2}}(B\bar{B}^\ast-B^\ast\bar{B})$ is around 10610
MeV. At present, $\Upsilon(10860)$ is the heaviest bottomonium
state. Thus we can only study the production of the hidden beauty
molecular(resonant) states via the radiative decays of the higher
radial excitations of bottomonium.

In Table \ref{tab:4}, we define
$$
y=\frac{H_{22}(M1)}{H_{12}(M1)},y'=\frac{H_{22}(E2)}{H_{12}(E2)},\beta=\frac{H_{12}(E2)}{H_{02}(E2)}
$$
where $H_{12}(M1)=\langle 1,1\|H_{eff}(M1)\|2\rangle$ and
$H_{22}(M1)=\langle 1,2\|H_{eff}(M1)\|2\rangle$ are the reduced
matrix elements for the $M1$ transition. $H_{02}(E2)=\langle
2,0\|H_{eff}(E2)\|2\rangle$, $H_{12}(E2)=\langle
2,1\|H_{eff}(E2)\|2\rangle$ and $H_{22}(E2)=\langle
2,2\|H_{eff}(E2)\|2\rangle$ are the reduced matrix elements of the
$E2$ transition. If comparing the results shown in Tables
\ref{tab:2} and \ref{tab:4}, we notice the similarity between these
ratios. For example, the
$\frac{1}{\sqrt{2}}(B_0\bar{B}^\ast-B^\ast\bar{B}_0)$
molecular(resonant) state can decay into $\chi_{bJ}(1^3P_J)\gamma$
($J=0,1,2$). In fact, if there exist the following relations between
the relevant reduced matrix elements
\begin{eqnarray*}
  &&H_{21}(M1)=-\frac{\sqrt{5}}{\sqrt{3}}H_{12}(M1) ,\,\,H_{21}(E2)=-\frac{\sqrt{5}}{\sqrt{3}}H_{12}(E2),\\
  &&H_{20}(E2)=\sqrt{5}H_{02}(E2),
\end{eqnarray*}
the obtained ratio
$\Gamma(\chi_{b0}\gamma(E1)):\Gamma(\chi_{b1}\gamma(E1)):\Gamma(\chi_{b2}\gamma(E1))$
is the same as that listed in the second row in Table \ref{tab:4},
where $\frac{1}{\sqrt{2}}(B_0\bar{B}^\ast-B^\ast\bar{B}_0)$ is
produced by the $E1$ radiative decays of $\chi_{bJ}(n^3P_J)$
($J=0,1,2$). This phenomenon reflects the crossing symmetry. In the
above discussion we do not consider the effect from the phase space
factor of the corresponding process.

\begin{table*}[htbp]
\begin{center}
\caption{\label{tab:4} The ratios between the $(b\bar{b})\to
\mathfrak{M}+\gamma$ decay widths. The parameters $y, y', \beta$ are
defined as $y=H_{22}(M1)/H_{12}(M1)$, $y'=H_{22}(E2)/H_{12}(E2)$,
and $\beta=H_{12}(E2)/H_{02}(E2)$, respectively.}
   \begin{tabular}{c|ccccccccccc} \toprule[1pt]
      %\multicolumn{2}{|c|}
      &$J^{PC}$ & Final state & Initial state \\\cline{2-4}
   \multirow{46}{*}{\rotatebox{90}{Initial \,\, state}}&&& $\chi_{b0}(n^3P_0):\chi_{b1}(n^3P_1):\chi_{b2}(n^3P_2)$  \\
     % \midrule[1pt]
      &\multirow{6}{*}{$1^{--}$} & $\frac{1}{\sqrt{2}}(B_0\bar{B}^\ast-B^\ast\bar{B}_0)\gamma(E1)$     & $4:3:5$   \\%\hline
      &&$\frac{1}{\sqrt{2}}(B_1'\bar{B}-B\bar{B}_1')\gamma(E1)$     & $4:3:5$    \\%\hline
      &&$\frac{1}{\sqrt{2}}(B_1\bar{B}-B\bar{B}_1)\gamma(E1)$     & $4:3:5$    \\%\hline
      &&$\frac{1}{\sqrt{2}}(B_1'\bar{B}^\ast+B^\ast\bar{B}_1')\gamma(E1)$     & $0:0:0$   \\%\hline
      &&$\frac{1}{\sqrt{2}}(B_1\bar{B}^\ast+B^\ast\bar{B}_1)\gamma(E1)$     & $4:3:5$    \\%\hline
      &&$\frac{1}{\sqrt{2}}(B_2\bar{B}^\ast-B^\ast\bar{B}_2)\gamma(E1)$     & $4:3:5$   \\\cline{2-4}
     && & $\chi_{b0}(n^3P_0):\chi_{b1}(n^3P_1):\chi_{b2}(n^3P_2)$  \\
     % \midrule[1pt]
     &\multirow{2}{*}{$1^{+-}$} & $\frac{1}{\sqrt{2}}(B\bar{B}^\ast-B^\ast\bar{B})\gamma(M1)$     & $1:3:5$ \\%\hline
      &&$B^\ast\bar{B}^\ast\gamma(M1)$     & $1:3:5$   \\\cline{2-4}
     &&& $\Upsilon(n^3D_1):\Upsilon(n^3D_2)$  \\
      %\midrule[1pt]
&\multirow{6}{*}{$1^{-+}$}     &  $\frac{1}{\sqrt{2}}(B_0\bar{B}^\ast+B^\ast\bar{B}_0)\gamma(M1)$     & $1:3$   \\%\hline
     && $\frac{1}{\sqrt{2}}(B_1'\bar{B}+B\bar{B}_1')\gamma(M1)$     & $1:3$   \\%\hline
      &&$\frac{1}{\sqrt{2}}(B_1\bar{B}+B\bar{B}_1)\gamma(M1)$     & $\frac{(1-3\sqrt{3}y)^2}{3(1+\sqrt{3}y)^2}$   \\%\hline
      &&$\frac{1}{\sqrt{2}}(B_1'\bar{B}^\ast-B^\ast\bar{B}_1')\gamma(M1)$     & $0:0$   \\%\hline
      &&$\frac{1}{\sqrt{2}}(B_1\bar{B}^\ast-B^\ast\bar{B}_1)\gamma(M1)$     & $\frac{(1+\sqrt{3}y)^2}{(\sqrt{3}-y)^2}$   \\%\hline
      &&$\frac{1}{\sqrt{2}}(B_2\bar{B}^\ast+B^\ast\bar{B}_2)\gamma(M1)$     & $\frac{(5-3\sqrt{3}y)^2}{3(5+\sqrt{3}y)^2}$   \\\cline{2-4}
    &&& $\Upsilon(n^3D_1):\Upsilon(n^3D_2)$  \\
      %\midrule[1pt]
     &\multirow{1}{*}{$1^{++}$} & $\frac{1}{\sqrt{2}}(B\bar{B}^\ast+B^\ast\bar{B})\gamma(E1)$     & $1:3$   \\\cline{2-4}
      && & $\chi_{b1}(n^3P_1):\chi_{b2}(n^3P_2)$  \\
     % \midrule[1pt]
    &\multirow{2}{*}{$1^{+-}$} &  $\frac{1}{\sqrt{2}}(B\bar{B}^\ast-B^\ast\bar{B})\gamma(E2)$     & $0:0$   \\%\hline
     && $B^\ast\bar{B}^\ast\gamma(E2)$     & $0:0$   \\\cline{2-4}
    &&& $\Upsilon(n^3D_1):\Upsilon(n^3D_2):\Upsilon(n^3D_3)$  \\
     % \midrule[1pt]
      &\multirow{6}{*}{$1^{-+}$} &$\frac{1}{\sqrt{2}}(B_0\bar{B}^\ast+B^\ast\bar{B}_0)\gamma(E2)$     & $27(1+\beta)^2:5(3+\beta)^2:7(3-2\beta)^2$   \\ %\hline
     && $\frac{1}{\sqrt{2}}(B_1'\bar{B}+B\bar{B}_1')\gamma(E2)$     & $27(1-\beta)^2:5(3-\beta)^2:7(3+2\beta)^2$   \\%\hline
      &&$\frac{1}{\sqrt{2}}(B_1\bar{B}+B\bar{B}_1)\gamma(E2)$     & $27(1-\sqrt{7}y')^2:5(1-3\sqrt{7}y')^2:4(\sqrt{7}+3y')^2$   \\%\hline
    && $\frac{1}{\sqrt{2}}(B_1'\bar{B}^\ast-B^\ast\bar{B}_1')\gamma(E2)$     & $3:5:7$   \\%\hline
    &&  $\frac{1}{\sqrt{2}}(B_1\bar{B}^\ast-B^\ast\bar{B}_1)\gamma(E2)$     & $3(3+\sqrt{7}y')^2:5(1-\sqrt{7}y')^2:4(\sqrt{7}-y')^2$  \\%\hline
     && $\frac{1}{\sqrt{2}}(B_2\bar{B}^\ast+B^\ast\bar{B}_2)\gamma(E2)$     & $27(5-\sqrt{7}y')^2:5(5+3\sqrt{7}y')^2:4(5\sqrt{7}+3y')^2$\\\cline{2-4}
      &&& $\chi_{b1}(n^3P_1):\chi_{b2}(n^3P_2)$   \\
      %\midrule[1pt]
    &\multirow{4}{*}{$2^{--}$} & $\frac{1}{\sqrt{2}}(B_1\bar{B}^\ast-B^\ast\bar{B}_1)\gamma(E1)$     & $1:3$    \\%\hline
      &&$\frac{1}{\sqrt{2}}(B_1'\bar{B}^\ast-B^\ast\bar{B}_1')\gamma(E1)$     & $1:3$    \\%\hline
      &&$\frac{1}{\sqrt{2}}(B_2\bar{B}-B\bar{B}_2)\gamma(E1)$     & $1:3$   \\%\hline
      &&$\frac{1}{\sqrt{2}}(B_2\bar{B}^\ast+B^\ast\bar{B}_2)\gamma(E1)$     & $1:3$  \\\cline{2-4}
      && & $\Upsilon(n^3D_1):\Upsilon(n^3D_2):\Upsilon(n^3D_3)$   \\
      %\midrule[1pt]
    &$2^{++}$&  $B^\ast\bar{B}^\ast\gamma(E1)$     & $1:15:84$    \\\cline{2-4}
      &&& $\Upsilon(n^3D_1):\Upsilon(n^3D_2):\Upsilon(n^3D_3)$  \\
      %\midrule[1pt]
     &\multirow{8}{*}{$2^{-+}$}&\multirow{1}{*}{$\frac{1}{\sqrt{2}}(B_1'\bar{B}^\ast+B^\ast\bar{B}_1')\gamma(M1)$ }    & $1:15:84$   \\%\midrule[1pt]
     && $\frac{1}{\sqrt{2}}(B_1\bar{B}^\ast+B^\ast\bar{B}_1)\gamma(M1)$ & $(1+9\sqrt{3}y)^2:15(1+5\sqrt{3}y)^2:84(1-\sqrt{3}y)^2$\\
     && $\frac{1}{\sqrt{2}}(B_2\bar{B}+B\bar{B}_2)\gamma(M1)$ & $(1-3\sqrt{3}y)^2:5(\sqrt{3}+5y)^2:28(\sqrt{3}+y)^2$\\
     && $\frac{1}{\sqrt{2}}(B_2\bar{B}^\ast-B^\ast\bar{B}_2)\gamma(M1)$ & $27(1+\sqrt{3}y)^2:5(9+5\sqrt{3}y)^2:28(9-\sqrt{3}y)^2$\\

     &&$\frac{1}{\sqrt{2}}(B_1'\bar{B}^\ast+B^\ast\bar{B}_1')\gamma(E2)$        & $9:35:56$  \\
     && $\frac{1}{\sqrt{2}}(B_1\bar{B}^\ast+B^\ast\bar{B}_1)\gamma(E2)$ & $9(1+3\sqrt{7}y')^2:5(\sqrt{7}+9y')^2:8(\sqrt{7}-9y')^2$\\
     && $\frac{1}{\sqrt{2}}(B_2\bar{B}+B\bar{B}_2)\gamma(E2)$ & $9(1-\sqrt{7}y')^2:5(\sqrt{7}-3y')^2:8(\sqrt{7}+3y')^2$\\
     && $\frac{1}{\sqrt{2}}(B_2\bar{B}^\ast-B^\ast\bar{B}_2)\gamma(E2)$ & $(3+\sqrt{7}y')^2:5(\sqrt{7}+y')^2:8(\sqrt{7}-y')^2$\\\bottomrule[1pt]
       \end{tabular}
\end{center}
\end{table*}

\begin{table*}[htbp]
\begin{center}
\caption{\label{tab:5} The typical ratios of the $(b\bar{b})\to
\mathfrak{M}+\gamma$ decay widths. The parameters $y, y', \beta$ are
defined as $y=H_{22}(M1)/H_{12}(M1)$, $y'=H_{22}(E2)/H_{12}(E2)$,
and $\beta=H_{12}(E2)/H_{02}(E2)$, respectively.}
   \begin{tabular}{c|ccccccccccc} \toprule[1pt]
      & &  & \multicolumn{2}{c}{Initial state} & &  &  & \multicolumn{3}{c}{Initial state}\\\cline{2-11}
      & & & $\eta_b(n^1S_0)$ & $\eta_{b2}(n^1D_2)$ & & & & $\chi_{b0}(n^3P_0)$ & $\chi_{b1}(n^3P_1)$  & $\chi_{b2}(n^3P_2)$ \\
      \multirow{14}{*}{\rotatebox{90}{Final state}}&\multirow{6}{*}{$1^{--}$} & $\frac{\frac{1}{\sqrt{2}}(B_0\bar{B}^\ast-B^\ast\bar{B}_0)\gamma(M1)}{\frac{1}{\sqrt{2}}(B_1'\bar{B}^\ast+B^\ast\bar{B}_1')\gamma(M1)}$     &$1:2$        &$1:2$    & & \multirow{6}{*}{$1^{--}$} & $\frac{\frac{1}{\sqrt{2}}(B_0\bar{B}^\ast-B^\ast\bar{B}_0)\gamma(E1)}{\frac{1}{\sqrt{2}}(B_1'\bar{B}^\ast+B^\ast\bar{B}_1')\gamma(E1)}$     &$4:0$   & $4:0$   & $4:0$  \\
      &&$\frac{\frac{1}{\sqrt{2}}(B_1\bar{B}^\ast+B^\ast\bar{B}_1)\gamma(M1)}{\frac{1}{\sqrt{2}}(B_2\bar{B}^\ast-B^\ast\bar{B}_2)\gamma(M1)}$ &$1:5$      &$1:5$ & & &$\frac{\frac{1}{\sqrt{2}}(B_1\bar{B}^\ast+B^\ast\bar{B}_1)\gamma(E1)}{\frac{1}{\sqrt{2}}(B_2\bar{B}^\ast-B^\ast\bar{B}_2)\gamma(E1)}$ &$9:5$   & $9:5$  & $9:5$  \\
      &&$\frac{\frac{1}{\sqrt{2}}(B_1'\bar{B}-B\bar{B}_1')\gamma(M1)}{\frac{1}{\sqrt{2}}(B_1'\bar{B}^\ast+B^\ast\bar{B}_1')\gamma(M1)}$      &$1:2$    &  $1:2$  & & & $\frac{\frac{1}{\sqrt{2}}(B_1'\bar{B}-B\bar{B}_1')\gamma(E1)}{\frac{1}{\sqrt{2}}(B_1'\bar{B}^\ast+B^\ast\bar{B}_1')\gamma(E1)}$      &$4:0$  & $4:0$  &  $4:0$   \\
      &&$\frac{\frac{1}{\sqrt{2}}(B_1\bar{B}-B\bar{B}_1)\gamma(M1)}{\frac{1}{\sqrt{2}}(B_1\bar{B}^\ast+B^\ast\bar{B}_1)\gamma(M1)}$           &$2:1$     &$2:1$  & & &$\frac{\frac{1}{\sqrt{2}}(B_1\bar{B}-B\bar{B}_1)\gamma(E1)}{\frac{1}{\sqrt{2}}(B_1\bar{B}^\ast+B^\ast\bar{B}_1)\gamma(E1)}$           &$2:9$   & $2:9$  & $2:9$     \\
      &&$\frac{\frac{1}{\sqrt{2}}(B_0\bar{B}^\ast-B^\ast\bar{B}_0)\gamma(M1)}{\frac{1}{\sqrt{2}}(B_1'\bar{B}-B\bar{B}_1')\gamma(M1)}$ &$1:1$ &$1:1$  & & & $\frac{\frac{1}{\sqrt{2}}(B_0\bar{B}^\ast-B^\ast\bar{B}_0)\gamma(E1)}{\frac{1}{\sqrt{2}}(B_1'\bar{B}-B\bar{B}_1')\gamma(E1)}$ &$1:1$ &$1:1$ &$1:1$ \\
      &&$\frac{\frac{1}{\sqrt{2}}(B_1\bar{B}-B\bar{B}_1)\gamma(M1)}{\frac{1}{\sqrt{2}}(B_2\bar{B}^\ast-B^\ast\bar{B}_2)\gamma(M1)}$ &$2:5$ &$2:5$ & & & $\frac{\frac{1}{\sqrt{2}}(B_1\bar{B}-B\bar{B}_1)\gamma(E1)}{\frac{1}{\sqrt{2}}(B_2\bar{B}^\ast-B^\ast\bar{B}_2)\gamma(E1)}$ &$2:5$ &$2:5$ &$2:5$ \\

      &&-- & -- & -- & & &-- & -- & --  & -- \\
      &$1^{+-}$ & $\frac{\frac{1}{\sqrt{2}}(B\bar{B}^\ast-B^\ast\bar{B})\gamma(E1)}{B^\ast\bar{B}^\ast\gamma(E1)}$     & $1:0$        &$1:1$   & & $1^{+-}$ & $\frac{\frac{1}{\sqrt{2}}(B\bar{B}^\ast-B^\ast\bar{B})\gamma(M1)}{B^\ast\bar{B}^\ast\gamma(M1)}$     &$1:1$        &$1:1$ &$1:1$ \\
      &&-- & -- & -- & & &-- & -- & --  & -- \\
      &\multirow{3}{*}{$2^{--}$} & $\frac{\frac{1}{\sqrt{2}}(B_1\bar{B}^\ast-B^\ast\bar{B}_1)\gamma(E2)}{\frac{1}{\sqrt{2}}(B_2\bar{B}^\ast+B^\ast\bar{B}_2)\gamma(E2)}$   & $1:1$ & $1:1$ & & \multirow{3}{*}{$2^{--}$} & $\frac{\frac{1}{\sqrt{2}}(B_1\bar{B}^\ast-B^\ast\bar{B}_1)\gamma(E1)}{\frac{1}{\sqrt{2}}(B_2\bar{B}^\ast+B^\ast\bar{B}_2)\gamma(E1)}$     &--& $1:9$  & $1:9$  \\
      &&$\frac{\frac{1}{\sqrt{2}}(B_2\bar{B}-B\bar{B}_2)\gamma(E2)}{\frac{1}{\sqrt{2}}(B_2\bar{B}^\ast+B^\ast\bar{B}_2)\gamma(E2)}$& $2:3$ & $2:3$ & & & $\frac{\frac{1}{\sqrt{2}}(B_2\bar{B}-B\bar{B}_2)\gamma(E1)}{\frac{1}{\sqrt{2}}(B_2\bar{B}^\ast+B^\ast\bar{B}_2)\gamma(E1)}$     &--& $2:3$  & $2:3$  \\
      &&$\frac{\frac{1}{\sqrt{2}}(B_1\bar{B}^\ast-B^\ast\bar{B}_1)\gamma(E2)}{\frac{1}{\sqrt{2}}(B_2\bar{B}-B\bar{B}_2)\gamma(E2)}$& $3:2$ & $3:2$ & & & $\frac{\frac{1}{\sqrt{2}}(B_1\bar{B}^\ast-B^\ast\bar{B}_1)\gamma(E1)}{\frac{1}{\sqrt{2}}(B_2\bar{B}-B\bar{B}_2)\gamma(E1)}$     &--& $1:6$  & $1:6$ \\
      &&-- & -- & -- & & &-- & -- & --  & -- \\
      &$0^{--}$ & $\frac{\frac{1}{\sqrt{2}}(B_0\bar{B}-B\bar{B}_0)\gamma(E2)}{\frac{1}{\sqrt{2}}(B_1'\bar{B}^\ast-B^\ast\bar{B}_1')\gamma(E2)}$     &--& $0:0$ & & $0^{--}$ & $\frac{\frac{1}{\sqrt{2}}(B_0\bar{B}-B\bar{B}_0)\gamma(E1)}{\frac{1}{\sqrt{2}}(B_1'\bar{B}^\ast-B^\ast\bar{B}_1')\gamma(E1)}$     &-- &$3:1$ &--\\\cline{2-11}

      &&&   $h_b(n^1P_1)$ &$\eta_{b2}(n^1D_2)$ & & & & $\Upsilon(n^3S_1)$ & $\Upsilon(n^3D_1)$ & $\Upsilon(n^3D_2)$ & \\
      \multirow{14}{*}{\rotatebox{90}{Final state}}&\multirow{6}{*}{$1^{-+}$} & $\frac{\frac{1}{\sqrt{2}}(B_0\bar{B}^\ast+B^\ast\bar{B}_0)\gamma(E1)}{\frac{1}{\sqrt{2}}(B_1'\bar{B}^\ast-B^\ast\bar{B}_1')\gamma(E1)}$     &$1:2$  &-- & & \multirow{6}{*}{$1^{-+}$} & $\frac{\frac{1}{\sqrt{2}}(B_0\bar{B}^\ast+B^\ast\bar{B}_0)\gamma(M1)}{\frac{1}{\sqrt{2}}(B_1'\bar{B}^\ast-B^\ast\bar{B}_1')\gamma(M1)}$     &$16:0$ &$16:0$ &$16:0$  \\
      && $\frac{\frac{1}{\sqrt{2}}(B_1\bar{B}^\ast-B^\ast\bar{B}_1)\gamma(E1)}{\frac{1}{\sqrt{2}}(B_2\bar{B}^\ast+B^\ast\bar{B}_2)\gamma(E1)}$ &$1:5$  &-- & & & $\frac{\frac{1}{\sqrt{2}}(B_1\bar{B}^\ast-B^\ast\bar{B}_1)\gamma(M1)}{\frac{1}{\sqrt{2}}(B_2\bar{B}^\ast+B^\ast\bar{B}_2)\gamma(M1)}$ &$9:5$ & $\frac{45(1+\sqrt{3}y)^2}{(5-3\sqrt{3}y)^2}$ & $\frac{5(3-\sqrt{3}y)^2}{(5+\sqrt{3}y)^2}$\\
      && $\frac{\frac{1}{\sqrt{2}}(B_1'\bar{B}+B\bar{B}_1')\gamma(E1)}{\frac{1}{\sqrt{2}}(B_1'\bar{B}^\ast-B^\ast\bar{B}_1')\gamma(E1)}$      &$1:2$   &-- & & & $\frac{\frac{1}{\sqrt{2}}(B_1'\bar{B}+B\bar{B}_1')\gamma(M1)}{\frac{1}{\sqrt{2}}(B_1'\bar{B}^\ast-B^\ast\bar{B}_1')\gamma(M1)}$      &$16:0$ &$16:0$ &$16:0$ \\
      && $\frac{\frac{1}{\sqrt{2}}(B_1\bar{B}+B\bar{B}_1)\gamma(E1)}{\frac{1}{\sqrt{2}}(B_1\bar{B}^\ast-B^\ast\bar{B}_1)\gamma(E1)}$           &$2:1$   &-- & & & $\frac{\frac{1}{\sqrt{2}}(B_1\bar{B}+B\bar{B}_1)\gamma(M1)}{\frac{1}{\sqrt{2}}(B_1\bar{B}^\ast-B^\ast\bar{B}_1)\gamma(M1)}$           &$2:9$ & $\frac{2(1-3\sqrt{3}y)^2}{9(1+\sqrt{3}y)^2}$ & $\frac{2(1+\sqrt{3}y)^2}{(3-\sqrt{3}y)^2}$\\
      && $\frac{\frac{1}{\sqrt{2}}(B_0\bar{B}^\ast+B^\ast\bar{B}_0)\gamma(E1)}{\frac{1}{\sqrt{2}}(B_1'\bar{B}+B\bar{B}_1')\gamma(E1)}$ &$1:1$  &-- & & & $\frac{\frac{1}{\sqrt{2}}(B_0\bar{B}^\ast+B^\ast\bar{B}_0)\gamma(M1)}{\frac{1}{\sqrt{2}}(B_1'\bar{B}+B\bar{B}_1')\gamma(M1)}$ &$1:1$ &$1:1$ &$1:1$ \\
      && $\frac{\frac{1}{\sqrt{2}}(B_1\bar{B}+B\bar{B}_1)\gamma(E1)}{\frac{1}{\sqrt{2}}(B_2\bar{B}^\ast+B^\ast\bar{B}_2)\gamma(E1)}$ &$2:5$  &-- & & & $\frac{\frac{1}{\sqrt{2}}(B_1\bar{B}+B\bar{B}_1)\gamma(M1)}{\frac{1}{\sqrt{2}}(B_2\bar{B}^\ast+B^\ast\bar{B}_2)\gamma(M1)}$ &$2:5$ & $\frac{10(1-3\sqrt{3}y)^2}{(5-3\sqrt{3}y)^2}$ & $\frac{10(1+\sqrt{3}y)^2}{(5+\sqrt{3}y)^2}$ \\
      &&-- & -- & -- & & &-- & -- & --  & -- \\
      &$0^{++}$ & $\frac{B\bar{B}\gamma(M1)}{B^\ast\bar{B}^\ast\gamma(M1)}$   & $1:3$ & & & $0^{++}$ & $\frac{B\bar{B}\gamma(E1)}{B^\ast\bar{B}^\ast\gamma(E1)}$  & $3:1$   & $3:1$ &--\\
      &&-- & -- & -- & & &-- & -- & --  & -- \\
      &$0^{-+}$ & $\frac{\frac{1}{\sqrt{2}}(B_0\bar{B}+B\bar{B}_0)\gamma(E1)}{\frac{1}{\sqrt{2}}(B_1'\bar{B}^\ast+B^\ast\bar{B}_1')\gamma(E1)}$  & $0:0$ &-- & & $0^{-+}$ & $\frac{\frac{1}{\sqrt{2}}(B_0\bar{B}+B\bar{B}_0)\gamma(M1)}{\frac{1}{\sqrt{2}}(B_1'\bar{B}^\ast+B^\ast\bar{B}_1')\gamma(M1)}$  & $3:1$  & $3:1$ &--\\
      &&-- & -- & -- & & &-- & -- & --  & -- \\
      &\multirow{3}{*}{$2^{--}$} & $\frac{\frac{1}{\sqrt{2}}(B_1\bar{B}^\ast-B^\ast\bar{B}_1)\gamma(M1)}{\frac{1}{\sqrt{2}}(B_2\bar{B}^\ast+B^\ast\bar{B}_2)\gamma(M1)}$     &--& $1:1$ & & $0^{-+}$ & $\frac{\frac{1}{\sqrt{2}}(B_0\bar{B}+B\bar{B}_0)\gamma(E2)}{\frac{1}{\sqrt{2}}(B_1'\bar{B}^\ast+B^\ast\bar{B}_1')\gamma(E2)}$  &-- &--& $3:1$ \\
      && $\frac{\frac{1}{\sqrt{2}}(B_2\bar{B}-B\bar{B}_2)\gamma(M1)}{\frac{1}{\sqrt{2}}(B_2\bar{B}^\ast+B^\ast\bar{B}_2)\gamma(M1)}$     &--& $2:3$ & \\
      && $\frac{\frac{1}{\sqrt{2}}(B_1\bar{B}^\ast-B^\ast\bar{B}_1)\gamma(M1)}{\frac{1}{\sqrt{2}}(B_2\bar{B}-B\bar{B}_2)\gamma(M1)}$     &--& $3:2$ & \\\midrule[1pt]

   \end{tabular}
\end{center}
\end{table*}

\begin{table}[htbp]
\begin{center}
\caption{\label{tab:6} The typical ratios between the
$(b\bar{b})\to\mathfrak{M}+\gamma$ decay widths. The parameters $y,
y', \beta$ are defined as $y=H_{22}(M1)/H_{12}(M1)$,
$y'=H_{22}(E2)/H_{12}(E2)$, and $\beta=H_{12}(E2)/H_{02}(E2)$,
respectively.}
   \begin{tabular}{c|cccccccccccccc} \toprule[1pt]
      & &  & \multicolumn{3}{c}{Initial state} \\\cline{2-6}
      & & & $\Upsilon(n^3D_1)$ & $\Upsilon(n^3D_2)$ & $\Upsilon(n^3D_3)$\\
      \multirow{16}{*}{\rotatebox{90}{Final state}} & \multirow{6}{*}{$1^{-+}$} &$\frac{\frac{1}{\sqrt{2}}(B_0\bar{B}^\ast+B^\ast\bar{B}_0)\gamma(E2)}{\frac{1}{\sqrt{2}}(B_1'\bar{B}^\ast-B^\ast\bar{B}_1')\gamma(E2)}$ & $1:2$ & $1:2$ & $1:2$\\
      &&$\frac{\frac{1}{\sqrt{2}}(B_1\bar{B}^\ast-B^\ast\bar{B}_1)\gamma(E2)}{\frac{1}{\sqrt{2}}(B_2\bar{B}^\ast+B^\ast\bar{B}_2)\gamma(E2)}$ & $\frac{5(3+\sqrt{7}y')^2}{(5-\sqrt{7}y')^2}$ & $\frac{45(1-\sqrt{7}y')^2}{(5+3\sqrt{7}y')^2}$ & $\frac{45(\sqrt{7}-y')^2}{(5\sqrt{7}+3y')^2}$\\
      &&$\frac{\frac{1}{\sqrt{2}}(B_1'\bar{B}+B\bar{B}_1')\gamma(E2)}{\frac{1}{\sqrt{2}}(B_1'\bar{B}^\ast-B^\ast\bar{B}_1')\gamma(E2)}$ & $1:2$ & $1:2$ & $1:2$\\
      &&$\frac{\frac{1}{\sqrt{2}}(B_1\bar{B}+B\bar{B}_1)\gamma(E2)}{\frac{1}{\sqrt{2}}(B_1\bar{B}^\ast-B^\ast\bar{B}_1)\gamma(E2)}$ & $\frac{2(1-\sqrt{7}y')^2}{(3+\sqrt{7}y')^2}$ & $\frac{2(1+3\sqrt{7}y')^2}{9(1-\sqrt{7}y')^2}$ & $\frac{2(\sqrt{7}+3y')^2}{9(\sqrt{7}-y')^2}$\\
      &&$\frac{\frac{1}{\sqrt{2}}(B_0\bar{B}^\ast+B^\ast\bar{B}_0)\gamma(E2)}{\frac{1}{\sqrt{2}}(B_1'\bar{B}+B\bar{B}_1')\gamma(E2)}$ & $\frac{(1+\beta)^2}{(1-\beta)^2}$ & $\frac{(3+\beta)^2}{(3-\beta)^2}$ & $\frac{(3-2\beta)^2}{(3+2\beta)^2}$\\
      &&$\frac{\frac{1}{\sqrt{2}}(B_1\bar{B}+B\bar{B}_1)\gamma(E2)}{\frac{1}{\sqrt{2}}(B_2\bar{B}^\ast+B^\ast\bar{B}_2)\gamma(E2)}$
      & $\frac{10(1-\sqrt{7}y')^2}{(5-\sqrt{7}y')^2}$ & $\frac{10(1+3\sqrt{7}y')^2}{(5+3\sqrt{7}y')^2}$ & $\frac{10(\sqrt{7}+3y')^2}{(5\sqrt{7}+3y')^2}$\\
      &&--&--&--&--\\
      &\multirow{3}{*}{$2^{-+}$} & $\frac{\frac{1}{\sqrt{2}}(B_1\bar{B}^\ast+B^\ast\bar{B}_1)\gamma(M1)}{\frac{1}{\sqrt{2}}(B_2\bar{B}^\ast-B^\ast\bar{B}_2)\gamma(M1)}$ & $\frac{(1+9\sqrt{3}y)^2}{9(1+\sqrt{3}y)^2}$ & $\frac{9(1+5\sqrt{3}y)^2}{(9+5\sqrt{3}y)^2}$ & $\frac{9(1-\sqrt{3}y)^2}{(9-\sqrt{3}y)^2}$\\
      &&$\frac{\frac{1}{\sqrt{2}}(B_2\bar{B}+B\bar{B}_2)\gamma(M1)}{\frac{1}{\sqrt{2}}(B_2\bar{B}^\ast-B^\ast\bar{B}_2)\gamma(M1)}$ & $\frac{2(1-3\sqrt{3}y)^2}{3(1+\sqrt{3}y)^2}$ & $\frac{6(\sqrt{3}+5y)^2}{(3\sqrt{3}+5y)^2}$ & $\frac{6(3+\sqrt{3}y)^2}{(9-\sqrt{3}y)^2}$\\
      &&$\frac{\frac{1}{\sqrt{2}}(B_1\bar{B}^\ast+B^\ast\bar{B}_1)\gamma(M1)}{\frac{1}{\sqrt{2}}(B_2\bar{B}+B\bar{B}_2)\gamma(M1)}$ & $\frac{(1+9\sqrt{3}y)^2}{6(1-3\sqrt{3}y)^2}$ & $\frac{3(1+5\sqrt{3}y)^2}{2(3+5\sqrt{3}y)^2}$ & $\frac{3(1+5\sqrt{3}y)^2}{2(3+5\sqrt{3}y)^2}$\\
      &&--&--&--&--\\
      &\multirow{3}{*}{$2^{-+}$} &$\frac{\frac{1}{\sqrt{2}}(B_1\bar{B}^\ast+B^\ast\bar{B}_1)\gamma(E2)}{\frac{1}{\sqrt{2}}(B_2\bar{B}^\ast-B^\ast\bar{B}_2)\gamma(E2)}$     & $\frac{(1+3\sqrt{7}y')^2}{(3+\sqrt{7}y')^2}$ & $\frac{(\sqrt{7}+9y')^2}{9(\sqrt{7}+y')^2}$ & $\frac{(\sqrt{7}-9y')^2}{9(\sqrt{7}-y')^2}$\\
      &&$\frac{\frac{1}{\sqrt{2}}(B_2\bar{B}+B\bar{B}_2)\gamma(E2)}{\frac{1}{\sqrt{2}}(B_2\bar{B}^\ast-B^\ast\bar{B}_2)\gamma(E2)}$     & $\frac{6(1-\sqrt{7}y')^2}{(3+\sqrt{7}y')^2}$ & $\frac{2(\sqrt{7}-3y')^2}{3(\sqrt{7}+y')^2}$ & $\frac{2(\sqrt{7}+3y')^2}{3(\sqrt{7}-y')^2}$\\
      &&$\frac{\frac{1}{\sqrt{2}}(B_1\bar{B}^\ast+B^\ast\bar{B}_1)\gamma(E2)}{\frac{1}{\sqrt{2}}(B_2\bar{B}+B\bar{B}_2)\gamma(E2)}$     & $\frac{(1+3\sqrt{7}y')^2}{6(1-\sqrt{7}y')^2}$ & $\frac{(\sqrt{7}+9y')^2}{6(\sqrt{7}-3y')^2}$ & $\frac{(\sqrt{7}-9y')^2}{6(\sqrt{7}+3y')^2}$\\\midrule[1pt]
      \end{tabular}
\end{center}
\end{table}

%%%%%%%%%%%%%%%%%%%%%%%%%%%%%%%%%%%%%%%%%%%%%%%%%%%%%%%%%%%%%%
\subsection{$\mathfrak{M} \to \mathfrak{M}^\prime+\gamma$}
%%%%%%%%%%%%%%%%%%%%%%%%%%%%%%%%%%%%%%%%%%%%%%%%%%%%%%%%%%%%%%

In this subsection, we discuss the $E1$ and $M1$ radiative
transition between two molecular(resonant) states. We only consider
the hidden beauty molecular(resonant) states with $J^{PC}=1^{--}$,
$1^{-+}$, $1^{+-}$ and $1^{++}$. The typical ratios of the
$\mathfrak{M} \to \mathfrak{M}^\prime+\gamma$ decay widths depend on
the following parameters
\begin{eqnarray*}
&&A=\frac{H_{10}(M1)}{H_{11}(M1)}, \,\,B=\frac{H_{12}(M1)}{H_{11}(M1)},\\
&&C=\frac{H_{01}(M1)}{H_{11}(M1)},\,\,D=\frac{H_{21}(M1)}{H_{11}(M1)},
\end{eqnarray*}
which are related to the reduced matrix elements of the $M1$
transitions, where $H_{10}(M1)=\langle 1,1\|H_{eff}(M1)\|0\rangle$,
$H_{01}(M1)=\langle 1,0\|H_{eff}(M1)\|1\rangle$, $H_{11}(M1)=\langle
1,1\|H_{eff}(M1)\|1\rangle$, $H_{12}(M1)=\langle
1,1\|H_{eff}(M1)\|2\rangle$, and $H_{21}(M1)=\langle
1,2\|H_{eff}(M1)\|1\rangle$.

The ratios of the $E1$ transitions between two molecular(resonant)
states depend on the following parameters
\begin{eqnarray}
 A'=\frac{H_{10}(E1)}{H_{11}(E1)}, \,\,C'=\frac{H_{01}(E1)}{H_{11}(E1)},
\end{eqnarray}
where $H_{10}(E1)=\langle 1,1\|H_{eff}(E1)\|0\rangle$,
$H_{11}(E1)=\langle 1,1\|H_{eff}(E1)\|1\rangle$, and
$H_{01}(E1)=\langle 1,0\|H_{eff}(E1)\|1\rangle$.

If the reduced matrix elements satisfy the following relations
\begin{eqnarray*}
  H_{01}(M1)&=&-\frac{1}{\sqrt{3}}H_{10}(M1),  \\
  H_{21}(M1)&=&-\frac{\sqrt{5}}{\sqrt{3}}H_{12}(M1),\\
  H_{01}(E1)&=&-\frac{1}{\sqrt{3}}H_{10}(E1),  \\
  H_{21}(E1)&=&-\frac{\sqrt{5}}{\sqrt{3}}H_{12}(E1),
\end{eqnarray*}
the ratios satisfy the crossing symmetry, which acts as an important
test of our calculation. We collect the obtained typical ratios of
the $\mathfrak{M} \to \mathfrak{M}^\prime+\gamma$ decay widths in
Tables \ref{tab:7} and \ref{tab:8}.

\begin{table*}[htbp]
\scriptsize
\begin{center}
\caption{\label{tab:7} The ratios of the $\gamma(M1)$ and
$\gamma(E1)$ transitions between two molecular(resonant) states.}
   \begin{tabular}{c|c|c ccccccc} \toprule[1pt]
    \multirow{46}{*}{\rotatebox{90}{Initial \,\, state}}&&  & \multicolumn{6}{c}{Final state\,\,$1^{--}$}  \\\cline{2-9}
    && & $\frac{1}{\sqrt{2}}(B_0\bar{B}^\ast-B^\ast\bar{B}_0)$ & $\frac{1}{\sqrt{2}}(B_1'\bar{B}-B\bar{B}_1')$ & $\frac{1}{\sqrt{2}}(B_1\bar{B}-B\bar{B}_1)$ & $\frac{1}{\sqrt{2}}(B_1'\bar{B}^\ast+B^\ast\bar{B}_1')$ & $\frac{1}{\sqrt{2}}(B_1\bar{B}^\ast+B^\ast\bar{B}_1)$ & $\frac{1}{\sqrt{2}}(B_2\bar{B}^\ast-B^\ast\bar{B}_2)$  \\
    && & $\gamma(M1)$ &$\gamma(M1)$ &$\gamma(M1)$ &$\gamma(M1)$ &$\gamma(M1)$ &$\gamma(M1)$  \\
    &\multirow{10}{*}{$1^{-+}$} & $\frac{\frac{1}{\sqrt{2}}(B_0\bar{B}^\ast+B^\ast\bar{B}_0)}{\frac{1}{\sqrt{2}}(B_1'\bar{B}^\ast-B^\ast\bar{B}_1')}$ & $\frac{(2-A)^2}{2(1+A)^2}$    & $\frac{A^2}{2(1-A)^2}$   & $\frac{A^2}{2(1-A)^2}$ & $1:2$  &$\frac{(2-3A)^2}{2(1-3A)^2}$    &$\frac{(2-A)^2}{2(1+A)^2}$  \\
    && $\frac{\frac{1}{\sqrt{2}}(B_1\bar{B}^\ast-B^\ast\bar{B}_1)}{\frac{1}{\sqrt{2}}(B_2\bar{B}^\ast+B^\ast\bar{B}_2)}$ &$\frac{(1-5B)^2}{5(5-B)^2}$     &$\frac{(7+5B)^2}{5(3+B)^2}$   &$\frac{(7+5B)^2}{5(3+B)^2}$ &$1:5$  &$\frac{(13+15B)^2}{5(1+3B)^2}$     &$\frac{(1-5B)^2}{5(5-B)^2}$  \\
    && $\frac{\frac{1}{\sqrt{2}}(B_1'\bar{B}+B\bar{B}_1')}{\frac{1}{\sqrt{2}}(B_1'\bar{B}^\ast-B^\ast\bar{B}_1')}$      &$\frac{1}{2(1+A)^2}$    &$\frac{(2+A)^2}{2(1-A)^2}$ &$\frac{(2+A)^2}{2(1-A)^2}$ &$1:2$   &$\frac{(4+3A)^2}{2(1-3A)^2}$    &$\frac{A^2}{2(1+A)^2}$  \\
    && $\frac{\frac{1}{\sqrt{2}}(B_1\bar{B}+B\bar{B}_1)}{\frac{1}{\sqrt{2}}(B_1\bar{B}^\ast-B^\ast\bar{B}_1)}$           &$\frac{2(3+5B)^2}{(1-5B)^2}$    &$\frac{50(1-B)^2}{(7+5B)^2}$  &$\frac{50(1-B)^2}{(7+5B)^2}$ &$2:1$ &$\frac{2(7-15B)^2}{(13+15B)^2}$    &$\frac{2(3+5B)^2}{(1-5B)^2}$  \\
    && $\frac{\frac{1}{\sqrt{2}}(B_0\bar{B}^\ast+B^\ast\bar{B}_0)}{\frac{1}{\sqrt{2}}(B_1'\bar{B}+B\bar{B}_1')}$ &$\frac{(2-A)^2}{A^2}$ &$\frac{A^2}{(2+A)^2}$  &$\frac{A^2}{(2+A)^2}$ &$1:1$ &$\frac{(2-3A)^2}{(4+3A)^2}$ &$\frac{(2-A)^2}{A^2}$  \\
    && $\frac{\frac{1}{\sqrt{2}}(B_1\bar{B}+B\bar{B}_1)}{\frac{1}{\sqrt{2}}(B_2\bar{B}^\ast+B^\ast\bar{B}_2)}$ &$\frac{2(3+5B)^2}{5(5-B)^2}$  &$\frac{10(1-B)^2}{(3+B)^2}$  &$\frac{10(1-B)^2}{(3+B)^2}$ &$2:5$ &$\frac{2(7-15B)^2}{5(1+3B)^2}$  &$\frac{2(3+5B)^2}{5(5-B)^2}$  \\\cline{2-9}
     &&  & \multicolumn{6}{c}{Final state\,\,$1^{-+}$}  \\\cline{2-9}
    && & $\frac{1}{\sqrt{2}}(B_0\bar{B}^\ast+B^\ast\bar{B}_0)$ & $\frac{1}{\sqrt{2}}(B_1'\bar{B}+B\bar{B}_1')$ & $\frac{1}{\sqrt{2}}(B_1\bar{B}+B\bar{B}_1)$ & $\frac{1}{\sqrt{2}}(B_1'\bar{B}^\ast-B^\ast\bar{B}_1')$ & $\frac{1}{\sqrt{2}}(B_1\bar{B}^\ast-B^\ast\bar{B}_1)$ & $\frac{1}{\sqrt{2}}(B_2\bar{B}^\ast+B^\ast\bar{B}_2)$ & \\
    && & $\gamma(M1)$ &$\gamma(M1)$ &$\gamma(M1)$ &$\gamma(M1)$ &$\gamma(M1)$ &$\gamma(M1)$  \\
    &\multirow{10}{*}{$1^{--}$} &$\frac{\frac{1}{\sqrt{2}}(B_0\bar{B}^\ast-B^\ast\bar{B}_0)}{\frac{1}{\sqrt{2}}(B_1'\bar{B}^\ast+B^\ast\bar{B}_1')}$     &$\frac{(2\sqrt{3}+3C)^2}{6}$        &$\frac{3C^2}{2}$    &$\frac{3(\sqrt{3}-\sqrt{5}D)^2}{32}$        &$\frac{(1-\sqrt{3}C)^2}{2}$    &$\frac{(\sqrt{3}-3\sqrt{5}D)^2}{96}$        &$\frac{(5\sqrt{15}+3D)^2}{480}$  \\
    && $\frac{\frac{1}{\sqrt{2}}(B_1\bar{B}^\ast+B^\ast\bar{B}_1)}{\frac{1}{\sqrt{2}}(B_2\bar{B}^\ast-B^\ast\bar{B}_2)}$ &$\frac{(2\sqrt{3}+9C)^2}{5(2\sqrt{3}-3C)^2}$      &$\frac{(4\sqrt{3}-9C)^2}{45C^2}$ &$\frac{4(7\sqrt{3}+9\sqrt{5}D)^2}{5(\sqrt{3}-\sqrt{5}D)^2}$      &$\frac{(1+3\sqrt{3}C)^2}{5(1-\sqrt{3}C)^2}$   &$\frac{(13\sqrt{3}-9\sqrt{5}D)^2}{5(\sqrt{3}+3\sqrt{5}D)^2}$      &$\frac{(4\sqrt{15}-9D)^2}{5(5\sqrt{15}+3D)^2}$\\
    && $\frac{\frac{1}{\sqrt{2}}(B_1'\bar{B}-B\bar{B}_1')}{\frac{1}{\sqrt{2}}(B_1'\bar{B}^\ast+B^\ast\bar{B}_1')}$      &$\frac{3C^2}{2}$    &  $\frac{(2\sqrt{3}-3C)^2}{6C^2}$  &$\frac{3(5\sqrt{3}+3\sqrt{5}D)^2}{288}$    &  $\frac{(1+\sqrt{3}C)^2}{2}$  &$\frac{(7\sqrt{3}-3\sqrt{5}D)^2}{96}$    &  $\frac{3(\sqrt{15}-D)^2}{160}$\\
    && $\frac{\frac{1}{\sqrt{2}}(B_1\bar{B}-B\bar{B}_1)}{\frac{1}{\sqrt{2}}(B_1\bar{B}^\ast+B^\ast\bar{B}_1)}$           &$\frac{18C^2}{(2\sqrt{3}+9C)^2}$     &$\frac{2(2\sqrt{3}-3C)^2}{(4\sqrt{3}-9C)^2}$   &$\frac{2(5\sqrt{3}+3\sqrt{5}D)^2}{(7\sqrt{3}+9\sqrt{5}D)^2}$     &$\frac{2(1+\sqrt{3}C)^2}{(1+3\sqrt{3}C)^2}$  &$\frac{2(7\sqrt{3}-3\sqrt{5}D)^2}{(13\sqrt{3}-9\sqrt{5}D)^2}$     &$\frac{18(\sqrt{15}-D)^2}{(4\sqrt{15}-9D)^2}$\\
    && $\frac{\frac{1}{\sqrt{2}}(B_0\bar{B}^\ast-B^\ast\bar{B}_0)}{\frac{1}{\sqrt{2}}(B_1'\bar{B}-B\bar{B}_1')}$ &$\frac{(2\sqrt{3}+3C)^2}{3C^2}$ &$\frac{3C^2}{(2\sqrt{3}-3C)^2}$  &$\frac{(\sqrt{3}-\sqrt{5}D)^2}{(5\sqrt{3}+3\sqrt{5}D)^2}$ &$\frac{(1-\sqrt{3}C)^2}{(1+\sqrt{3}C)^2}$      &$\frac{(\sqrt{3}+3\sqrt{5}D)^2}{(7\sqrt{3}-3\sqrt{5}D)^2}$ &$\frac{(5\sqrt{15}+3D)^2}{3(\sqrt{15}-D)^2}$\\
    && $\frac{\frac{1}{\sqrt{2}}(B_1\bar{B}-B\bar{B}_1)}{\frac{1}{\sqrt{2}}(B_2\bar{B}^\ast-B^\ast\bar{B}_2)}$ &$\frac{18C^2}{5(2\sqrt{3}+3C)^2}$ &$\frac{2(2\sqrt{3}-3C)^2}{45C^2}$    &$\frac{2(5\sqrt{3}+3\sqrt{5}D)^2}{5(\sqrt{3}-\sqrt{5}D)^2}$ &$\frac{2(1+\sqrt{3}C)^2}{5(1-\sqrt{3}C)^2}$    &$\frac{2(7\sqrt{3}-3\sqrt{5}D)^2}{5(\sqrt{3}+3\sqrt{5}D)^2}$ &$\frac{18(\sqrt{15}-D)^2}{5(5\sqrt{15}+3D)^2}$&\\\cline{2-9}
     &&  & \multicolumn{6}{c}{Final state\,\,$1^{-+}$}  \\\cline{2-9}
    && & $\frac{1}{\sqrt{2}}(B_0\bar{B}^\ast+B^\ast\bar{B}_0)$ & $\frac{1}{\sqrt{2}}(B_1'\bar{B}+B\bar{B}_1')$ & $\frac{1}{\sqrt{2}}(B_1\bar{B}+B\bar{B}_1)$ & $\frac{1}{\sqrt{2}}(B_1'\bar{B}^\ast-B^\ast\bar{B}_1')$ & $\frac{1}{\sqrt{2}}(B_1\bar{B}^\ast-B^\ast\bar{B}_1)$ & $\frac{1}{\sqrt{2}}(B_2\bar{B}^\ast+B^\ast\bar{B}_2)$  \\
    && & $\gamma(E1)$ & $\gamma(E1)$ & $\gamma(E1)$ & $\gamma(E1)$ & $\gamma(E1)$ & $\gamma(E1)$  \\
    &$1^{+-}$& $\frac{\frac{1}{\sqrt{2}}(B\bar{B}^\ast-B^\ast\bar{B})}{B^\ast\bar{B}^\ast}$ & $\frac{(\sqrt{3}+2A')^2}{(\sqrt{3}-2A')^2}$ & $\frac{(\sqrt{3}-2A')^2}{(\sqrt{3}+2A')^2}$ & $\frac{(2\sqrt{3}-A')^2}{(2\sqrt{3}+A')^2}$ & $\frac{1}{1}$ & $\frac{(2\sqrt{3}+A')^2}{(2\sqrt{3}-A')^2}$ & $\frac{(2\sqrt{3}-3A')^2}{(2\sqrt{3}+3A')^2}$  \\\cline{2-9}
     &&  & \multicolumn{6}{c}{Final state\,\,$1^{++}$} \\\cline{2-9}
    && & \multicolumn{6}{c}{$\frac{1}{\sqrt{2}}(B\bar{B}^\ast+B^\ast\bar{B})$}    \\
    && & \multicolumn{6}{c}{$\gamma(M1)$}  \\
    &$1^{+-}$ & $\frac{\frac{1}{\sqrt{2}}(B\bar{B}^\ast-B^\ast\bar{B})}{B^\ast\bar{B}^\ast}$  & \multicolumn{6}{c}{$1:1$}  \\\cline{2-9}

     &&  & \multicolumn{2}{c}{Final state\,\,$1^{+-}$} &  &  &  & {Final state\,\,$1^{++}$} \\\cline{2-9}
    && & $\frac{1}{\sqrt{2}}(B\bar{B}^\ast-B^\ast\bar{B})$ & $B^\ast\bar{B}^\ast$ &  & & & $\frac{1}{\sqrt{2}}(B\bar{B}^\ast+B^\ast\bar{B})$   \\
    && & $\gamma(E1)$ & $\gamma(E1)$ & & & & $\gamma(E1)$  \\
    &\multirow{11}{*}{$1^{-+}$} & $\frac{\frac{1}{\sqrt{2}}(B_0\bar{B}^\ast+B^\ast\bar{B}_0)}{\frac{1}{\sqrt{2}}(B_1'\bar{B}^\ast-B^\ast\bar{B}_1')}$     &$\frac{(1-2C')^2}{2}$    &$\frac{(1+2C')^2}{2}$ & & \multirow{6}{*}{$1^{--}$} & $\frac{\frac{1}{\sqrt{2}}(B_0\bar{B}^\ast-B^\ast\bar{B}_0)}{\frac{1}{\sqrt{2}}(B_1'\bar{B}^\ast+B^\ast\bar{B}_1')}$     &$4:0$  \\
    && $\frac{\frac{1}{\sqrt{2}}(B_1\bar{B}^\ast-B^\ast\bar{B}_1)}{\frac{1}{\sqrt{2}}(B_2\bar{B}^\ast+B^\ast\bar{B}_2)}$ &$\frac{(2+3C')^2}{5(5-C')^2}$     &$\frac{(2-3C')^2}{5(5+C')^2}$ & & & $\frac{\frac{1}{\sqrt{2}}(B_1\bar{B}^\ast+B^\ast\bar{B}_1)}{\frac{1}{\sqrt{2}}(B_2\bar{B}^\ast-B^\ast\bar{B}_2)}$ &$9:5$  \\
    && $\frac{\frac{1}{\sqrt{2}}(B_1'\bar{B}+B\bar{B}_1')}{\frac{1}{\sqrt{2}}(B_1'\bar{B}^\ast-B^\ast\bar{B}_1')}$      &$\frac{(1-2C')^2}{2}$    &$\frac{(1+2C')^2}{2}$ & & & $\frac{\frac{1}{\sqrt{2}}(B_1'\bar{B}-B\bar{B}_1')}{\frac{1}{\sqrt{2}}(B_1'\bar{B}^\ast+B^\ast\bar{B}_1')}$      &$4:0$  \\
    && $\frac{\frac{1}{\sqrt{2}}(B_1\bar{B}+B\bar{B}_1)}{\frac{1}{\sqrt{2}}(B_1\bar{B}^\ast-B^\ast\bar{B}_1)}$           &$\frac{2(2+C')^2}{(2+3C')^2}$    &$\frac{2(2-C')^2}{(2-3C')^2}$ & & & $\frac{\frac{1}{\sqrt{2}}(B_1\bar{B}-B\bar{B}_1)}{\frac{1}{\sqrt{2}}(B_1\bar{B}^\ast+B^\ast\bar{B}_1)}$           &$2:9$  \\
    && $\frac{\frac{1}{\sqrt{2}}(B_0\bar{B}^\ast+B^\ast\bar{B}_0)}{\frac{1}{\sqrt{2}}(B_1'\bar{B}+B\bar{B}_1')}$ &$1:1$ &$1:1$ & & & $\frac{\frac{1}{\sqrt{2}}(B_0\bar{B}^\ast-B^\ast\bar{B}_0)}{\frac{1}{\sqrt{2}}(B_1'\bar{B}-B\bar{B}_1')}$ &$1:1$  \\
    && $\frac{\frac{1}{\sqrt{2}}(B_1\bar{B}+B\bar{B}_1)}{\frac{1}{\sqrt{2}}(B_2\bar{B}^\ast+B^\ast\bar{B}_2)}$ &$\frac{2(2+C')^2}{5(2-C')^2}$ &$\frac{2(2-C')^2}{5(2+C')^2}$ & & &$\frac{\frac{1}{\sqrt{2}}(B_1\bar{B}-B\bar{B}_1)}{\frac{1}{\sqrt{2}}(B_2\bar{B}^\ast-B^\ast\bar{B}_2)}$ &$2:5$  \\\bottomrule[1pt]
    \end{tabular}
\end{center}
\end{table*}

\begin{table*}[htbp]
\scriptsize
\begin{center}
\caption{\label{tab:8} The ratios of the $\gamma(M1)$ and
$\gamma(E1)$ transitions between two molecular(resonant) states.}
   \begin{tabular}{c|c|c ccccccc} \toprule[1pt]
    \multirow{40}{*}{\rotatebox{90}{Initial \,\, state}}&&  & \multicolumn{6}{c}{Final state\,\,$1^{--}$}   \\\cline{2-9}
   && & $\frac{\frac{1}{\sqrt{2}}(B_0\bar{B}^\ast-B^\ast\bar{B}_0)\gamma(M1)}{\frac{1}{\sqrt{2}}(B_1'\bar{B}^\ast+B^\ast\bar{B}_1')\gamma(M1)}$ & $\frac{\frac{1}{\sqrt{2}}(B_1\bar{B}^\ast+B^\ast\bar{B}_1)\gamma(M1)}{\frac{1}{\sqrt{2}}(B_2\bar{B}^\ast-B^\ast\bar{B}_2)\gamma(M1)}$ & $\frac{\frac{1}{\sqrt{2}}(B_1'\bar{B}-B\bar{B}_1')\gamma(M1)}{\frac{1}{\sqrt{2}}(B_1'\bar{B}^\ast+B^\ast\bar{B}_1')\gamma(M1)}$ & $\frac{\frac{1}{\sqrt{2}}(B_1\bar{B}-B\bar{B}_1)\gamma(M1)}{\frac{1}{\sqrt{2}}(B_1\bar{B}^\ast+B^\ast\bar{B}_1)\gamma(M1)}$ & $\frac{\frac{1}{\sqrt{2}}(B_0\bar{B}^\ast-B^\ast\bar{B}_0)\gamma(M1)}{\frac{1}{\sqrt{2}}(B_1'\bar{B}-B\bar{B}_1')\gamma(M1)}$ & $\frac{\frac{1}{\sqrt{2}}(B_1\bar{B}-B\bar{B}_1)\gamma(M1)}{\frac{1}{\sqrt{2}}(B_2\bar{B}^\ast-B^\ast\bar{B}_2)\gamma(M1)}$ \\
   &\multirow{6}{*}{$1^{-+}$}& $\frac{1}{\sqrt{2}}(B_0\bar{B}^\ast+B^\ast\bar{B}_0)$ & $\frac{(2-A)^2}{2}$ &$\frac{(2-3A)^2}{5(2-A)^2}$ &$\frac{A^2}{2}$ &$\frac{2A^2}{(2-3A)^2}$ &$\frac{(2-3A)^2}{A^2}$ &$\frac{2A^2}{5(2-A)^2}$   \\
   && $\frac{1}{\sqrt{2}}(B_1'\bar{B}+B\bar{B}_1')$  &$\frac{A^2}{2}$ &$\frac{(4+3A)^2}{5A^2}$ &  $\frac{(2+A)^2}{2}$ &$\frac{2(2+A)^2}{(4+3A)^2}$ &$\frac{A^2}{(2+A)^2}$ &$\frac{2(2+A)^2}{5A^2}$  \\
   && $\frac{1}{\sqrt{2}}(B_1\bar{B}+B\bar{B}_1)$  &$\frac{(3+5B)^2}{32}$ &$\frac{(7-15B)^2}{5(3+5B)^2}$ &$\frac{25(1-B)^2}{36}$ &$\frac{50(1-B)^2}{(7-15B)^2}$ &$\frac{(3+5B)^2}{25(1-B)^2}$ &$\frac{10(1-B)^2}{(3+5B)^2}$ \\
   && $\frac{1}{\sqrt{2}}(B_1'\bar{B}^\ast-B^\ast\bar{B}_1')$ &$\frac{(1+A)^2}{2}$ &$\frac{(1-3A)^2}{5(1+A)^2}$ &  $\frac{(1-A)^2}{2}$ &$\frac{2(1-A)^2}{(1-3A)^2}$ &$\frac{(1+A)^2}{(1-A)^2}$ &$\frac{2(1-A)^2}{5(1+A)^2}$ \\
   && $\frac{1}{\sqrt{2}}(B_1\bar{B}^\ast-B^\ast\bar{B}_1)$ &$\frac{(1-5B)^2}{32}$ &$\frac{(13+15B)^2}{5(1-5B)^2}$ &$\frac{(7+5B)^2}{32}$ &$\frac{2(7+5B)^2}{(13+15B)^2}$ &$\frac{(1-5B)^2}{(7+5B)^2}$ &$\frac{2(7+5B)^2}{5(1-5B)^2}$ \\
   && $\frac{1}{\sqrt{2}}(B_2\bar{B}^\ast+B^\ast\bar{B}_2)$ &$\frac{(5-B)^2}{32}$ &$\frac{(1+3B)^2}{5(5-B)^2}$ &  $\frac{(3+B)^2}{32}$ &$\frac{2(3+B)^2}{(1+3B)^2}$ &$\frac{(5-B)^2}{(3+B)^2}$ &$\frac{2(3+B)^2}{5(5-B)^2}$ \\\cline{2-9}
    &&  & \multicolumn{6}{c}{Final state\,\,$1^{-+}$}   \\\cline{2-9}
   && & $\frac{\frac{1}{\sqrt{2}}(B_0\bar{B}^\ast+B^\ast\bar{B}_0)\gamma(M1)}{\frac{1}{\sqrt{2}}(B_1'\bar{B}^\ast-B^\ast\bar{B}_1')\gamma(M1)}$ & $\frac{\frac{1}{\sqrt{2}}(B_1\bar{B}^\ast-B^\ast\bar{B}_1)\gamma(M1)}{\frac{1}{\sqrt{2}}(B_2\bar{B}^\ast+B^\ast\bar{B}_2)\gamma(M1)}$ & $\frac{\frac{1}{\sqrt{2}}(B_1'\bar{B}+B\bar{B}_1')\gamma(M1)}{\frac{1}{\sqrt{2}}(B_1'\bar{B}^\ast-B^\ast\bar{B}_1')\gamma(M1)}$ & $\frac{\frac{1}{\sqrt{2}}(B_1\bar{B}+B\bar{B}_1)\gamma(M1)}{\frac{1}{\sqrt{2}}(B_1\bar{B}^\ast-B^\ast\bar{B}_1)\gamma(M1)}$ & $\frac{\frac{1}{\sqrt{2}}(B_0\bar{B}^\ast+B^\ast\bar{B}_0)\gamma(M1)}{\frac{1}{\sqrt{2}}(B_1'\bar{B}+B\bar{B}_1')\gamma(M1)}$ & $\frac{\frac{1}{\sqrt{2}}(B_1\bar{B}+B\bar{B}_1)\gamma(M1)}{\frac{1}{\sqrt{2}}(B_2\bar{B}^\ast+B^\ast\bar{B}_2)\gamma(M1)}$ \\
   &\multirow{6}{*}{$1^{--}$} & $\frac{1}{\sqrt{2}}(B_0\bar{B}^\ast-B^\ast\bar{B}_0)$ &$\frac{(2\sqrt{3}+3C)^2}{6(1-\sqrt{3}C)^2}$ &$\frac{(\sqrt{3}+3\sqrt{5}D)^2}{(5\sqrt{15}+3D)^2}$ &$\frac{3C^2}{2(1-\sqrt{3}C)^2}$ &$\frac{18(\sqrt{3}-\sqrt{5}D)^2}{(\sqrt{3}-3\sqrt{5}D)^2}$ &$\frac{(2\sqrt{3}+3C)^2}{3C^2}$ &$\frac{18(\sqrt{3}-\sqrt{5}D)^2}{(5\sqrt{15}+3D)^2}$  \\
   && $\frac{1}{\sqrt{2}}(B_1'\bar{B}-B\bar{B}_1')$ &$\frac{3C^2}{2(1+\sqrt{3}C)^2}$ &$\frac{(7\sqrt{3}-3\sqrt{5}D)^2}{(3\sqrt{15}-3D)^2}$ &$\frac{(2\sqrt{3}-3C)^2}{6(1+\sqrt{3})^2}$ &$\frac{2(5\sqrt{3}+3\sqrt{5}D)^2}{(7\sqrt{3}-3\sqrt{5}D)^2}$ &$\frac{3C^2}{(2\sqrt{3}-3C)^2}$ &$\frac{2(5\sqrt{3}+3\sqrt{5}D)^2}{(3\sqrt{15}-3D)^2}$  \\
   && $\frac{1}{\sqrt{2}}(B_1\bar{B}-B\bar{B}_1)$ &$\frac{3C^2}{2(1+\sqrt{3}C)^2}$ &$\frac{(7\sqrt{3}-3\sqrt{5}D)^2}{9(\sqrt{15}-D)^2}$ &$\frac{(2\sqrt{3}-3C)^2}{6(1+\sqrt{3}C)^2}$ &$\frac{2(5\sqrt{3}+3\sqrt{5}D)^2}{(7\sqrt{3}-3\sqrt{5}D)^2}$ &$\frac{3C^2}{(2\sqrt{3}-3C)^2}$ &$\frac{2(5\sqrt{3}+3\sqrt{5}D)^2}{9(\sqrt{15}-D)^2}$  \\
   && $\frac{1}{\sqrt{2}}(B_1'\bar{B}^\ast+B^\ast\bar{B}_1')$ & $1:2$ &$1:5$ &$1:2$ &$2:1$ &$1:1$&$2:5$ \\
   && $\frac{1}{\sqrt{2}}(B_1\bar{B}^\ast+B^\ast\bar{B}_1)$ &$\frac{(2\sqrt{3}+9C)^2}{6(1+3\sqrt{3}C)^2}$ &$\frac{(13\sqrt{3}+9\sqrt{5}D)^2}{(4\sqrt{15}-9D)^2}$ &$\frac{(4\sqrt{3}-9C)^2}{6(1+3\sqrt{3}C)^2}$  &$\frac{2(7\sqrt{3}+9\sqrt{5}D)^2}{(13\sqrt{3}-9\sqrt{5}D)^2}$ &$\frac{(2\sqrt{3}+9C)^2}{(4\sqrt{3}-9C)^2}$ &$\frac{2(7\sqrt{3}+9\sqrt{5}D)^2}{(4\sqrt{15}-9D)^2}$  \\
   && $\frac{1}{\sqrt{2}}(B_2\bar{B}^\ast-B^\ast\bar{B}_2)$ &$\frac{(2\sqrt{3}+3C)^2}{6(1-\sqrt{3}C)^2}$ &$\frac{(\sqrt{3}+3\sqrt{5}D)^2}{(5\sqrt{15}+3D)^2}$ &$\frac{3C^2}{2(1-\sqrt{3}C)^2}$ &$\frac{18(\sqrt{3}-\sqrt{5}D)^2}{(\sqrt{3}+3\sqrt{5}D)^2}$ &$\frac{(2\sqrt{3}+3C)^2}{3C^2}$ &$\frac{18(\sqrt{3}-\sqrt{5}D)^2}{(5\sqrt{15}+3D)^2}$  \\\cline{2-9}
    &&  & \multicolumn{6}{c}{Final state\,\,$1^{-+}$}   \\\cline{2-9}
   && & $\frac{\frac{1}{\sqrt{2}}(B_0\bar{B}^\ast+B^\ast\bar{B}_0)\gamma(E1)}{\frac{1}{\sqrt{2}}(B_1'\bar{B}^\ast-B^\ast\bar{B}_1')\gamma(E1)}$ & $\frac{\frac{1}{\sqrt{2}}(B_1\bar{B}^\ast-B^\ast\bar{B}_1)\gamma(E1)}{\frac{1}{\sqrt{2}}(B_2\bar{B}^\ast+B^\ast\bar{B}_2)\gamma(E1)}$ & $\frac{\frac{1}{\sqrt{2}}(B_1'\bar{B}+B\bar{B}_1')\gamma(E1)}{\frac{1}{\sqrt{2}}(B_1'\bar{B}^\ast-B^\ast\bar{B}_1')\gamma(E1)}$ & $\frac{\frac{1}{\sqrt{2}}(B_1\bar{B}+B\bar{B}_1)\gamma(E1)}{\frac{1}{\sqrt{2}}(B_1\bar{B}^\ast-B^\ast\bar{B}_1)\gamma(E1)}$ & $\frac{\frac{1}{\sqrt{2}}(B_0\bar{B}^\ast+B^\ast\bar{B}_0)\gamma(E1)}{\frac{1}{\sqrt{2}}(B_1'\bar{B}+B\bar{B}_1')\gamma(E1)}$ & $\frac{\frac{1}{\sqrt{2}}(B_1\bar{B}+B\bar{B}_1)\gamma(E1)}{\frac{1}{\sqrt{2}}(B_2\bar{B}^\ast+B^\ast\bar{B}_2)\gamma(E1)}$ \\
   &\multirow{2}{*}{$1^{+-}$} & $\frac{1}{\sqrt{2}}(B\bar{B}^\ast-B^\ast\bar{B})$ &$\frac{(\sqrt{3}+2A')^2}{6}$ &$\frac{(2\sqrt{3}-3A')^2}{(2\sqrt{15}+\sqrt{5}A')^2}$  &$\frac{(\sqrt{3}-2A')^2}{6}$  &$\frac{(2\sqrt{6}-\sqrt{2}A')^2}{(2\sqrt{3}+3A')^2}$ &$\frac{(\sqrt{3}+2A')^2}{(\sqrt{3}-2A')^2}$ &$\frac{(2\sqrt{6}-\sqrt{2}A')^2}{(2\sqrt{15}+\sqrt{5}A')^2}$  \\
   && $B^\ast\bar{B}^\ast$ &$\frac{(\sqrt{3}-2A')^2}{6}$ &$\frac{(2\sqrt{3}+3A')^2}{(2\sqrt{15}-\sqrt{5}A')^2}$ &$\frac{(\sqrt{3}+2A')^2}{6}$ &$\frac{(2\sqrt{6}+\sqrt{2}A')^2}{(2\sqrt{3}-3A')^2}$  &$\frac{(\sqrt{3}-2A')^2}{(\sqrt{3}+2A')^2}$ &$\frac{(2\sqrt{6}+\sqrt{2}A')^2}{(2\sqrt{15}-\sqrt{5}A')^2}$ \\\cline{2-9}
    &&  & \multicolumn{6}{c}{Final state\,\,$1^{--}$}  \\\cline{2-9}
   && & $\frac{\frac{1}{\sqrt{2}}(B_0\bar{B}^\ast-B^\ast\bar{B}_0)\gamma(E1)}{\frac{1}{\sqrt{2}}(B_1'\bar{B}^\ast+B^\ast\bar{B}_1')\gamma(E1)}$ & $\frac{\frac{1}{\sqrt{2}}(B_1\bar{B}^\ast+B^\ast\bar{B}_1)\gamma(E1)}{\frac{1}{\sqrt{2}}(B_2\bar{B}^\ast-B^\ast\bar{B}_2)\gamma(E1)}$ & $\frac{\frac{1}{\sqrt{2}}(B_1'\bar{B}-B\bar{B}_1')\gamma(E1)}{\frac{1}{\sqrt{2}}(B_1'\bar{B}^\ast+B^\ast\bar{B}_1')\gamma(E1)}$ & $\frac{\frac{1}{\sqrt{2}}(B_1\bar{B}-B\bar{B}_1)\gamma(E1)}{\frac{1}{\sqrt{2}}(B_1\bar{B}^\ast+B^\ast\bar{B}_1)\gamma(E1)}$ & $\frac{\frac{1}{\sqrt{2}}(B_0\bar{B}^\ast-B^\ast\bar{B}_0)\gamma(E1)}{\frac{1}{\sqrt{2}}(B_1'\bar{B}-B\bar{B}_1')\gamma(E1)}$ & $\frac{\frac{1}{\sqrt{2}}(B_1\bar{B}-B\bar{B}_1)\gamma(E1)}{\frac{1}{\sqrt{2}}(B_2\bar{B}^\ast-B^\ast\bar{B}_2)\gamma(E1)}$ \\
   &$1^{++}$ & $\frac{1}{\sqrt{2}}(B\bar{B}^\ast+B^\ast\bar{B})$ &$4:0$   &$9:5$ &$4:0$  &$2:9$ &$1:1$  &$2:5$ \\\cline{2-9}

    &&  & {Final state\,\,$1^{+-}$}&   & &  &  & {Final state\,\,$1^{+-}$}   \\\cline{2-9}
   & && $\frac{\frac{1}{\sqrt{2}}(B\bar{B}^\ast-B^\ast\bar{B})\gamma(E1)}{B^\ast\bar{B}^\ast\gamma(E1)}$ &  && & & $\frac{\frac{1}{\sqrt{2}}(B\bar{B}^\ast-B^\ast\bar{B})\gamma(M1)}{B^\ast\bar{B}^\ast\gamma(M1)}$   \\
   &\multirow{6}{*}{$1^{-+}$} & $\frac{1}{\sqrt{2}}(B_0\bar{B}^\ast+B^\ast\bar{B}_0)$     & $\frac{(1-2C')^2}{(1+2C')^2}$ & & & $1^{++}$ & $\frac{1}{\sqrt{2}}(B\bar{B}^\ast+B^\ast\bar{B})$ &$1:1$   \\
   && $\frac{1}{\sqrt{2}}(B_1'\bar{B}+B\bar{B}_1')$     & $\frac{(1-2C')^2}{(1+2C')^2}$ &&&&&\\
   && $\frac{1}{\sqrt{2}}(B_1\bar{B}+B\bar{B}_1)$     & $\frac{(2+C')^2}{(2-C')^2}$ &&&&&\\
   && $\frac{1}{\sqrt{2}}(B_1'\bar{B}^\ast-B^\ast\bar{B}_1')$     & $\frac{1}{1}$ &&&&&\\
   && $\frac{1}{\sqrt{2}}(B_1\bar{B}^\ast-B^\ast\bar{B}_1)$     & $\frac{(2+3C')^2}{(2-3C')^2}$ &&&&&\\
   && $\frac{1}{\sqrt{2}}(B_2\bar{B}^\ast+B^\ast\bar{B}_2)$     & $\frac{(2-C')^2}{(2+C')^2}$ &&&&&\\\bottomrule[1pt]

   \end{tabular}
\end{center}
\end{table*}

%%%%%%%%%%%%%%%%%%%%%%%%%%%%%%%%%%%%%%%%
\section{The radiative decays of the hidden-charm molecules}
\label{sec4}
%%%%%%%%%%%%%%%%%%%%%%%%%%%%%%%%%%%%%%%%

In the previous sections, we focus on the radiative decays of the
hidden beauty molecular(resonant) states. However, we can easily
extend our approach to study the radiative decays of the hidden
charm molecular(resonant) states. Readers should keep in mind that
the heavy quark symmetry is expected to work well for the hidden
beauty molecular(resonant) system since the bottom quark mass is
much larger than $\Lambda_{\mbox{QCD}}$. For the hidden charm
molecular(resonant) states, the $1/m_Q$ correction may be
non-negligible in certain cases since the charm quark mass is not so
heavy compared to the bottom mass \cite{Zhao:2014gqa}. Especially the $1/m_Q$ correction
plays the dominant role in those radiative channels which are
forbidden in the heavy quark limit. For the other allowed channels,
one may naively expect that the decay pattern and ratios of the
radiative decays of the hidden charm molecular(resonant) states are
roughly the same as those in the heavy quark limit.

As mentioned in Sec. \ref{sec1}, some observed charmonium-like
states were considered as the candidates of the hidden-charm
molecular(resonant) states. In the following, we discuss the
radiative decays of several charmonium-like states under the
molecular(resonant) assignment.

%%%%%%%%%%%%%%%%%%%%%%%%%%%%%%%%%%%%%%
\subsection{$Y(4260)$ and $Y(4360)$}
%%%%%%%%%%%%%%%%%%%%%%%%%%%%%%%%%%%%%%

%%%%%%%%%%%%%%%%%%%%%%%%%%%%%%%%%%%%%%
\subsubsection{$Y(4260)\to \chi_{cJ}\gamma$ }
%%%%%%%%%%%%%%%%%%%%%%%%%%%%%%%%%%%%%%

The $J^{PC}=1^{--}$ state $Y(4260)$ was reported by the BaBar
Collaboration in the $e^+e^-\to \pi^+\pi^- J/\psi$ mode
\cite{Aubert:2005rm}. Assuming $Y(4260)$ is the
$\frac{1}{\sqrt{2}}(D_1\bar{D}-D\bar{D}_1)$ molecular state
\cite{zhu-review,Ding:2008gr,Wang:2013kra}, we can write down the
rearranged spin structure of $Y(4260)$,
\begin{eqnarray*}
  |Y(4260)\rangle &=& %\frac{1}{\sqrt{2}}(|D_1\bar{D}\rangle-|D\bar{D}_1\rangle ) \\
  %&\to&
  \Big[\frac{\sqrt{6}}{6}(0_H^{-+}\otimes 1_l^{++})|_{J=1}^{-+}-\frac{\sqrt{3}}{6}(0_H^{-+}\otimes 1_l^{+-})|_{J=1}^{--}\\
  &&+\frac{\sqrt{3}}{6}(1_H^{--}\otimes 1_l^{++})|_{J=1}^{--}-\frac{\sqrt{6}}{12}(1_H^{--}\otimes 1_l^{+-})|_{J=1}^{-+}\\
  &&+\frac{\sqrt{10}}{4}(1_H^{--}\otimes 1_l^{+-})|_{J=1}^{-+}\Big]|c\bar{q};\bar{c}q\rangle \\
  &&+\Big[\frac{\sqrt{6}}{6}(0_H^{-+}\otimes 1_l^{++})|_{J=1}^{-+}+\frac{\sqrt{3}}{6}(0_H^{-+}\otimes 1_l^{+-})|_{J=1}^{--}\\
  &&-\frac{\sqrt{3}}{6}(1_H^{--}\otimes 1_l^{++})|_{J=1}^{--}-\frac{\sqrt{6}}{12}(1_H^{--}\otimes 1_l^{+-})|_{J=1}^{-+}\\
  &&+\frac{\sqrt{10}}{4}(1_H^{--}\otimes 1_l^{+-})|_{J=1}^{-+}\Big]|\bar{c}q;c\bar{q}\rangle.
\end{eqnarray*}
After performing the spin rearrangement, the allowed final states
can be expressed as
\begin{eqnarray*}
  |\chi_{c0}\gamma(E1)\rangle &=& \Big[\frac{1}{3}(1_H^{--}\otimes 0_l^{++})|_{J=1}^{--}-\frac{\sqrt{3}}{3}(1_H^{--}\otimes 1_l^{++})|_{J=1}^{--}\\
  &&+\frac{\sqrt{5}}{3}(1_H^{--}\otimes 2_l^{++})|_{J=1}^{--}\Big]|(c\bar{c})\rangle|\gamma\rangle, \\
  |\chi_{c1}\gamma(E1)\rangle &=& \Big[-\frac{\sqrt{3}}{3}(1_H^{--}\otimes 0_l^{++})|_{J=1}^{--}+\frac{1}{2}(1_H^{--}\otimes 1_l^{++})|_{J=1}^{--}\\
  &&+\frac{\sqrt{15}}{6}(1_H^{--}\otimes 2_l^{++})|_{J=1}^{--}\Big]|(c\bar{c})\rangle|\gamma\rangle,  \\
  |\chi_{c2}\gamma(E1)\rangle &=&\Big[\frac{\sqrt{5}}{3}(1_H^{--}\otimes 0_l^{++})|_{J=1}^{--}-\frac{\sqrt{15}}{6}(1_H^{--}\otimes 1_l^{++})|_{J=1}^{--}\\
  &&+\frac{1}{6}(1_H^{--}\otimes 2_l^{++})|_{J=1}^{--}\Big]|(c\bar{c})\rangle|\gamma\rangle.
\end{eqnarray*}
In the heavy quark limit, we get the ratio of the $E1$ transitions
$Y(4260)\rightarrow \chi_{cJ}\gamma(E1)$ ($J=0,1,2$),
\begin{eqnarray*}
  &&\Gamma(\chi_{c0}\gamma(E1)):\Gamma(\chi_{c1}\gamma(E1)):\Gamma(\chi_{c2}\gamma(E1))\\
  &&= 4:3:5,
\end{eqnarray*}
where we ignore the phase space factor. If considering the
difference of the phase space of the three decay channels, we have
\begin{eqnarray*}
  &&\Gamma(\chi_{c0}\gamma(E1)):\Gamma(\chi_{c1}\gamma(E1)):\Gamma(\chi_{c2}\gamma(E1))\\
  &&= 1.8:1:1.4.
\end{eqnarray*}
The above ratios can be used to test the molecular assignment of
$Y(4260)$ if the BES collaboration can measure these decay widths.
$Z_c(4248)$ may be the isovector molecular partner of $Y(4260)$
\cite{zhu-review}. The radiative decay pattern of its neutral
component is the same as that of $Y(4260)$.

%%%%%%%%%%%%%%%%%%%%%%%%%%%%%%%%%%%%%%
\subsubsection{$Y(4360)\to \chi_{cJ}\gamma$ }
%%%%%%%%%%%%%%%%%%%%%%%%%%%%%%%%%%%%%%

$Y(4360)$ was observed by the Belle collaboration in the
$\psi(2S)\pi^+\pi^-$ invariant mass spectrum in the $e^+e^-\to
\psi(2S)\pi^+\pi^-$ process \cite{Aubert:2007zz}. Since the mass of
$Y(4360)$ is close to the threshold of $D_1\bar{D}^*$, $Y(4360)$ may
be a $\frac{1}{\sqrt{2}}(D_1\bar{D}^\ast+D^\ast\bar{D}_1)$ molecular
state \cite{zhu-review}. If so, its spin structure can be written as
\begin{eqnarray*}
&&  |Y(4360)\rangle\\ &&= %\frac{1}{\sqrt{2}}(|D_1\bar{D}^\ast\rangle+|D^\ast\bar{D}_1\rangle ) \\
  %&=&
   \Big[\frac{\sqrt{3}}{6}(0_H^{-+}\otimes 1_l^{++})|_{J=1}^{-+}-\frac{\sqrt{6}}{12}(0_H^{-+}\otimes 1_l^{+-})|_{J=1}^{--}\\
  &&\quad+\frac{\sqrt{6}}{4}(1_H^{--}\otimes 1_l^{++})|_{J=1}^{--}-\frac{\sqrt{3}}{4}(1_H^{--}\otimes 1_l^{+-})|_{J=1}^{-+}\\
  &&\quad-\frac{\sqrt{5}}{4}(1_H^{--}\otimes 1_l^{+-})|_{J=1}^{-+}\Big]|c\bar{q};\bar{c}q\rangle \\
  &&\quad+\Big[-\frac{\sqrt{3}}{6}(0_H^{-+}\otimes 1_l^{++})|_{J=1}^{-+}-\frac{\sqrt{6}}{12}(0_H^{-+}\otimes 1_l^{+-})|_{J=1}^{--}\\
  &&\quad+\frac{\sqrt{6}}{4}(1_H^{--}\otimes 1_l^{++})|_{J=1}^{--}+\frac{\sqrt{3}}{4}(1_H^{--}\otimes 1_l^{+-})|_{J=1}^{-+}\\
  &&\quad+\frac{\sqrt{5}}{4}(1_H^{--}\otimes 1_l^{+-})|_{J=1}^{-+}\Big]|\bar{c}q;c\bar{q}\rangle.
\end{eqnarray*}
$Y(4360)$ can decay into $\chi_{cJ}$ with the following ratio
\begin{eqnarray*}
  &&\Gamma(\chi_{c0}\gamma(E1)):\Gamma(\chi_{c1}\gamma(E1)):\Gamma(\chi_{c2}\gamma(E1))\\
  &&=4:3:5\, (1.8:1:1.4),
  \end{eqnarray*}
where the value listed in the bracket is the result considering the
phase space correction.

$Y(4660)$ may be the hidden-strangeness partner of $Y(4360)$ which
is composed of $D_{s1}\bar{D}_s^*$ \cite{zhu-review}. Its radiative
decay pattern is the same as that of $Y(4360)$.

%%%%%%%%%%%%%%%%%%%%%%%%%%%%%%%%%%%%%%
\subsubsection{Ratios between the radiative decay widths of $Y(4260)$ and $Y(4360)$ }
%%%%%%%%%%%%%%%%%%%%%%%%%%%%%%%%%%%%%%

$Y(4260)$ and $Y(4360)$ can also decay into $\eta_c$ and
$\eta_{c2}(1^1D_2)$ via the M1 transition. The spin-rearranged final
state are
\begin{eqnarray*}
  |\eta_c\gamma(M1)\rangle &=& (0_H^{-+}\otimes 1_l^{--})|_{J=1}^{+-}|(c\bar{c})\rangle|\gamma\rangle, \\
  |\eta_{c2}(1^1D_2)\gamma(M1)\rangle &=& (0_H^{-+}\otimes 1_l^{--})|_{J=1}^{+-}|(c\bar{c})\rangle|\gamma\rangle.
\end{eqnarray*}
Under the above molecular assumption, $Y(4260)$ and $Y(4360)$ have
the same spatial wave functions. Therefore, we obtain the following
ratios
\begin{eqnarray*}
  \frac{\Gamma(Y(4260)\rightarrow\eta_c\gamma(M1))}{\Gamma(Y(4360)\rightarrow\eta_c\gamma(M1))}=2:1,\\
  \frac{\Gamma(Y(4260)\rightarrow\eta_{c2}(1^1D_2)\gamma(M1))}{\Gamma(Y(4360)\rightarrow\eta_{c2}(1^1D_2)\gamma(M1))}=2:1,\\
  \frac{\Gamma(Y(4260)\rightarrow\chi_{bJ}\gamma(E1))}{\Gamma(Y(4360)\rightarrow\chi_{bJ}\gamma(E1))}=2:9,
\end{eqnarray*}
where the $J=0,1,2$. Considering the phase factors, we have
\begin{eqnarray*}
  \frac{\Gamma(Y(4260)\rightarrow\eta_c\gamma(M1))}{\Gamma(Y(4360)\rightarrow\eta_c\gamma(M1))}=1.7:1,\\
  \frac{\Gamma(Y(4260)\rightarrow\chi_{b0}\gamma(E1))}{\Gamma(Y(4360)\rightarrow\chi_{b0}\gamma(E1))}=1:5.8,\\
  \frac{\Gamma(Y(4260)\rightarrow\chi_{b1}\gamma(E1))}{\Gamma(Y(4360)\rightarrow\chi_{b1}\gamma(E1))}=1:6.1,\\
  \frac{\Gamma(Y(4260)\rightarrow\chi_{b2}\gamma(E1))}{\Gamma(Y(4360)\rightarrow\chi_{b2}\gamma(E1))}=1:6.2.
\end{eqnarray*}

$Y(4260)$ was suggested as a conventional $c\bar{c}$ state
$\psi(4S)$ \cite{Li:2009zu} or the isoscalar partner of the
$\frac{1}{\sqrt{2}}(D_1\bar{D}-D\bar{D}_1)$ molecular state
\cite{Ding:2008gr}. The spin structure of $\psi(4S)$ is $(1_H\otimes
1_l)_{J=1}$. In contrast, the spin structure of
$\frac{1}{\sqrt{2}}(D_1\bar{D}-D\bar{D}_1)$ not only contains
$(1_H\otimes 1_l)_{J=1}$ but also $(0_H\otimes 1_l)_{J=1}$, which
governs the decay modes $\eta_c\gamma(M1)$ and
$\eta_{c2}(1^1D_2)\gamma(M1)$. However, these decay modes are
suppressed for $\psi(4S)$ if the heavy quark symmetry works well.
Thus, further experimental measurement of these decay modes can
distinguish these two assignments.

%%%%%%%%%%%%%%%%%%%%%%%%%%%%%%%%%%%%%%
\subsection{$X(3872)$ }
%%%%%%%%%%%%%%%%%%%%%%%%%%%%%%%%%%%%%%

There are extensive discussions of the possibility of $X(3872)$ as a
$D\bar{D}^\ast$ molecular state with $J^{pc}=1^{++}$
\cite{Close:2003sg,Voloshin:2003nt,Wong:2003xk,Swanson:2003tb,Tornqvist:2004qy,Suzuki:2005ha,Liu:2008fh,Thomas:2008ja,Lee:2009hy,Li:2012cs}.
Under this molecular state assignment, the spin structure of
$X(3872)$ reads
\begin{eqnarray*}
  |X(3872)\rangle &=& %\frac{1}{\sqrt{2}}(|D\bar{D}^\ast\rangle+|D^\ast\bar{D}\rangle)\\
  %&=&
  \Big[\frac{1}{2}(0_H^{-+}\otimes 1_l^{--})|_{J=1}^{+-}-\frac{1}{2}(1_H^{--}\otimes 0_l^{-+})|_{J=1}^{+-}\\
  &&+\frac{1}{\sqrt{2}}(1_H^{--}\otimes 1_l^{--})|_{J=1}^{++}\Big]|c\bar{q};\bar{c}q\rangle \\
  &&+\Big[-\frac{1}{2}(0_H^{-+}\otimes 1_l^{--})|_{J=1}^{+-}+\frac{1}{2}(1_H^{--}\otimes 0_l^{-+})|_{J=1}^{+-}\\
  &&+\frac{1}{\sqrt{2}}(1_H^{--}\otimes 1_l^{--})|_{J=1}^{++}\Big]|\bar{c}q;c\bar{q}\rangle.
\end{eqnarray*}
If we ignore the heavy quark symmetry, the radiative decay modes of
$X(3872)$ are $\psi(1^3D_1)\gamma(E1)$, $\psi(1^3D_2)\gamma(E1)$,
$J/\psi\gamma(E1)$, $h_c\gamma(M1)$ and $h_c\gamma(E2)$, whose spin
structures are
\begin{eqnarray*}
  |\psi(1^3D_1)\gamma(E1)\rangle &=& \Big[-\frac{1}{2}(1_H^{--}\otimes1_l^{--})|_{J=1}^{++}\\
  &&+\frac{\sqrt{3}}{2}(1_H^{--}\otimes2_l^{--})|_{J=1}^{++}\Big]|(c\bar{c})\rangle|\gamma\rangle, \\
  |\psi(1^3D_2)\gamma(E1)\rangle &=& \Big[\frac{\sqrt{3}}{2}(1_H^{--}\otimes1_l^{--})|_{J=1}^{++}\\
  &&+\frac{1}{2}(1_H^{--}\otimes2_l^{--})|_{J=1}^{++}\Big]|(c\bar{c})\rangle|\gamma\rangle.
\end{eqnarray*}
\begin{eqnarray*}
  |J/\psi\gamma(E1)\rangle &=& (1_H^{--}\otimes 1_l^{--})|_{J=1}^{++}|(c\bar{c})\rangle|\gamma\rangle, \\
  |h_c\gamma(M1)\rangle &=& (0_H^{-+}\otimes 1_l^{-+})|_{J=1}^{++}|(c\bar{c})\rangle|\gamma\rangle, \\
  |h_c\gamma(E2)\rangle &=& (0_H^{-+}\otimes 1_l^{-+})|_{J=1}^{++}|(c\bar{c})\rangle|\gamma\rangle.
\end{eqnarray*}
However, the decay modes $h_c\gamma(M1)$ and $h_c\gamma(E2)$ are
suppressed while the other three ones are allowed in the heavy quark
limit. The decay widths of $X(3872)\rightarrow J/\psi\gamma(E1)$ and
$X(3872)\rightarrow \psi'\gamma(E1)$ depend on the reduced matrix
element $|\langle1,0\|H_{eff}(E1)\|1\rangle|^2$. For the decay modes
$\psi(1^3D_1)\gamma(E1)$ and $\psi(1^3D_2)\gamma(E1)$, we have the
ratio
\begin{equation}
\Gamma(\psi(1^3D_1)\gamma(E1)):\Gamma(\psi(1^3D_2)\gamma(E1))=1:3
(1:2.9)
\end{equation}
without and with the phase space correction, respectively.

%%%%%%%%%%%%%%%%%%%%%%%%%%%%%%%%%%%%%%
\subsection{$Y(4260)\to X(3872) \gamma$ }
%%%%%%%%%%%%%%%%%%%%%%%%%%%%%%%%%%%%%%

There may also exist the $E1$ transition between the two molecular
candidates $Y(4260)$ and $X(3872)$. The spin-recoupled final state
can be expressed as
\begin{eqnarray*}
 && |X(3872)\gamma(E1)\rangle \\&&=
  \Big[\frac{1}{2}(0_H^{-+}\otimes 1_l^{++})|_{J=1}^{-+}-\frac{1}{2}(1_H^{--}\otimes 1_l^{+-})|_{J=1}^{-+}\\
  &&\quad-\frac{\sqrt{6}}{6}(1_H^{--}\otimes 0_l^{++})|_{J=1}^{--}+\frac{\sqrt{2}}{4}(1_H^{--}\otimes 1_l^{++})|_{J=1}^{--}\\
  &&\quad+\frac{\sqrt{30}}{12}(1_H^{--}\otimes 2_l^{++})|_{J=1}^{--}\Big]|c\bar{q};\bar{c}q\rangle|\gamma\rangle \\
  &&\quad+\Big[-\frac{1}{2}(0_H^{-+}\otimes 1_l^{++})|_{J=1}^{-+}+\frac{1}{2}(1_H^{--}\otimes 1_l^{+-})|_{J=1}^{-+}\\
  &&\quad-\frac{\sqrt{6}}{6}(1_H^{--}\otimes 0_l^{++})|_{J=1}^{--}+\frac{\sqrt{2}}{4}(1_H^{--}\otimes 1_l^{++})|_{J=1}^{--}\\
  &&\quad+\frac{\sqrt{30}}{12}(1_H^{--}\otimes 2_l^{++})|_{J=1}^{--}\Big]|\bar{c}q;c\bar{q}\rangle|\gamma\rangle.
\end{eqnarray*}
Clearly the decay width depends on the reduced matrix element
$|\langle1,1\|H_{eff}(E1)\|1\rangle|$ and does not vanish in the
heavy quark limit.

%%%%%%%%%%%%%%%%%%%%%%%%%%%%%%%%%%%%%%
\subsection{$Z_c(3900)$ and $Z_c(4020)$ }
%%%%%%%%%%%%%%%%%%%%%%%%%%%%%%%%%%%%%%

$Z_c(3900)$ was first observed in the channel $e^+e^-\rightarrow
J/\psi\pi^+\pi^-$ at $\sqrt{s}=4.26\,\,\textmd{GeV}$ by BESIII
\cite{Ablikim:2013mio}. $Z_c(3900)$ or $Z_c(3885)$ may be the
candidate of the charged $D\bar{D}^\ast$ molecular state with
$I^G(J^p)=1^+(1^+)$. $Z_c(4020)$ was reported in the $h_c\pi^{\pm}$
invariant mass spectrum of $e^+e^-\rightarrow h_c\pi^+\pi^-$ at
$\sqrt{s}=4.26\,\,\textmd{GeV}$ \cite{Ablikim:2013wzq}. The similar
state $Z_c(4025)$ was observed in the $e^+e^-\rightarrow
(D^\ast\bar{D}^\ast)^{\pm}\pi^{\mp}$ at
$\sqrt{s}=4.26\,\,\textmd{GeV}$ \cite{Ablikim:2013emm}. $Z_c(4020)$
or $Z_c(4025)$ may be the $D^\ast\bar{D}^\ast$ molecular state. The
quantum number of the neutral partner of $Z_c(3900)$ and $Z_c(4020)$
is $I^GJ^{pc}=1^+1^{+-}$ \cite{He:2013nwa}, whose spin structures
are
\begin{eqnarray*}
  |Z_c(3900)\rangle &=& %\frac{1}{\sqrt{2}}(|D\bar{D}^\ast\rangle+|D^\ast\bar{D}\rangle)\\
  %&=&
  \Big[\frac{1}{2}(0_H^{-+}\otimes 1_l^{--})|_{J=1}^{+-}-\frac{1}{2}(1_H^{--}\otimes 0_l^{-+})|_{J=1}^{+-}\\
  &&+\frac{1}{\sqrt{2}}(1_H^{--}\otimes 1_l^{--})|_{J=1}^{++}\Big]|c\bar{q};\bar{c}q\rangle \\
  &&-\Big[-\frac{1}{2}(0_H^{-+}\otimes 1_l^{--})|_{J=1}^{+-}+\frac{1}{2}(1_H^{--}\otimes 0_l^{-+})|_{J=1}^{+-}\\
  &&+\frac{1}{\sqrt{2}}(1_H^{--}\otimes 1_l^{--})|_{J=1}^{++}\Big]|\bar{c}q;c\bar{q}\rangle,\\
  |Z_c(4020)\rangle &=&
  \Big[\frac{1}{\sqrt{2}}(0_H^{-+}\otimes 1_l^{--})|_{J=1}^{+-}+\frac{1}{\sqrt{2}}(1_H^{--}\otimes 0_l^{-+})|_{J=1}^{+-}
  \Big]\\
  &&\times\frac{1}{\sqrt{2}}(|c\bar{q};\bar{c}q\rangle+|\bar{c}q;c\bar{q}\rangle).
\end{eqnarray*}
The $M1$ transitions of $Z_c(3900)$ and $Z_c(4020)$ into $\chi_{cJ}$
result in the simple ratio
$\Gamma(\chi_{c0}\gamma(M1)):\Gamma(\chi_{c1}\gamma(M1)):\Gamma(\chi_{c2}\gamma(M1))=1:3:5$
if we ignore the phase space correction
\cite{Ohkoda:2012rj,He:2013nwa}.

The $E2$ transition modes $\chi_{c1}\gamma(E2), \chi_{c2}\gamma(E2)$
of $Z_c(3900)$ and $Z_c(4020)$ are suppressed in the heavy quark
limit, which is manifest from their spin structures
\begin{eqnarray*}
  |\chi_{c1}\gamma(E2)\rangle &=& \Big[-\frac{1}{2}(1_H^{--}\otimes 1_l^{-+})|_{J=1}^{+-}\\
  &&+\frac{\sqrt{3}}{2}(1_H^{--}\otimes 2_l^{-+})|_{J=1}^{+-}\Big]|(c\bar{c})\rangle|\gamma\rangle,  \\
  |\chi_{c2}\gamma(E2)\rangle &=&\Big[\frac{\sqrt{3}}{2}(1_H^{--}\otimes 1_l^{-+})|_{J=1}^{+-}\\
  &&+\frac{1}{2}(1_H^{--}\otimes 2_l^{-+})|_{J=1}^{+-}\Big]|(c\bar{c})\rangle|\gamma\rangle.
\end{eqnarray*}
Their $E1$ modes $\eta_c\gamma(E1)$ and
$\eta_{c2}(1^1D_2)\gamma(E1)$ are allowed due to their spin
structure
\begin{eqnarray*}
  |\eta_c\gamma(E1)\rangle &=& (0_H^{-+}\otimes 1_l^{+-})|_{J=1}^{--}|(c\bar{c})\rangle|\gamma\rangle, \\
  |\eta_{c2}(1^1D_2)\gamma(E1)\rangle &=& (0_H^{-+}\otimes
  1_l^{+-})|_{J=1}^{--}|(c\bar{c})\rangle|\gamma\rangle,
\end{eqnarray*}
which leads to the ratio
$\Gamma_{Z_c(3900)}(\eta_c\gamma(E1)):\Gamma_{Z_c(4020)}(\eta_c\gamma(E1))=1:1\,
(1:1.30)$.

%%%%%%%%%%%%%%%%%%%%%%%%%%%%%%%%%%%%%%
\subsection{$Y(4274)$}
%%%%%%%%%%%%%%%%%%%%%%%%%%%%%%%%%%%%%%

$Y(4274)$ was reported by CDF in the $J/\psi\phi$ invariant mass
spectrum \cite{Yi:2010aa}. It was proposed to be a S-wave
$D_s\bar{D}_{s0}(2317)$ molecular state with $J^{PC}=0^{-+}$
\cite{Liu:2010hf}. Its spin structure is
\begin{eqnarray*}
  |Y(4274)\rangle &=& %\frac{1}{\sqrt{2}}(|D_{s0}\bar{D}_s\rangle+|D_s\bar{D}_{s0}\rangle)\\
  %&=&
  \Big[-\frac{1}{2}(0_H^{-+}\otimes 0_l^{+-})|_{J=0}^{--}+\frac{1}{2}(1_H^{--}\otimes 1_l^{++})|_{J=0}^{--}\\
  &&+\frac{\sqrt{2}}{2}(1_H^{--}\otimes 1_l^{+-})|_{J=0}^{-+}\Big]|c\bar{s};\bar{c}s\rangle \\
  &&-\Big[\frac{1}{2}(0_H^{-+}\otimes 0_l^{+-})|_{J=0}^{--}-\frac{1}{2}(1_H^{--}\otimes 1_l^{++})|_{J=0}^{--}\\
  &&+\frac{\sqrt{2}}{2}(1_H^{--}\otimes 1_l^{+-})|_{J=0}^{-+}\Big]|\bar{c}s;c\bar{s}\rangle.\\
\end{eqnarray*}
With the heavy quark spin symmetry, the $h_c\gamma(E1)$ is
suppressed while $J/\psi\gamma(M1)$, $\psi(1^3D_1)\gamma(M1)$ and
$\psi(1^3D_2)\gamma(E2)$ are allowed, whose spin structures are
\begin{eqnarray*}
  |J/\psi\gamma(M1)\rangle &=& (1_H^{--}\otimes 1_l^{+-})|_{J=0}^{-+}|(c\bar{c})\rangle|\gamma\rangle, \\
  |h_c\gamma(E1)\rangle &=& (0_H^{-+}\otimes 0_l^{++})|_{J=0}^{-+}|(c\bar{c})\rangle|\gamma\rangle, \\
  |\psi(1^3D_1)\gamma(M1)\rangle &=& (1_H^{--}\otimes 1_l^{+-})|_{J=0}^{-+}|(c\bar{c})\rangle|\gamma\rangle,\\
  |\psi(1^3D_2)\gamma(E2)\rangle &=& (1_H^{--}\otimes 1_l^{+-})|_{J=0}^{-+}|(c\bar{c})\rangle|\gamma\rangle.
\end{eqnarray*}
The decay widths depend on the reduced matrix elements
$|\langle1,0\|H_{eff}(M1)\|1\rangle|$,
$|\langle1,2\|H_{eff}(M1)\|1\rangle|$,
$|\langle2,2\|H_{eff}(E2)\|1\rangle|$ respectively.

%%%%%%%%%%%%%%%%%%%%%%%%%%%%%%%%%%%%%%
\subsection{$Z(4430)$}
%%%%%%%%%%%%%%%%%%%%%%%%%%%%%%%%%%%%%%

$Z(4430)$ was suggested as a $D_1\bar{D}^\ast$ molecular state with
$J^p=0^-$ or a $D_1'\bar{D}^\ast$ with $J^P=0^-,1^-,2^-$
\cite{Liu:2008xz}. In the case that the neutral partner of $Z(4430)$
is a $D_1'\bar{D}^\ast$ molecular state with $2^{-+}$, its spin
structure is
\begin{eqnarray*}
  |Z(4430)\rangle &=&
  \Big[\frac{\sqrt{3}}{3}(1_H^{--}\otimes 1_l^{++})|_{J=2}^{--}+\frac{\sqrt{6}}{3}(1_H^{--}\otimes 1_l^{+-})|_{J=2}^{-+}\Big]\\
  &&|c\bar{q};\bar{c}q\rangle \\
  &&+\Big[-\frac{\sqrt{3}}{3}(1_H^{--}\otimes 1_l^{++})|_{J=2}^{--}+\frac{\sqrt{6}}{3}(1_H^{--}\otimes 1_l^{+-})|_{J=2}^{-+}\Big]\\
  &&|\bar{c}q;c\bar{q}\rangle.\\
\end{eqnarray*}
In the heavy quark spin symmetry limit, the allowed decay modes are
\begin{eqnarray*}
  |J/\psi\gamma(M1)\rangle &=& (1_H^{--}\otimes
  1_l^{+-})|_{J=2}^{-+}|(c\bar{c})\rangle|\gamma\rangle,
\end{eqnarray*}
\begin{eqnarray*}
  &&|\psi(1^3D_1)\gamma(M1)\rangle \\&&=
  \Big[\frac{1}{10}(1_H^{--}\otimes 1_l^{+-})|_{J=2}^{-+}-\frac{\sqrt{15}}{10}(1_H^{--}\otimes 2_l^{+-})|_{J=2}^{-+}\\
  &&\quad+\frac{\sqrt{21}}{5}(1_H^{--}\otimes 3_l^{+-})|_{J=2}^{-+}\Big]|(c\bar{c})\rangle|\gamma\rangle,\\
 && |\psi(1^3D_2)\gamma(M1)\rangle \\&&=
  \Big[-\frac{\sqrt{15}}{10}(1_H^{--}\otimes 1_l^{+-})|_{J=2}^{-+}+\frac{5}{6}(1_H^{--}\otimes 2_l^{+-})|_{J=2}^{-+}\\
  &&\quad+\frac{\sqrt{35}}{15}(1_H^{--}\otimes 3_l^{+-})|_{J=2}^{-+}\Big]|(c\bar{c})\rangle|\gamma\rangle,\\
 && |\psi(1^3D_3)\gamma(M1)\rangle \\&&=
  \Big[\frac{\sqrt{21}}{5}(1_H^{--}\otimes 1_l^{+-})|_{J=2}^{-+}+\frac{\sqrt{35}}{15}(1_H^{--}\otimes 2_l^{+-})|_{J=2}^{-+}\\
  &&\quad+\frac{1}{15}(1_H^{--}\otimes 3_l^{+-})|_{J=2}^{-+}\Big]|(c\bar{c})\rangle|\gamma\rangle.
\end{eqnarray*}
\begin{eqnarray*}
 && |\psi(1^3D_1)\gamma(E2)\rangle \\&&=
  \Big[\frac{3}{10}(1_H^{--}\otimes 1_l^{+-})|_{J=2}^{-+}-\frac{\sqrt{35}}{10}(1_H^{--}\otimes 2_l^{+-})|_{J=2}^{-+}\\
  &&\quad+\frac{\sqrt{14}}{5}(1_H^{--}\otimes 3_l^{+-})|_{J=2}^{-+}\Big]|(c\bar{c})\rangle|\gamma\rangle,\\
 && |\psi(1^3D_2)\gamma(E2)\rangle \\&&=
  \Big[-\frac{\sqrt{35}}{10}(1_H^{--}\otimes 1_l^{+-})|_{J=2}^{-+}+\frac{1}{2}(1_H^{--}\otimes 2_l^{+-})|_{J=2}^{-+}\\
  &&\quad+\frac{\sqrt{10}}{5}(1_H^{--}\otimes 3_l^{+-})|_{J=2}^{-+}\Big]|(c\bar{c})\rangle|\gamma\rangle,\\
  &&|\psi(1^3D_3)\gamma(E2)\rangle \\&&=
  \Big[\frac{\sqrt{14}}{5}(1_H^{--}\otimes 1_l^{+-})|_{J=2}^{-+}+\frac{\sqrt{10}}{5}(1_H^{--}\otimes 2_l^{+-})|_{J=2}^{-+}\\
  &&\quad+\frac{1}{5}(1_H^{--}\otimes 3_l^{+-})|_{J=2}^{-+}\Big]|(c\bar{c})\rangle|\gamma\rangle.
\end{eqnarray*}
We obtain the following $M1$ and $E2$ transition ratios
\begin{eqnarray*}
  &&\Gamma(\psi(1^3D_1)\gamma(M1)):\Gamma(\psi(1^3D_2)\gamma(M1)):\Gamma(\psi(1^3D_3)\gamma(M1))\\
  &&=1:15:84, \\
  &&\Gamma(\psi(1^3D_1)\gamma(E2)):\Gamma(\psi(1^3D_2)\gamma(E2)):\Gamma(\psi(1^3D_3)\gamma(E2))\\
  &&=9:35:56.
\end{eqnarray*}

%%%%%%%%%%%%%%%%%%%%%%%%%%%%%%%%%%%%%%
\subsection{$Y(3940)$ and $Y(4140)$}
%%%%%%%%%%%%%%%%%%%%%%%%%%%%%%%%%%%%%%

$Y(3940)$ was announced by BaBar collaboration \cite{Abe:2004zs} in
$B\to K J/\psi \omega$ while $Y(4140)$ was observed by CDF
collaboration \cite{Aaltonen:2009tz}. Due to their similarity,
$Y(3940)$ and $Y(4140)$ were proposed as the candidates of the
$D^\ast \bar{D}^\ast$ and $D_s^\ast \bar{D}_s^\ast$ molecular states
\cite{Liu:2009ei,Liu:2008tn}, respectively, where their quantum
numbers are either $J^{PC}=0^{++}$ or $2^{++}$
\cite{Liu:2009ei,Liu:2008tn}.

%%%%%%%%%%%%%%%%%%%%%%%%%%%%%%%%%%%%%%
\subsubsection{$J^{PC}=0^{++}$}
%%%%%%%%%%%%%%%%%%%%%%%%%%%%%%%%%%%%%%

If the quantum numbers of $Y(3940)$ and $Y(4140)$ are $0^{++}$,
their spin structures are
\begin{eqnarray*}
  |Y(3940)\rangle &=&
  \Big[\frac{\sqrt{3}}{2}(0_H^{-+}\otimes 0_l^{-+})|_{J=0}^{++}-\frac{1}{2}(1_H^{--}\otimes 1_l^{--})|_{J=0}^{++}\Big]\\
  &&|c\bar{q};\bar{c}q\rangle,\\
  |Y(4140)\rangle &=&
  \Big[\frac{\sqrt{3}}{2}(0_H^{-+}\otimes 0_l^{-+})|_{J=0}^{++}-\frac{1}{2}(1_H^{--}\otimes 1_l^{--})|_{J=0}^{++}\Big]\\
  &&|c\bar{s};\bar{c}s\rangle.
\end{eqnarray*}
In the heavy quark limit, the allowed decay modes are
\begin{eqnarray*}
  |J/\psi\gamma(E1)\rangle &=& (1_H^{--}\otimes 1_l^{--})|_{J=0}^{++}|(c\bar{c})\rangle|\gamma\rangle, \\
  |h_c\gamma(M1)\rangle &=& (0_H^{-+}\otimes 0_l^{-+})|_{J=0}^{++}|(c\bar{c})\rangle|\gamma\rangle, \\
  |\psi(1^3D_2)\gamma(E1)\rangle &=& (1_H^{--}\otimes 1_l^{--})|_{J=0}^{++}|(c\bar{c})\rangle|\gamma\rangle.
\end{eqnarray*}
We have
\begin{eqnarray*}
  &&\Gamma_{Y(3940)}(J/\psi\gamma(E1)):\Gamma_{Y(4140)}(J/\psi\gamma(E1))\\
  &&=1:1\, (1:1.6),\\
  &&\Gamma_{Y(3940)}(h_c\gamma(M1)):\Gamma_{Y(4140)}(h_c\gamma(M1))\\
  &&=1:1\, (1:2.8),\\
  &&\Gamma_{Y(3940)}(\psi(1^3D_2)\gamma(E1)):\Gamma_{Y(4140)}(\psi(1^3D_2)\gamma(E1))\\
  &&=1:1.
  \end{eqnarray*}

%%%%%%%%%%%%%%%%%%%%%%%%%%%%%%%%%%%%%%
\subsubsection{$J^{PC}=2^{++}$}
%%%%%%%%%%%%%%%%%%%%%%%%%%%%%%%%%%%%%%

If $Y(3940)$ and $Y(4140)$ are the molecular candidates with
$J^{PC}=2^{++}$, their spin structures are
\begin{eqnarray*}
  |Y(3940)\rangle &=& (1_H^{--}\otimes1_l^{--})|_{J=2}^{++}|c\bar{q};\bar{c}q\rangle, \\
  |Y(4140)\rangle &=&
  (1_H^{--}\otimes1_l^{--})|_{J=2}^{++}|c\bar{s};\bar{c}s\rangle,
\end{eqnarray*}
respectively. In the heavy quark symmetry limit, the allowed decay
modes are $J/\psi\gamma(E1)$, $\psi(1^3D_1)\gamma(E1)$,
$\psi(1^3D_2)\gamma(E1)$ and $\psi(1^3D_3)\gamma(E1)$. Their spin
structures are
\begin{eqnarray*}
  |J/\psi\gamma(E1)\rangle &=& (1_H^{--}\otimes 1_l^{--})|_{J=2}^{++}|(c\bar{c})\rangle|\gamma\rangle,
\end{eqnarray*}
\begin{eqnarray*}
 && |\psi(1^3D_1)\gamma(E1)\rangle\nonumber\\ &&= [\frac{1}{10}(1_H^{--}\otimes1_l^{--})|_{J=2}^{++}-\frac{\sqrt{15}}{10}(1_H^{--}\otimes2_l^{--})|_{J=2}^{++}\\
  &&\quad+\frac{\sqrt{21}}{5}(1_H^{--}\otimes3_l^{--})|_{J=2}^{++}]|(c\bar{c})\rangle|\gamma\rangle, \\
 && |\psi(1^3D_2)\gamma(E1)\rangle \\&&= [-\frac{\sqrt{15}}{10}(1_H^{--}\otimes1_l^{--})|_{J=2}^{++}+\frac{5}{6}(1_H^{--}\otimes2_l^{--})|_{J=2}^{++}\\
  &&\quad+\frac{\sqrt{35}}{15}(1_H^{--}\otimes3_l^{--})|_{J=2}^{++}]|(c\bar{c})\rangle|\gamma\rangle, \\
 && |\psi(1^3D_3)\gamma(E1)\rangle \\&&= [\frac{\sqrt{21}}{5}(1_H^{--}\otimes1_l^{--})|_{J=2}^{++}-\frac{\sqrt{35}}{15}(1_H^{--}\otimes2_l^{--})|_{J=2}^{++}\\
  &&\quad+\frac{1}{15}(1_H^{--}\otimes3_l^{--})|_{J=2}^{++}]|(c\bar{c})\rangle|\gamma\rangle.
\end{eqnarray*}
We obtain the following $E1$ transition ratio of $Y(3940)$ and
$Y(4140)$
\begin{eqnarray*}
  &&\Gamma(\psi(1^3D_1)\gamma(E1)):\Gamma(\psi(1^3D_2)\gamma(E1)):\Gamma(\psi(1^3D_3)\gamma(E1))\\
  &&=1:15:84.
\end{eqnarray*}

%%%%%%%%%%%%%%%%%%%%%%%%%%%%%%%%%%%%%%%%%
\section{Summary}\label{sec5}
%%%%%%%%%%%%%%%%%%%%%%%%%%%%%%%%%%%%%%%%%

In the past decade many charmonium-like and bottomonium-like states
were reported experimentally. These states are sometimes called the
$XYZ$ states. Many $XYZ$ states are very close to the open-charm or
open-bottom threshold. Some of them are even charged. Right now, it
is difficult to accommodate these $XYZ$ states within the simple
quark model spectrum. Especially those charged states can not be a
charmonium or bottomonium.

Many theoretical speculations were proposed to explain the inner
structures of these $XYZ$ states. Among them, the hadronic molecular
picture becomes quite popular due to the closeness of these $XYZ$
states to the open-charm or open-bottom threshold. The molecule
scheme seems quite natural since we all know that the deuteron is a
very loosely bound molecular state composed of a proton and neutron.
It is very intriguing to explore whether the loosely bound di-meson
molecular states exist or not. The dynamical calculation of the
di-meson system within the framework of the meson exchange model may
partly answer whether there exists attraction between the two mesons
and whether the interaction is strong enough. The decay and
production processes will provide additional information of these
$XYZ$ state.

In the heavy quark limit, the interaction of the heavy quark with
both the gluon and photon does not flip the heavy quark spin. The
conservation of the heavy quark spin greatly simplifies the analysis
of the decay and production processes of heavy flavored hadrons. In
this work we employ the heavy quark symmetry and adopt the spin
rearrangement approach to study the radiative decay pattern of the
possible molecular(resonant) states composed of a pair of heavy
mesons. We use the hidden-beauty molecules(resonances) to illustrate
the formalism. We have extensively investigated three classes of the
radiative decays: $\mathfrak{M}\to (b\bar{b})+\gamma$,
$(b\bar{b})\to \mathfrak{M}+\gamma$, $ \mathfrak{M} \to
\mathfrak{M}^\prime+\gamma$, corresponding to the electromagnetic
transitions between one molecular(resonant) state and bottomonium,
one bottomonium and molecular(resonant) state, and two
molecular(resonant) states respectively. We have also extended the
same formalism to study the radiative decays of the
molecular(resonant) states with hidden charm.

If the initial or final states belong to the same spin flavor
multiplet, their spatial wave functions are the same. Then we can
derive some model independent ratios of the radiative decay widths
between different channels in the heavy quark limit. These ratios
are different under different assumptions of the underlying
structures of these $XYZ$ states. Experimental measurement of the
radiative decay ratios of the $XYZ$ states may test different
theoretical scenarios and help unveil their inner structures.

The application of the method adopted in this work is sometimes
limited \cite{Voloshin:2012dk}. The spin assignments are valid for
the states near the thresholds. If the mass of the state is distant
from the thresholds, the mixing with the states which carry the same
quantum numbers but have different spin structures is
non-negligible. The isoscalar molecular states may couple to the
usual heavy quarkonia such as $\Upsilon$ and $\eta_b$. In addition,
the simple spin assignments of the bottomonia in Eqs. (3)-(12)
sometimes fail. Some excited charmonia and bottomonia seem to
violate the spin selection rule based on such simple assignments
\cite{Voloshin:2012dk} \footnote{We would like to thank the
anonymous referee for indicating this point.}

There exist speculations that some of the $XYZ$ states may arise
from either the pure kinematical effect or final state interaction
(FSI). For example, if $Z_c(3900)$ or $Z_c(4020)$ is a kinematical
artifact, their neutral component will not emit a photon and decay
into the charmonium. If $Z_c(3900)$ or $Z_c(4020)$ arises from the
final state interaction through the triangle diagram where a charmed
meson is exchanged, the spin decomposition of this FSI signal may be
different from the spin decomposition of the initial two mesons or
$Z_c(3900)$ or $Z_c(4020)$ as the molecular candidates. The
resulting radiative pattern of this FSI signal will deviate from
that of the molecular states discussed in this work.

In this work, we focus on the radiative decays of the $XYZ$ states.
In fact, the strong decay behavior is also very important to reveal
their underlying structures. In a recent work, we have carried out
the calculation of the strong decay of the $XYZ$ states
\cite{Ma:2014zva}.

In short summary, the radiative decay ratios of the $XYZ$ states may
encode important information of their underlying spin structures,
which can reflect their possible inner structures to some extent.
Systematical experimental measurement of these ratios will help us
to understand the properties of these observed $XYZ$ states.
Hopefully the present extensive investigations will be useful to the
understanding of the future radiative data.

\subsection*{Acknowledgments}

This project is supported by the National Natural Science Foundation
of China under Grants No. 11222547, No. 11175073, No. 11035006, No.
11375240 and No. 11261130311, the Ministry of Education of China
(FANEDD under Grant No. 200924, SRFDP under Grant No. 2012021111000,
and NCET), the China Postdoctoral Science Foundation under Grant No.
2013M530461, the Fok Ying Tung Education Foundation (Grant No.
131006).

\end{document}